\documentclass[aps,reprint,amsmath,amssymb,pre,longbibliography,superscriptaddress]{revtex4-2}
\usepackage[utf8]{inputenc}
\usepackage{bm}
\usepackage{graphicx}
    \usepackage{color}
\usepackage{hyperref}
\usepackage[hyphenbreaks]{breakurl}

\newcommand\beq{\begin{equation}}
\newcommand\eeq{\end{equation}}
\newcommand\beqa{\begin{eqnarray}}
\newcommand\eeqa{\end{eqnarray}}
\newcommand{\nn}{\nonumber\\}
\def\bal#1\eal{\begin{align}#1\end{align}}
\def\zero{{(0)}}
\def\one{{(1)}}
\def\two{{(2)}}

\newcommand{\alphab}{\xi}
\newcommand{\nq}{{n_s}}

\newcommand\Nc{N_c}

\newcommand{\q}{q}

\newcommand{\MD}{\text{MD}}
\newcommand{\talpha}{\alpha}
\newcommand{\tomega}{\omega}
\newcommand{\tdelta}{\delta}

\newcommand{\tA}{A}
\newcommand{\tr}{r}

\newcommand{\dd}{d}

\begin{document}

\title{Structural properties of additive binary hard-sphere mixtures. II.
Asymptotic behavior and structural crossovers}
\author{S{\l}awomir Pieprzyk}
\email{slawomir.pieprzyk@ifmpan.poznan.pl}
\affiliation{Institute of Molecular Physics, Polish Academy of Sciences, M. Smoluchowskiego
17, 60-179 Pozna\'n, Poland}
\author{Santos B. Yuste}
\email{santos@unex.es}
\homepage{http://www.unex.es/eweb/fisteor/santos/}
\author{Andr\'es Santos}
\email{andres@unex.es}
\homepage{http://www.unex.es/eweb/fisteor/andres/}
\affiliation{Departamento de F\'{\i}sica  and Instituto de Computaci\'on Cient\'{\i}fica Avanzada (ICCAEx), Universidad de
Extremadura, Badajoz, E-06006, Spain}
\author{Mariano L\'{o}pez de Haro}
\email{malopez@unam.mx}
\affiliation{Instituto de Energ\'{\i}as Renovables, Universidad Nacional Aut\'onoma de M\'exico (U.N.A.M.),
Temixco, Morelos 62580, M{e}xico}
\author{Arkadiusz C. Bra\'nka}
\email{branka@ifmpan.poznan.pl}
\affiliation{Institute of Molecular Physics, Polish Academy of Sciences, M. Smoluchowskiego
17, 60-179 Pozna\'n, Poland}

\begin{abstract}
The structural properties of additive binary hard-sphere mixtures are addressed as a follow-up of a previous paper [S. Pieprzyk \emph{et al.}, Phys.\ Rev.\ E \textbf{101}, 012117 (2020)]. The so-called rational-function
approximation method and an approach combining accurate molecular dynamics simulation data, the pole structure representation of the total correlation functions, and the Ornstein--Zernike equation are considered. The density, composition, and size-ratio dependencies of the leading poles of the Fourier transforms of the total correlation functions $h_{ij} (r)$ of such mixtures are presented, those poles  accounting for the asymptotic decay of $h_{ij} (r)$ for large $r$.
Structural crossovers,  in which the asymptotic  wavelength of the oscillations of the total correlation functions  changes discontinuously, are investigated. The behavior of the structural crossover lines as the size ratio and densities of the two species  are changed is also discussed.
\end{abstract}

\date{\today}


\maketitle

\section{Introduction}
\label{intro}
The close connection between the thermodynamic properties and the structural correlation functions of fluids in the statistical-mechanical formulation of liquid state theory is well known. In particular, for a fluid mixture of $\Nc$ components, the virial route to the equation of state  leads to  \cite{BH76,HM13,S16}
\bal
\label{presMul}
Z\equiv &\frac{p}{\rho k_B T}\nn
=&1-\frac{\rho}{6k_BT}
\sum_{i,j=1}^{\Nc}x_i x_j\int \dd\mathbf{r}\, g_{ij}(r) r \frac{\partial \phi_{ij}(r)}{\partial r},
\eal
where $Z$ is the compressibility factor, $p$ is the pressure, $\rho$ is the number density, $k_B$ is the Boltzmann constant,  $T$ is the absolute temperature, $x_i$ is the mole fraction of molecules of species $i$, $\phi_{ij}(r)$ is the interaction potential (assumed to be  spherically symmetric and  pairwise additive) between a particle of species $i$ and a particle of species $j$,  and $g_{ij}(r)$ is the radial distribution function
(RDF), which is a measure of the probability of finding a molecule of species $i$ at a distance $r$ from another molecule of species $j$.
In the case of hard spheres, Eq.\ \eqref{presMul} reduces to
\begin{equation}
Z=1+\frac{2}{3}\pi \rho\sum_{i,j=1}^{\Nc}
x_{i}x_{j}{\sigma_{ij} } g_{ij}(\sigma_{ij}),
\label{1}
\end{equation}
where $\sigma_{ij}$ is the hard-core diameter of the interaction between a sphere of species $i$ and another sphere of species $j$, and $g_{ij}(\sigma_{ij})$ is the contact value of the RDF.

On the other hand, from the compressibility route to the equation of state one gets the relationship  \cite{S16}
\bal
\label{CompresMul}
\chi^{-1}\equiv&\frac{1}{k_BT}\left(\frac{\partial p}{\partial \rho}\right)_T\nn
=&\sum_{i,j=1}^{\Nc}\sqrt{x_i x_j}\left[\mathsf{I}+\widehat{\mathsf{h}}(0)\right]^{-1}_{ij}\nn
=&
1-\rho\sum_{i,j=1}^{\Nc}{x_i x_j}\widetilde{c}_{ij}(0),
\eal
where $\mathsf{I}$ is the $\Nc\times \Nc$ identity matrix and the element
$\widehat{h}_{ij}(k)\equiv \rho \sqrt{x_i x_j}\widetilde{h}_{ij}(k) $ of the matrix $\widehat{\mathsf{h}}(k)$ is proportional to the Fourier transform
\beq
\label{htildeMul}
\widetilde{h}_{ij}(k)=\int \dd\mathbf{r}\,e^{-\imath\mathbf{k}\cdot\mathbf{r}}h_{ij}\left({r}\right)
\eeq
of the total correlation functions $h_{ij}({r})\equiv g_{ij}(r)-1$, \linebreak $\imath$ being the imaginary unit. Further, in the last equality of
Eq.\ (\ref{CompresMul}),  $\widetilde{c}_{ij}(0)$ is the zero wavenumber limit of the Fourier transform
\beq
\label{htildec}
\widetilde{c}_{ij}(k)=\int \dd\mathbf{r}\,e^{-\imath\mathbf{k}\cdot\mathbf{r}}c_{ij}\left({r}\right)
\eeq
of the direct correlation function $c_{ij}({r})$. The latter is defined through the Ornstein--Zernike (OZ) relation, namely,
\begin{subequations}
\begin{equation}
 \label{OZMulti}
  h_{ij}(r_{12})=c_{ij}(r_{12})+\rho\sum_{\ell=1}^{\Nc} x_\ell \int \dd \mathbf{r}_3\, c_{i\ell}(r_{13})h_{\ell j}(r_{23}),
  \end{equation}
\begin{equation}
\label{Eq:OZ1}
\widetilde{h}_{ij}(k)  = \widetilde{c}_{ij}(k) +  \rho \sum_{\ell=1}^{\Nc}  x_\ell\widetilde{c}_{i\ell}({k}) \widetilde{h}_{\ell j}({k}),
\end{equation}
  \end{subequations}
in real and Fourier spaces, respectively.

Apart from this thermodynamic connection, it has been further established
that abrupt changes in the structural correlation functions of a fluid may also show up in its phase behavior. This is the case of the Fisher--Widom line of simple fluids  \cite{FW69,EHHPS93,VRL95,B96,DE00,TCV03,SBHPG16,HRYS18}, which  distinguishes between the  region  where  the  large-$r$
behavior of the total  correlation  function shows damped oscillatory decay (typical of  dense  and/or high-temperature liquids)  and the region where  the nature of the decay  is  monotonic (typical of low-density gases, low-temperature liquids, and near-critical fluids). Structural transitions such as the one related to the Fisher--Widom line, including the physics behind them, are clearly of interest but their study is hampered by the lack of exact results for the structural correlation functions and, despite the availability of interesting work on this subject, additional efforts are clearly required, especially in the case of mixtures.

In a previous paper \cite{PBYSH20}, hereafter referred to as paper I,  we presented a method (denoted as the WM scheme) allowing us to obtain an accurate representation of the structural correlation functions of additive binary hard-sphere (BHS) mixtures. The WM method successfully combines molecular dynamics (MD)  simulation data, residue-theorem analysis, and the OZ relations, additionally taking into account the tail parts of the structural correlation functions, without using any approximate closures. In particular, by considering a mixture with a fixed diameter ratio of $0.648$ and a fixed total packing fraction of $0.5$ (which was the system analyzed previously theoretically and through experimental data by Statt \emph{et al.} \cite{SPTER16}),  we confirmed in paper I the presence of structural crossovers in such a mixture and examined the role played in the crossover by the first two poles of the Fourier transforms of the total correlation functions. We also found very good agreement between the results of
the new WM method and those obtained from the use of the rational-function approximation (RFA)  \cite{YSH98,HYS08,S16,SYH20} to compute analytically the total correlation functions, as well as an improvement of the agreement between the RFA and WM results and the ones derived from experimental data when compared to the analysis performed in Ref.\  \cite{SPTER16}.

The aim of the present paper is two-fold. On the one hand, to consolidate the RFA approach as a valuable tool to investigate asymptotic behavior and structural crossover issues, by testing such approach against the results of the WM method. On the other hand, to carry out a more thorough analysis of the role of the pole structure of $\widetilde{h}_{ij}(k)$ on the asymptotic behavior of $h_{ij}(r)$ and the structural crossovers in these functions by considering various BHS mixtures.

The paper is organized as follows. The system of interest (BHS mixtures) is briefly described in Sec.\ \ref{sec2}, where also succinct accounts of
the RFA approach and of the WM  scheme developed in paper I, as well as some  details of our MD simulations, are presented.  In Sec.\ \ref{sec3} we provide the basics of the analysis of the poles of the total correlation functions, while in Sec.\ \ref{sec4} we provide the results of our calculations and an illustration of our main findings. The paper closes in
Sec.\ \ref{concl} with some concluding remarks.

\section{Structural properties of a binary hard-sphere mixture} \label{sec2}

\subsection{System}
Let us consider a binary ($\Nc=2$) fluid mixture  of ``small'' (label $s$) and ``big'' (label $b$) hard spheres.   The {additive}  hard core of the interaction between a sphere of species $i$ and a sphere of species $j$ ($i,j=s,b$) is $\sigma_{ij}=\frac{1}{2
}(\sigma _{i}+\sigma _{j})$, where the diameter of a sphere of
species $i$ is $\sigma _{ii}=\sigma _{i}$. Let the number density of
the mixture be $\rho $, the mole fraction of species $i$ be
$x_{i}=\rho _{i}/\rho $ (where $\rho_i=N_i/V$ is the partial number density, $N_i$ and $V$ being the number of particles of species $i$ and  the volume of the system,  respectively), and let the size ratio be $\q=\sigma_s/\sigma_b\leq 1$. From these quantities one can define the partial packing fractions $\eta_i =\frac{\pi}{6}\rho_i \sigma_i^3$ and the total packing fraction  $\eta =\eta_s+\eta_b=\frac{\pi}{6}\rho \sigma_b^3 (x_s \q^3  + x_b)$. Note that in this system  there are three characteristic separations between particles at contact, namely the small-small particle separation, $\sigma_s=q\sigma_b$, the small-big particle separation, $\sigma_{sb}=\frac{1}{2}\sigma_b(1+q)$, and the big-big particle separation, $\sigma_b$.

\subsection{The rational-function approximation method}
In order to examine the structural properties, we shall now sketch the RFA approach to obtain the structural properties of additive hard-sphere mixtures. The detailed description of such an approach  may be found in Refs.\  \cite{YSH98,HYS08,S16,SYH20}. First, the Laplace transforms of $r g_{ij}(r)$ are introduced:
\beq
\label{3.1}
G_{ij}(z)\equiv\int_0^\infty d r\, e^{-zr}r g_{ij}(r).
\eeq
Next, an explicit form for $G_{ij}(z)$ in terms of a free parameter $\alphab$ and an $\Nc\times\Nc$ matrix $\mathsf{L}(z)=\mathsf{L}^\zero+\mathsf{L}^\one z+\mathsf{L}^\two z^2$  is proposed. Then, by imposing certain
consistency conditions, the elements
of the matrices $\mathsf{L}^\zero$, $\mathsf{L}^\one$, $\mathsf{L}^\two$ are expressed as linear functions of $\alphab$. In particular, $L_{ij}^\two=2\pi\alphab\sigma_{ij} g_{ij}(\sigma_{ij})$.

Interestingly, the simple choice $\alphab=0$, and hence $L_{ij}^\two=0$, gives the Percus--Yevick (PY) solution  \cite{L64,BH76}, which is known to yield different equations of state via the virial [Eq.\ \eqref{1}] and compressibility [Eq.\ \eqref{CompresMul}] routes. However, by an appropriate determination of $\alphab\neq 0$, the RFA can be made thermodynamically consistent  and, additionally, allows one to freely choose the contact values $g_{ij}(\sigma_{ij})$. A convenient choice is provided by the
 popular BGHLL expression  proposed independently by Boubl\'ik \cite{B70}, Grundke and Henderson \cite{GH72}, and Lee and Levesque \cite{LL73}.

Clearly, once $G_{ij}(z)$ has been determined, inverse Laplace transformation
directly yields $rg_{ij}(r)$ and hence also $h_{ij}(r)$. On the other hand, explicit knowledge of $G_{ij}(z)$ also allows us to determine
the Fourier transform $\widetilde{h}_{ij}(k)$ through the relation
\beq
\widetilde{h}_{ij}(k)=-2\pi \left.\frac{G_{ij}(z)-G_{ij}(-z)}{z}
\right|_{z=\imath k}.
\label{1.7}
\end{equation}

In the RFA (as well as in the PY approximation), $G_{ij}(z)$ is obtained from the inner product of the matrix $\mathsf{L}(z)$ and the inverse of another related matrix $\mathsf{B}(z)$. Therefore,  the Laplace transforms
$G_{ij}(z)$ for all the pairs $ij$ share the same poles, namely the zeros
of the determinant $D(z)$ of $\mathsf{B}(z)$. In the particular case of a
binary mixture, the functional form of $D(z)$ is
\bal
\label{D(s)}
D(z)=&\mathcal{P}_6^{(0)}(z)+\mathcal{P}_4^{(s)}(z)e^{-\sigma_s z}+\mathcal{P}_4^{(b)}(z)e^{-\sigma_b z}\nn
&+\mathcal{P}_2^{(sb)}(z)e^{-2\sigma_{sb}z},
\eal
where $\mathcal{P}_6^{(0)}(z)$, $\mathcal{P}_4^{(s)}(z)$, $\mathcal{P}_4^{(b)}(z)$, and $\mathcal{P}_2^{(sb)}(z)$ are polynomials of degrees $6$, $4$, $4$, and $2$, respectively.
In the PY case ($\alphab=0$), the degrees of those polynomials decrease
in two units, i.e., $\mathcal{P}_6^{(0)}(z)\to \mathcal{P}_4^{(0)}(z)$, $\mathcal{P}_4^{(s)}(z)\to \mathcal{P}_2^{(s)}(z)$, $\mathcal{P}_4^{(b)}(z)\to \mathcal{P}_2^{(b)}(z)$, and $\mathcal{P}_2^{(sb)}(z)\to \text{const}$. A basic property of $D(z)$ is $D(0)=0$, which is tied to the physical condition $\lim_{r\to\infty}g_{ij}(r)=1$.

It should be stressed that perhaps the most valuable asset of the RFA approach is that, apart from ensuring thermodynamic consistency, it leads to
explicit analytic expressions for all the structural properties of the BHS mixture  \cite{note_21_03}. Additionally, the asymptotic long-range behavior of $h_{ij}(r)$ is directly obtained from the  roots of Eq.\ \eqref{D(s)}.

\subsection{The WM method}

The WM method proposed in paper I \cite{PBYSH20} allows us to obtain the structural properties of additive BHS mixtures by  combining accurate MD simulation data, the pole structure representation of the total correlation functions, and the OZ equation.

In the method, a semi-empirical approximation for the structural properties of additive BHS mixtures  is constructed by considering the following analytic form of $h_{ij}({r})$:
\begin{equation}
\label{Eq:hrWM}
h^{WM}_{ij}({r}) =
\begin{cases}
-1,  \qquad 0 < \tr < \sigma_{ij}, \\
\sum\limits_{n=1}^{W} b_{ij}^{(n)} \tr^{n-1},\qquad \sigma_{ij} < \tr \leq \tr_{ij}^{{\min}}, \\
\sum\limits_{n=1}^{M} \frac{\tA_{ij}^{(n)}}{\tr} e^{-\talpha_{n} \tr} \sin\left(\tomega_{n} \tr + \tdelta_{ij}^{(n)}\right) ,\quad  r\geq\tr_{ij}^{{\min}}.
\end{cases}
\end{equation}

The parameters $\{ b_{ij}^{(1)}, b_{ij}^{(2)}, \ldots, b_{ij}^{(W)} \}$ and $\{ \tA_{ij}^{(1)}, \talpha_{1}, \tomega_{1}, \tdelta_{ij}^{(1)}, \ldots, \tA_{ij}^{(M)}, \talpha_{M}, \tomega_{M},\tdelta_{ij}^{(M)} \}$ are obtained by enforcing the BGHLL contact values and the continuity conditions at $r=r_{ij}^{{\min}}$, and  by a nonlinear fitting procedure based on the minimization  of
$\left\vert h_{ij}^{WM}(r)-h_{ij}^{\MD}(r) \right\vert$ for each $r/\sigma_{ij} \in (1,r_c^*)$, where $h_{ij}^{\MD}(r)$ is obtained from our MD simulations, the details of which will be specified below.
In our fitting procedure, the values of $|h_{ij}^{WM}(r)-h_{ij}^\text{MD}(r)|$ for $1<r/\sigma_{ij}<r_c^*$ were required to be smaller than $10^{-3}$.
The suitable value of $r_c^*$ depends on the size ratio $q$ and we took $r_c^*=5$, $4$, $3$, and $3$ for $q= 0.648$, $0.4$, $0.3$, and $0.2$,
respectively. This value is connected with the half-length of the simulation box. Note that decreasing $q$ causes a decrease of the available space (smaller simulation box). Therefore, in order to carry out simulations and increase the simulation box with the assumed $r_c^*$ values, it was necessary to add more spheres in the cases $q= 0.3$ and $0.2$.
For our calculations an appropriate choice for $\tr_{ij}^{{\min}}$ was the position of the first minimum of $h_{ij}({r})$.
Also, as in paper I, for the subsequent calculations we will usually take
$W=15$ and $M=10$.

\begin{table}
   \caption{BHS mixtures investigated by MD in this work.}\label{table1}
\begin{ruledtabular}
\begin{tabular}{lll}
$q$&$\eta_b$&$\eta_s$\\
\hline
$0.648$&$0.10$&$0.20$, $0.25$, $0.30$, $0.35$, $0.40$\\
&$0.20$&$0.15$, $0.20$, $0.25$, $0.30$, $0.35$\\
$0.4$&$0.05$&$0.05$, $0.10$, $0.15$, $0.20$, $0.25$, $0.30$, $0.35$\\
&$0.20$&$0.05$, $0.10$, $0.15$, $0.20$, $0.25$\\
$0.3$&$0.05$&$0.10$, $0.15$, $0.20$, $0.25$, $0.30$, $0.35$\\
&$0.20$&$0.10$, $0.15$, $0.20$\\
$0.2$&$0.02$&$0.25$\\
&$0.20$&$0.10$\\
\end{tabular}
\end{ruledtabular}
\end{table}

\subsection{Details of the molecular dynamics simulations}

\begin{figure*}
\hspace{0.6cm} $q = 0.2$ \hspace{3.1cm} $q = 0.3$ \hspace{3.1cm} $q =
0.4$ \hspace{3.1cm} $q = 0.648$\\
\vspace{0.2cm}
\includegraphics[width=0.22\textwidth]{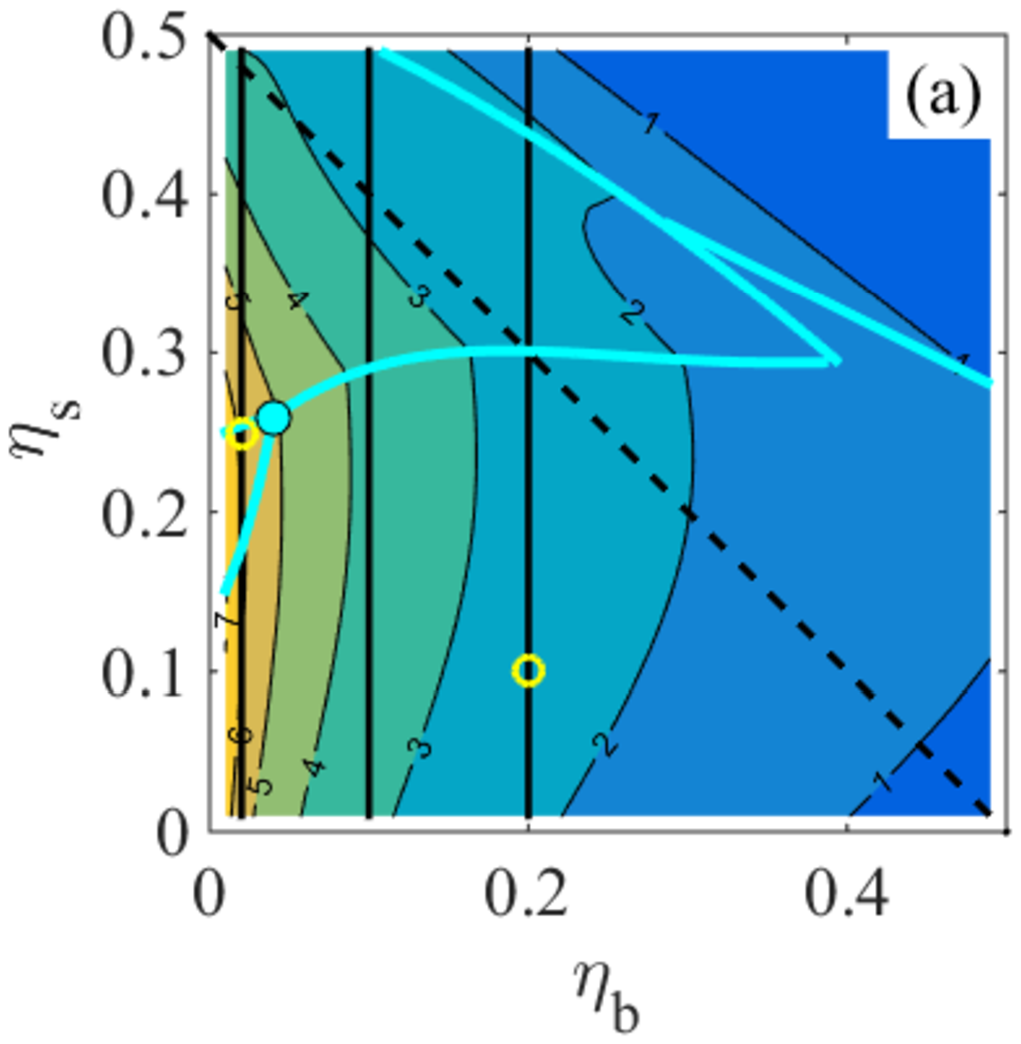}\hspace{0.5cm}
\includegraphics[width=0.22\textwidth]{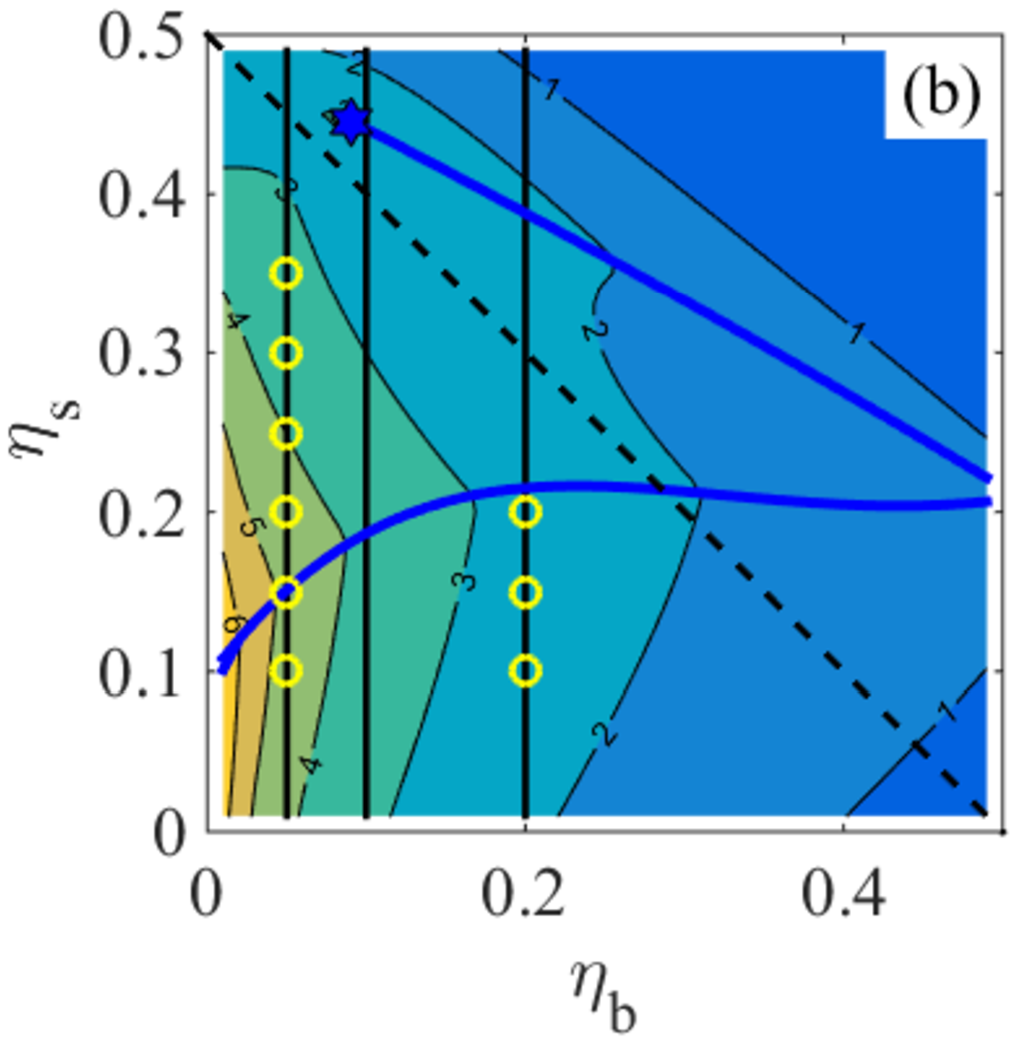}\hspace{0.5cm}
\includegraphics[width=0.22\textwidth]{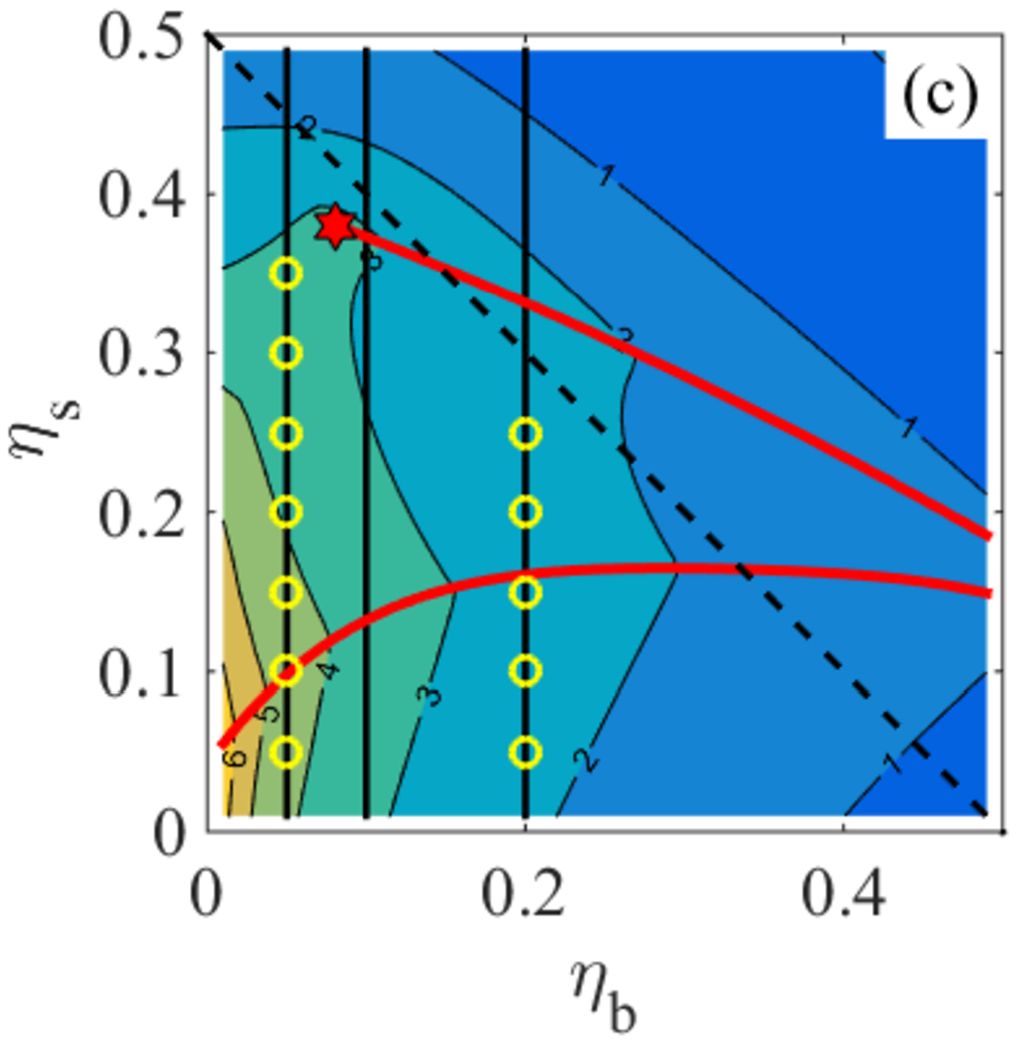}\hspace{0.5cm}
\includegraphics[width=0.22\textwidth]{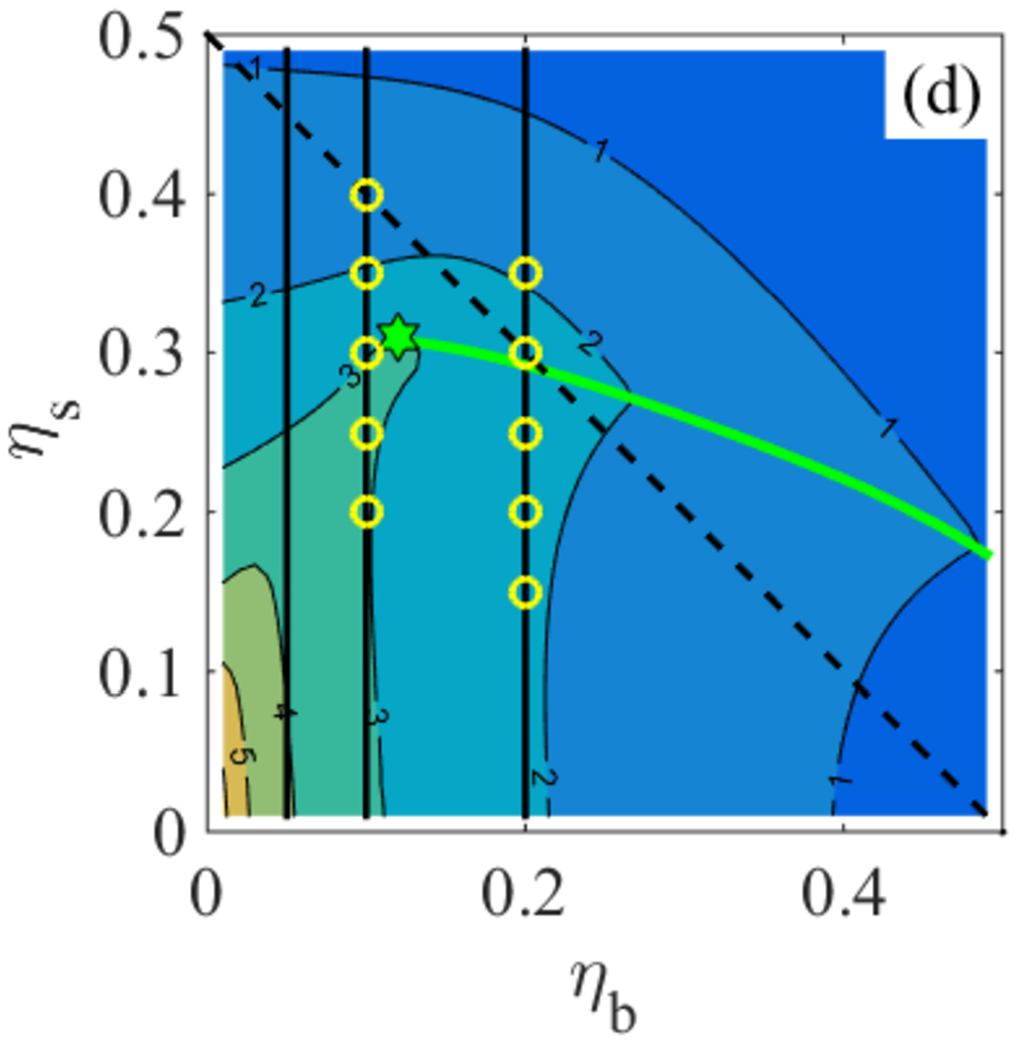}
\vspace{0.2cm}\\
\includegraphics[width=0.22\textwidth]{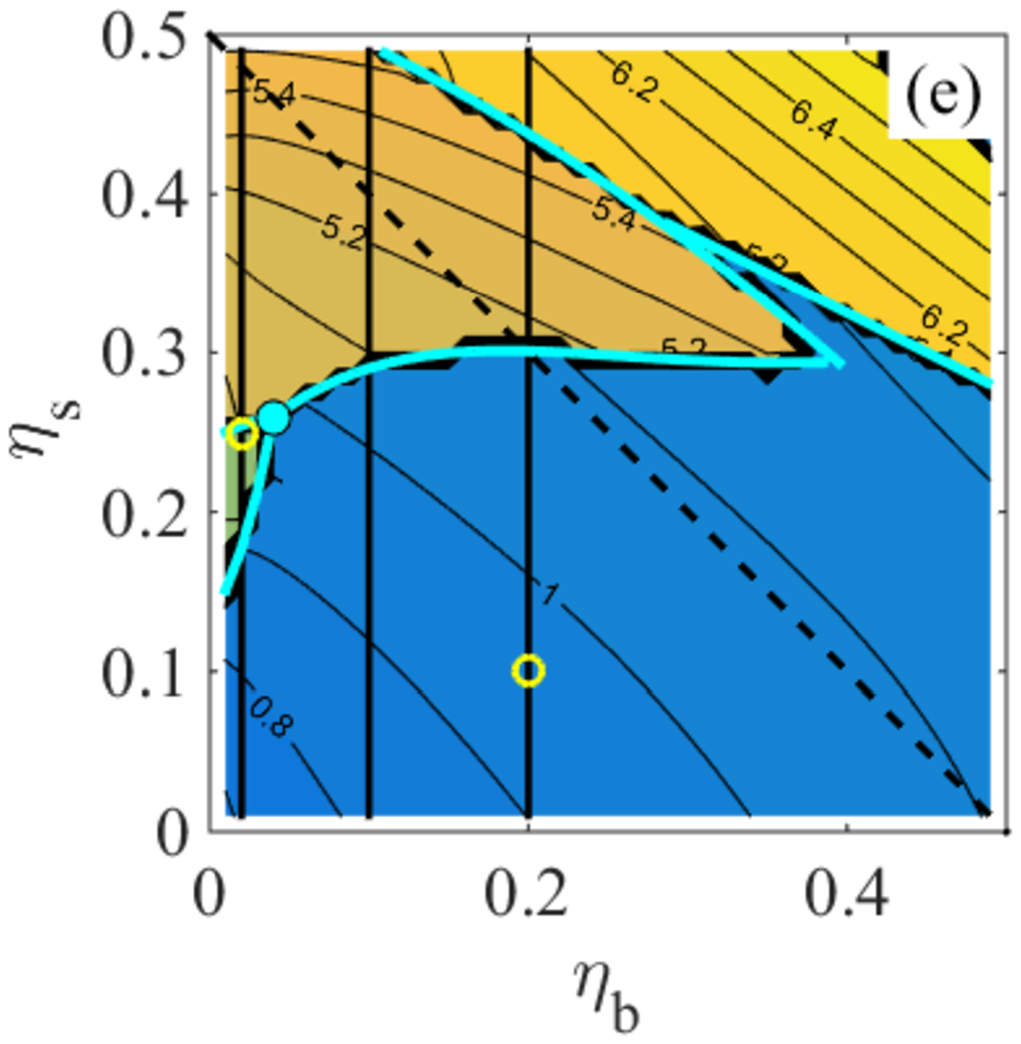}\hspace{0.5cm}
\includegraphics[width=0.22\textwidth]{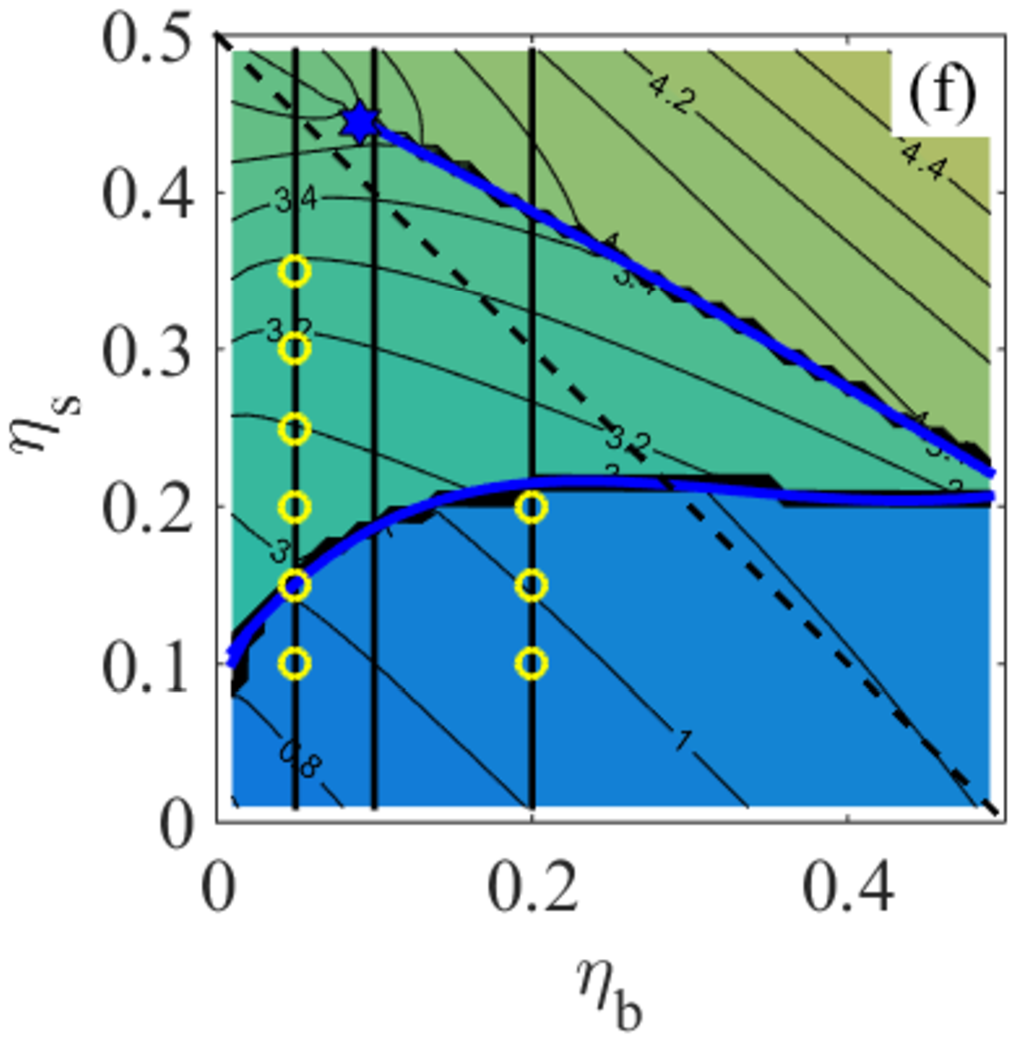}\hspace{0.5cm}
\includegraphics[width=0.22\textwidth]{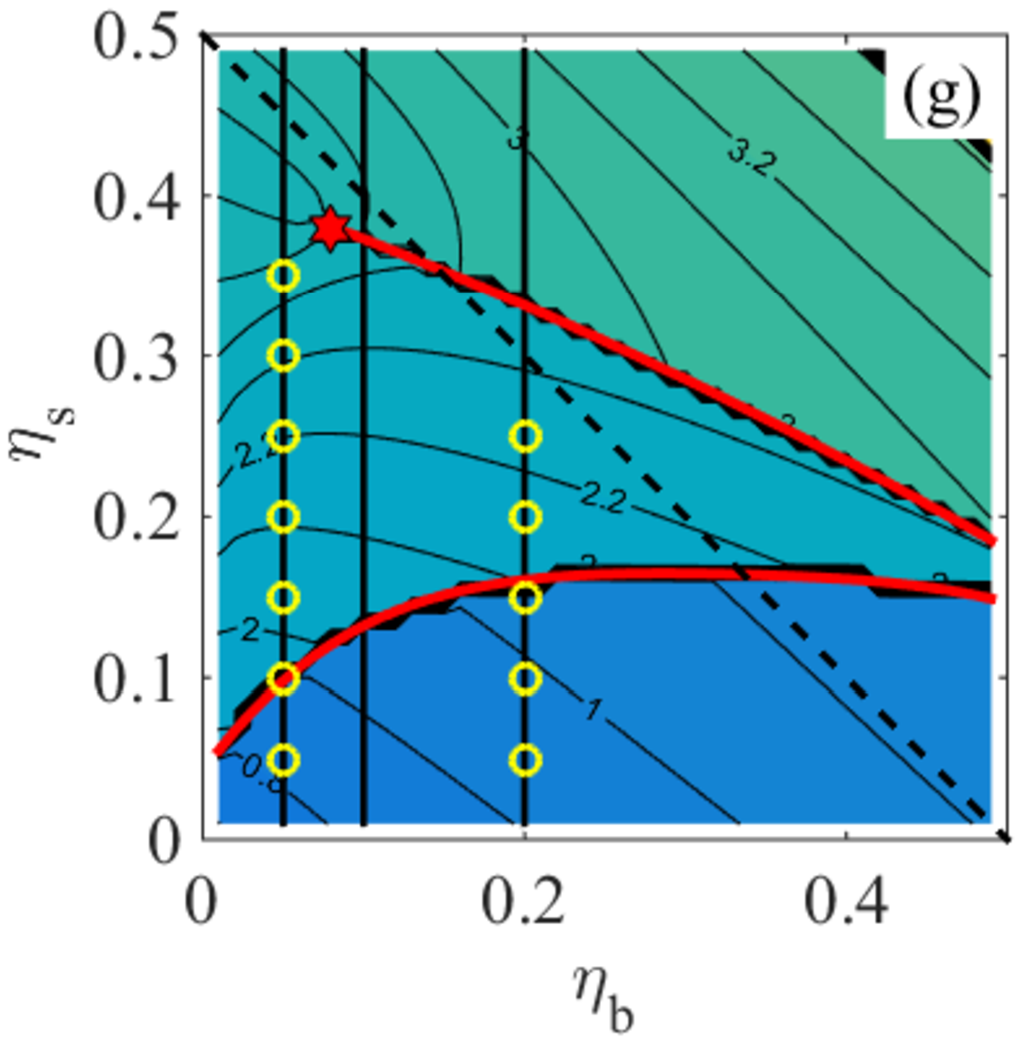}\hspace{0.5cm}
\includegraphics[width=0.22\textwidth]{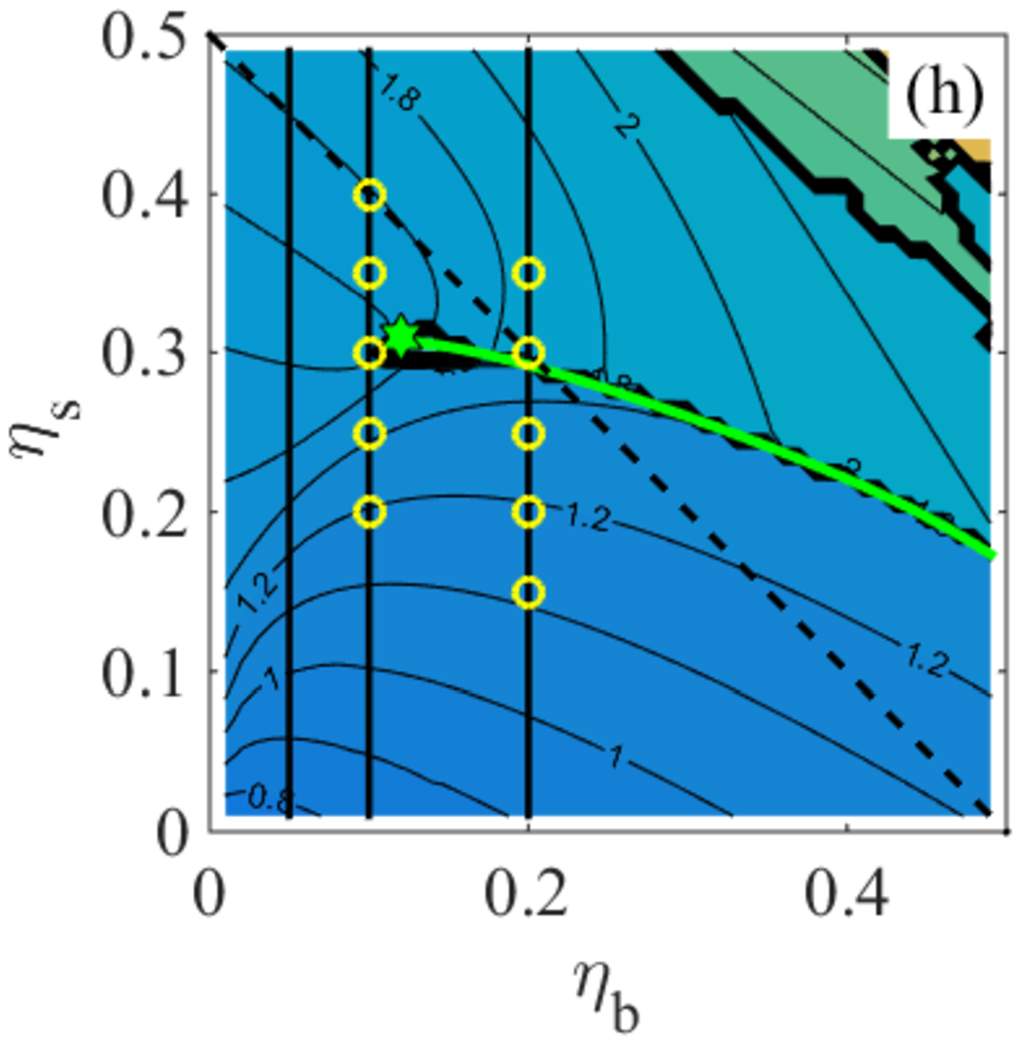}
\vspace{0.2cm}\\
\includegraphics[width=0.3\textwidth]{mapcolor.eps}
\caption{RFA predictions for the contour plots of the (reduced) damping coefficient $\alpha \sigma_b$ (top panels) and the (reduced) oscillation frequency $\omega\sigma_b/2\pi\equiv\sigma_b/\lambda$ (bottom panels) corresponding to the leading pole for a size ratio (a, e) $q=0.2$, (b, f) $q=0.3$, (c, g) $q=0.4$, and (d, h) $q=0.648$.
The colormap in the bottom bar indicates the color code for the values of $\alpha \sigma_b$ and $\omega\sigma_b/2\pi$. In each panel, the (black) dashed diagonal line represents the locus $\eta_s+\eta_b=0.5$, the (black) solid vertical lines represent the values $\eta_b= 0.02$ (for $q=0.2$) or $0.05$ (for $q=0.3$, $0.4$, and $0.648$), $0.10$, and $0.20$ considered in Figs.\ \ref{fig:q02}--\ref{fig:q0648}, and the (yellow) circles denote those cases where MD simulations have been performed.
The (colored) thick solid lines represent the crossover lines, while the solid (colored) stars and circles indicate the end points and the splitting points, respectively.}
\label{fig:plane}
\end{figure*}

The computation of $h_{ij}^{\MD}(r)$ was performed with the DynamO program \cite{BSL11} for the size ratios  $q=0.648$, $0.4$, $0.3$ and $0.2$, and different values of the partial packing fractions $(\eta_b,\eta_s)$, which were chosen according to the investigated size ratio to examine a substantial part of the phase diagram (including low, moderate, and high total densities) of the BHS mixture.
More specifically, the values of the partial packing fractions studied by
MD are given in Table \ref{table1} and denoted in Fig.\ \ref{fig:plane} as open yellow circles.

In order to reduce sufficiently  finite-size effects  and the statistical
errors in the simulations, the data of $h_{ij}^{\MD}(r)$ for $r/\sigma_{ij} <r_c^*$ must be obtained from long simulations with a large number of particles $(N \sim 10^4)$. It has been checked that $16\,384$ particles were sufficient to obtain reasonably accurate data for the size ratios $q=0.648$ and $0.4$. In the cases $q=0.3$ and $0.2$, due to the reduction of the size of the simulation box, the systems  were investigated with $48\,668$ particles.
The histogram grid size of $h_{ij}^{\MD}(r)$ was set to $\delta r/\sigma_{ij} = 0.01$, which was found to be a suitable  choice to balance finite-size effects and statistical errors.

The MD simulations were carried out typically for a total number of $2 \times 10^9$ collisions, and the statistical uncertainty of  $h_{ij}^{\MD}(r)$  was obtained with the block averaging method  \cite{AT17}. For each density, and in the whole range  $r/\sigma_{ij} \in (1,r_c^*)$, the accuracy of $h_{ij}^{\MD}(r)$ was such that the estimated uncertainty was typically smaller than $10^{-3}$, being up to $0.005$ near contact (for the highest densities) and becoming less than $0.0002$ at larger particle separations. For large systems, the finite-size effects in the MD calculations of the RDF arise mainly from fixing the particle number, i.e., from the
relation between canonical and grand-canonical ensembles.
The corrections required to convert data from the MD simulations to the canonical ensemble are of $\mathcal{O}(1/N^2)$  \cite{SDE96,BH01}, which are negligible here. Also, it was checked for
a few densities that the remaining part of the correction factor involving density derivatives was smaller than the obtained data accuracy and therefore could be neglected.

\begin{figure*}
\hspace{0.25cm} $q = 0.648$ \hspace{2.75cm} $q = 0.625$ \hspace{2.75cm} $q = 0.600$ \hspace{2.75cm} $q = 0.500$ \hspace{1.4cm}\\
\vspace{0.2cm}
\hspace{-0.3cm}\includegraphics[width=0.22\textwidth]{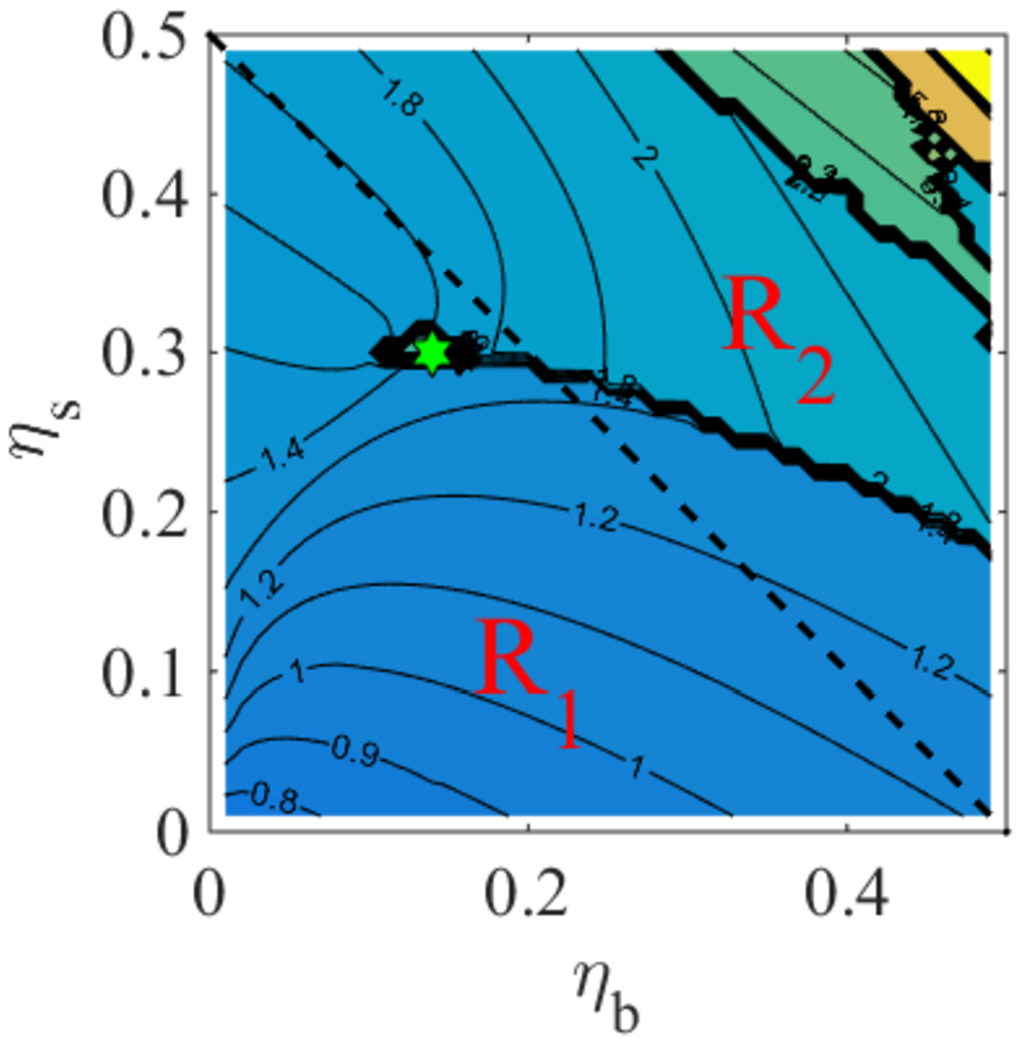}\hspace{0.5cm}
\includegraphics[width=0.22\textwidth]{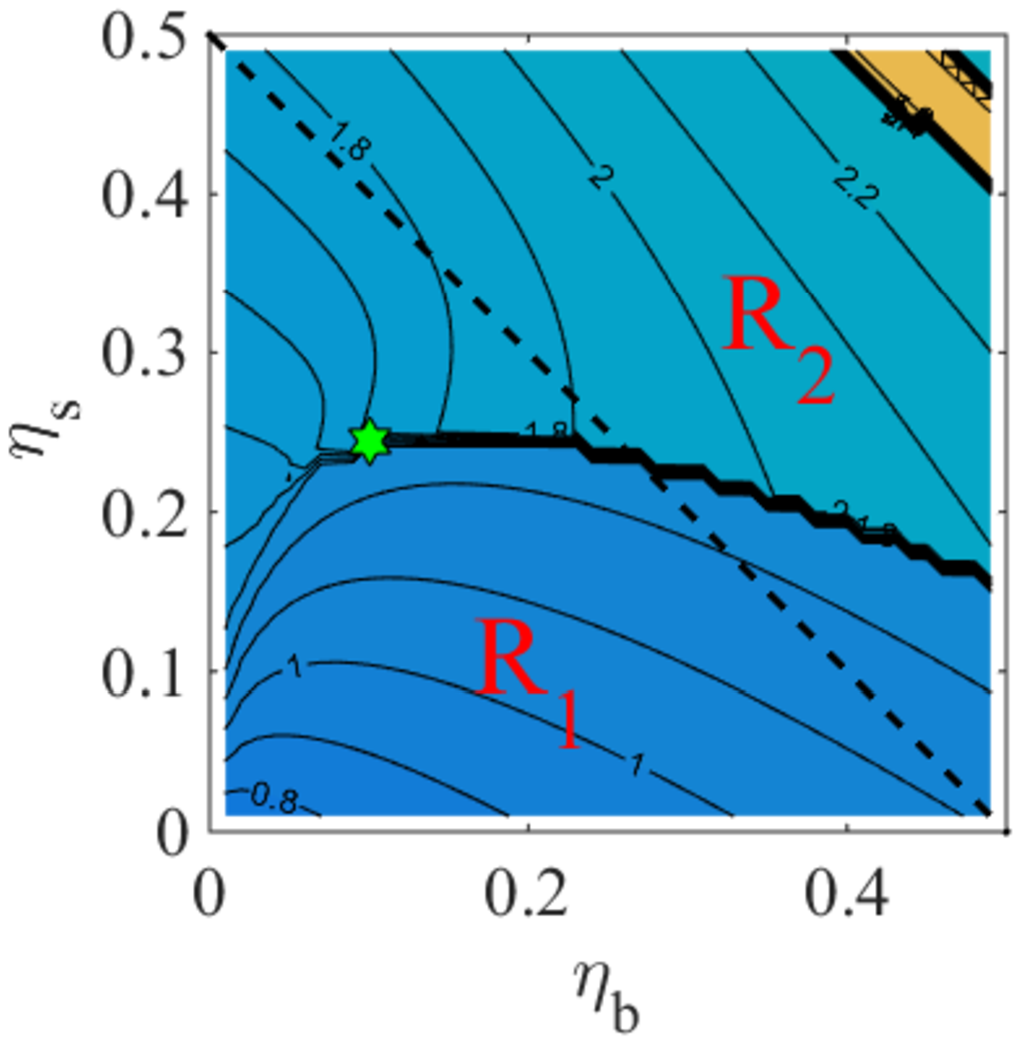}\hspace{0.5cm}
\includegraphics[width=0.22\textwidth]{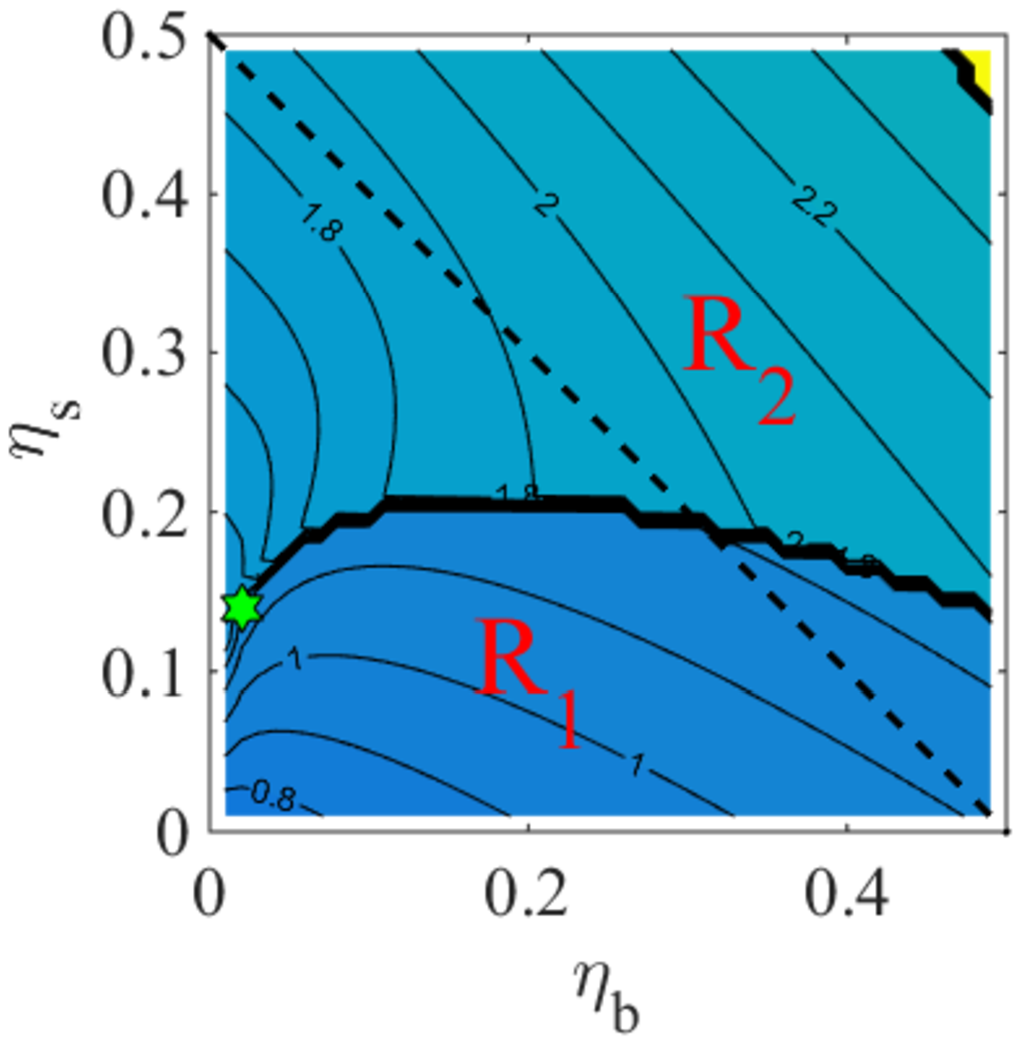}\hspace{0.5cm}
\includegraphics[width=0.22\textwidth]{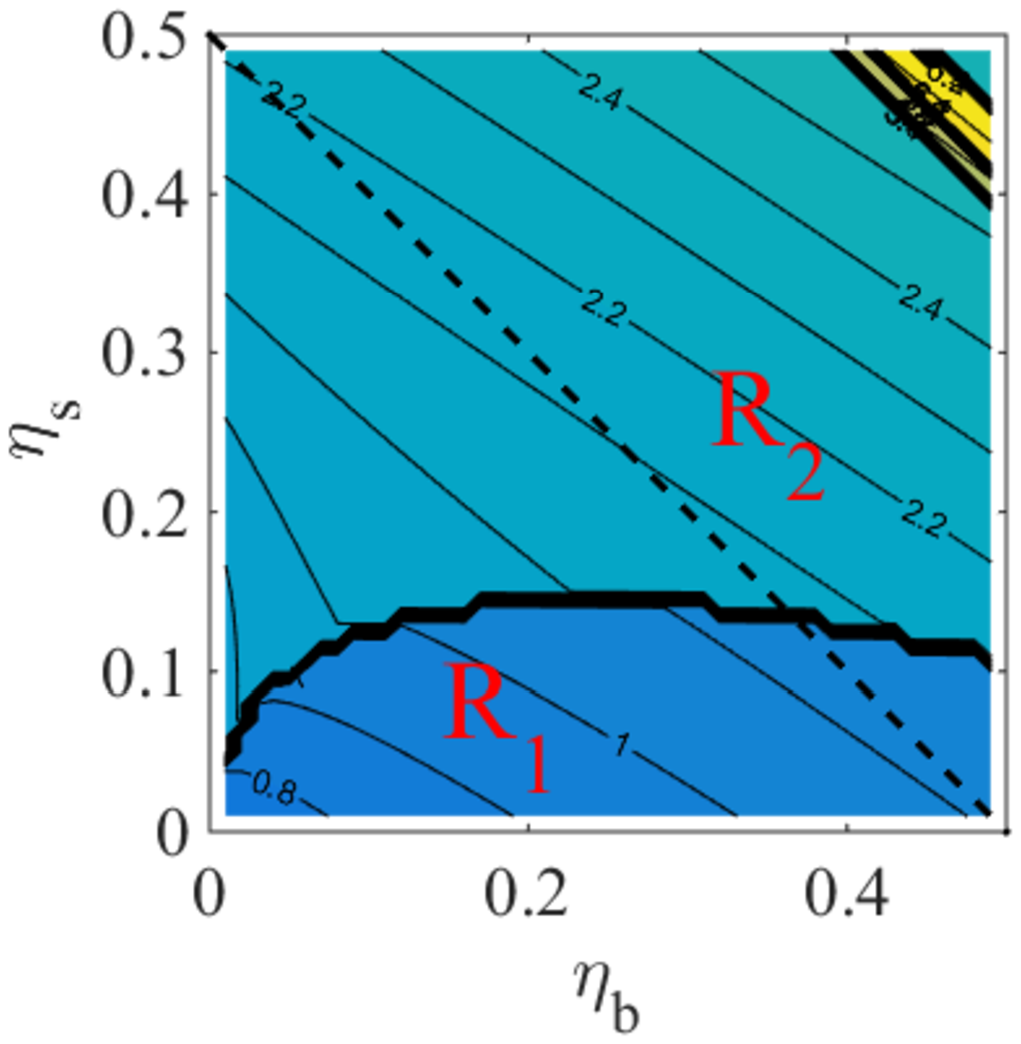}
\vspace{0.2cm}\\
\hspace{0.25cm} $q = 0.425$ \hspace{2.75cm} $q = 0.400$ \hspace{2.75cm} $q = 0.375$ \hspace{2.75cm} $q = 0.350$ \hspace{1.4cm}\\
\vspace{0.2cm}
\hspace{-0.3cm}\includegraphics[width=0.22\textwidth]{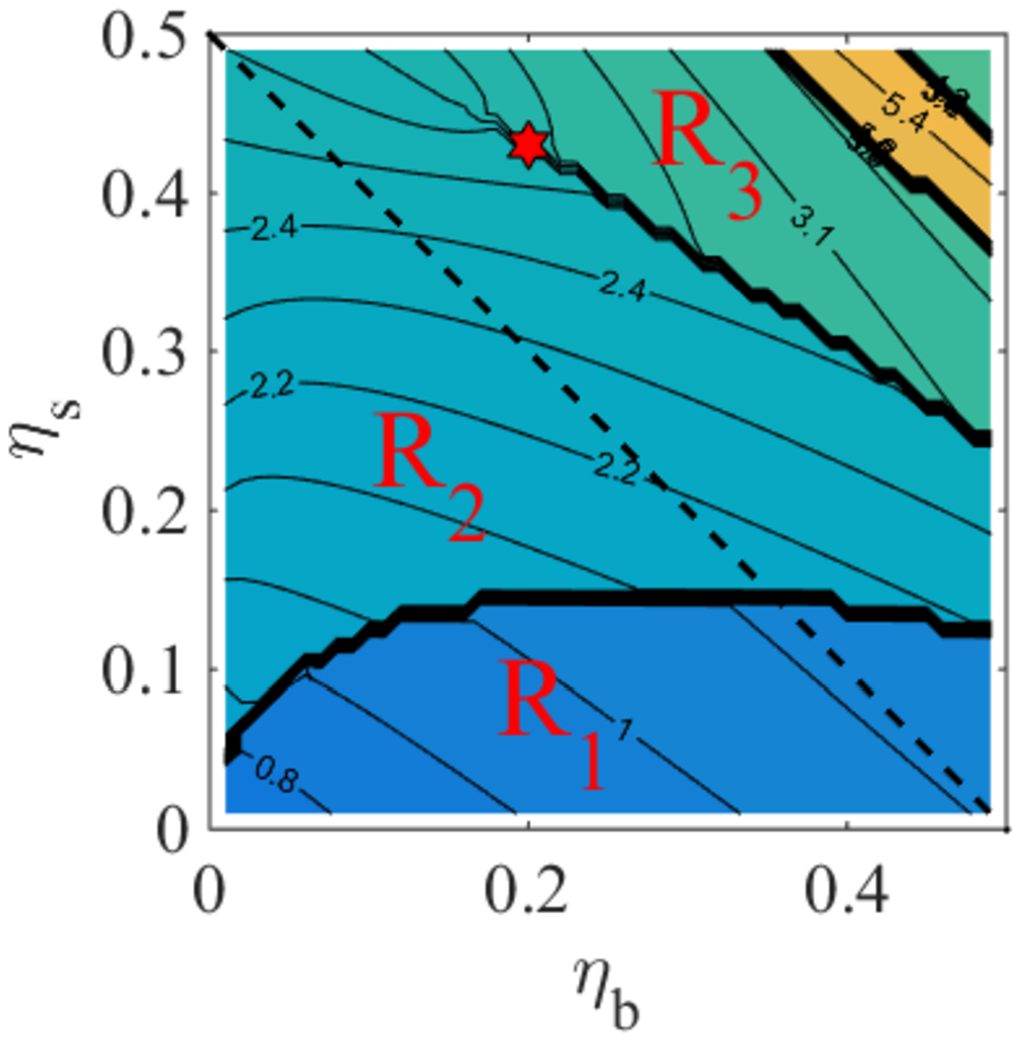}\hspace{0.5cm}
\includegraphics[width=0.22\textwidth]{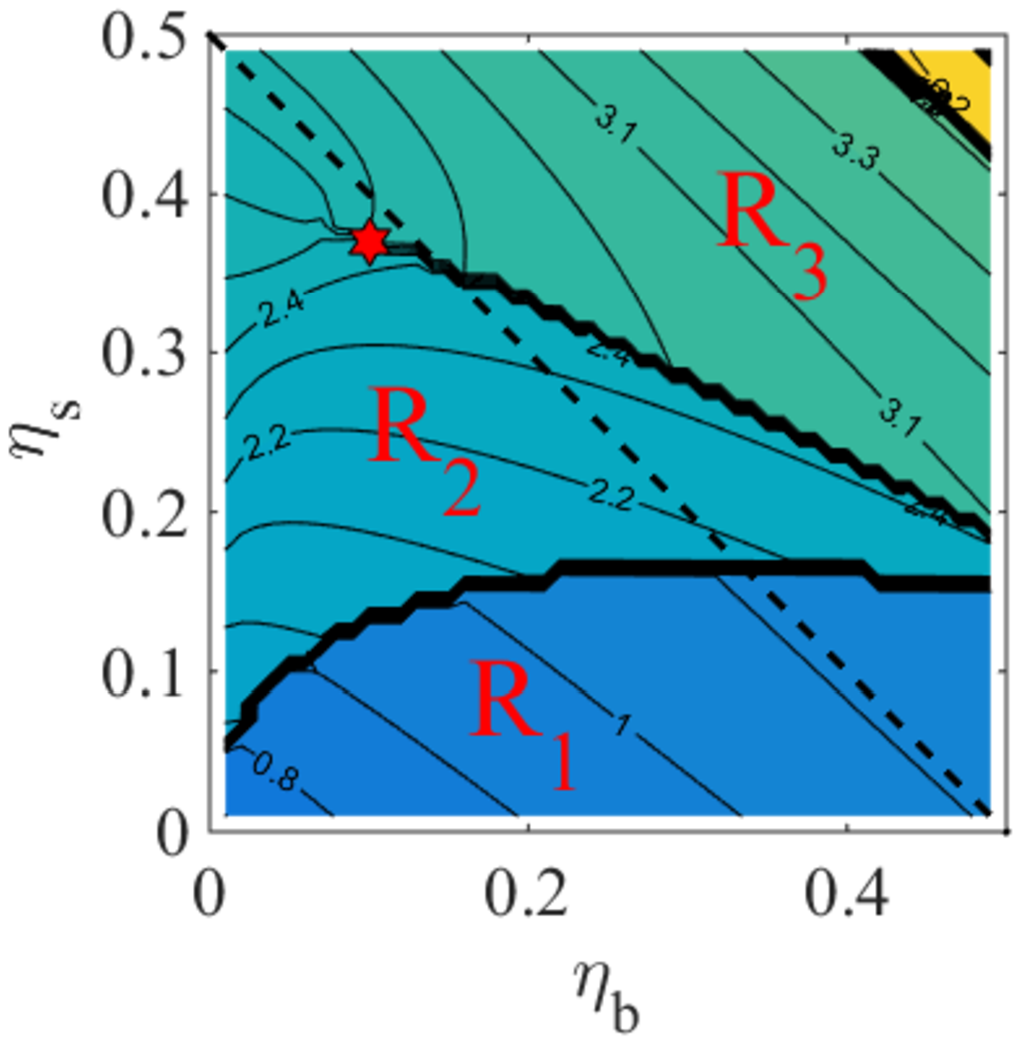}\hspace{0.5cm}
\includegraphics[width=0.22\textwidth]{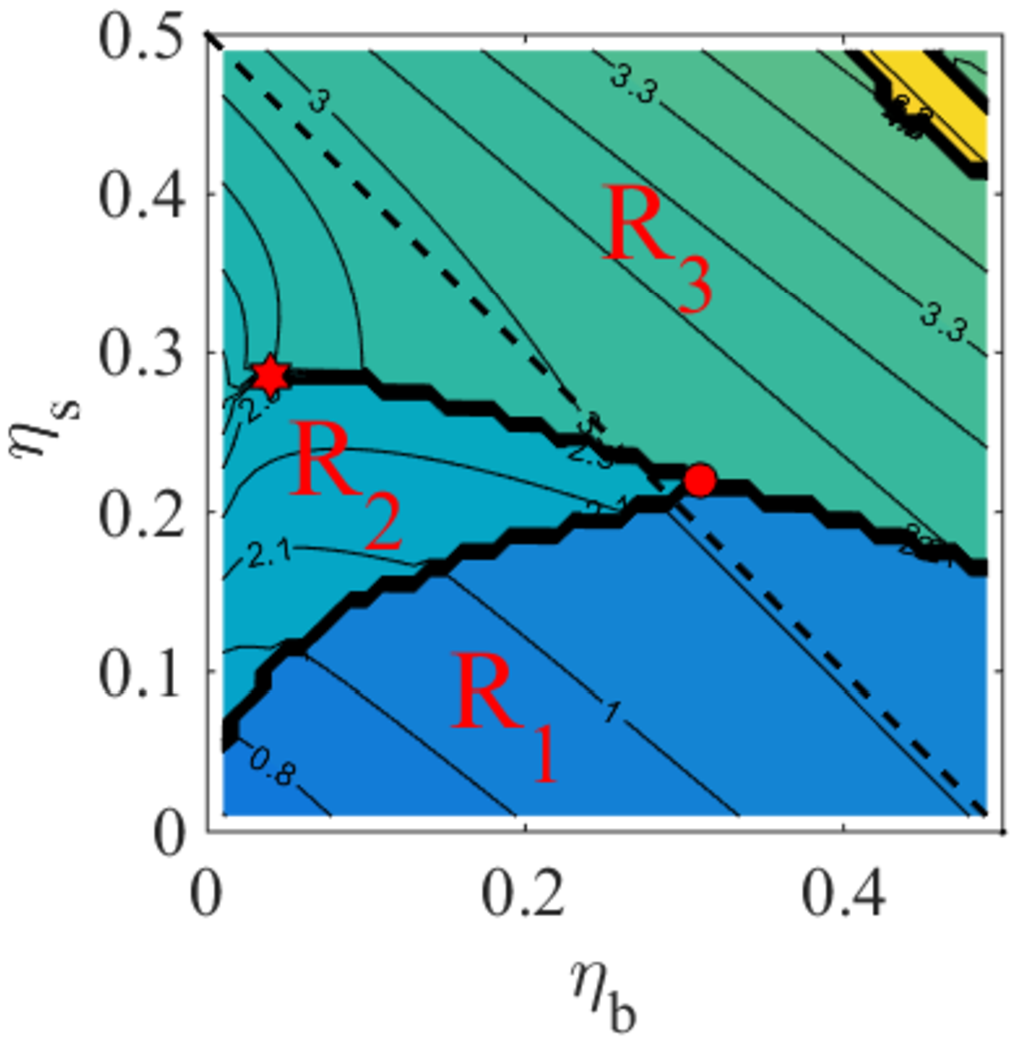}\hspace{0.5cm}
\includegraphics[width=0.22\textwidth]{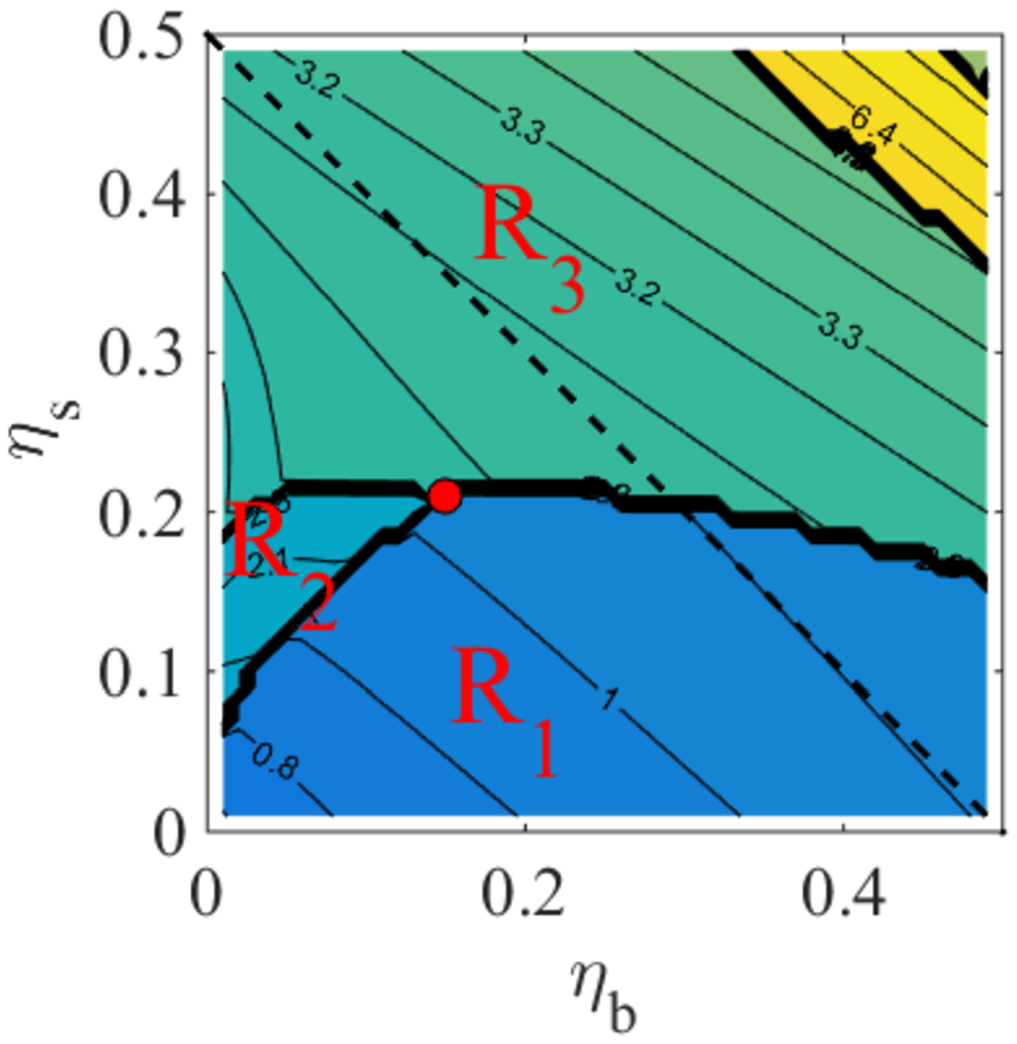}
\vspace{0.2cm}\\
\hspace{0.25cm} $q = 0.315$ \hspace{2.75cm} $q = 0.300$ \hspace{2.75cm} $q = 0.275$ \hspace{2.75cm} $q = 0.250$ \hspace{1.4cm}\\
\vspace{0.2cm}
\hspace{-0.3cm}\includegraphics[width=0.22\textwidth]{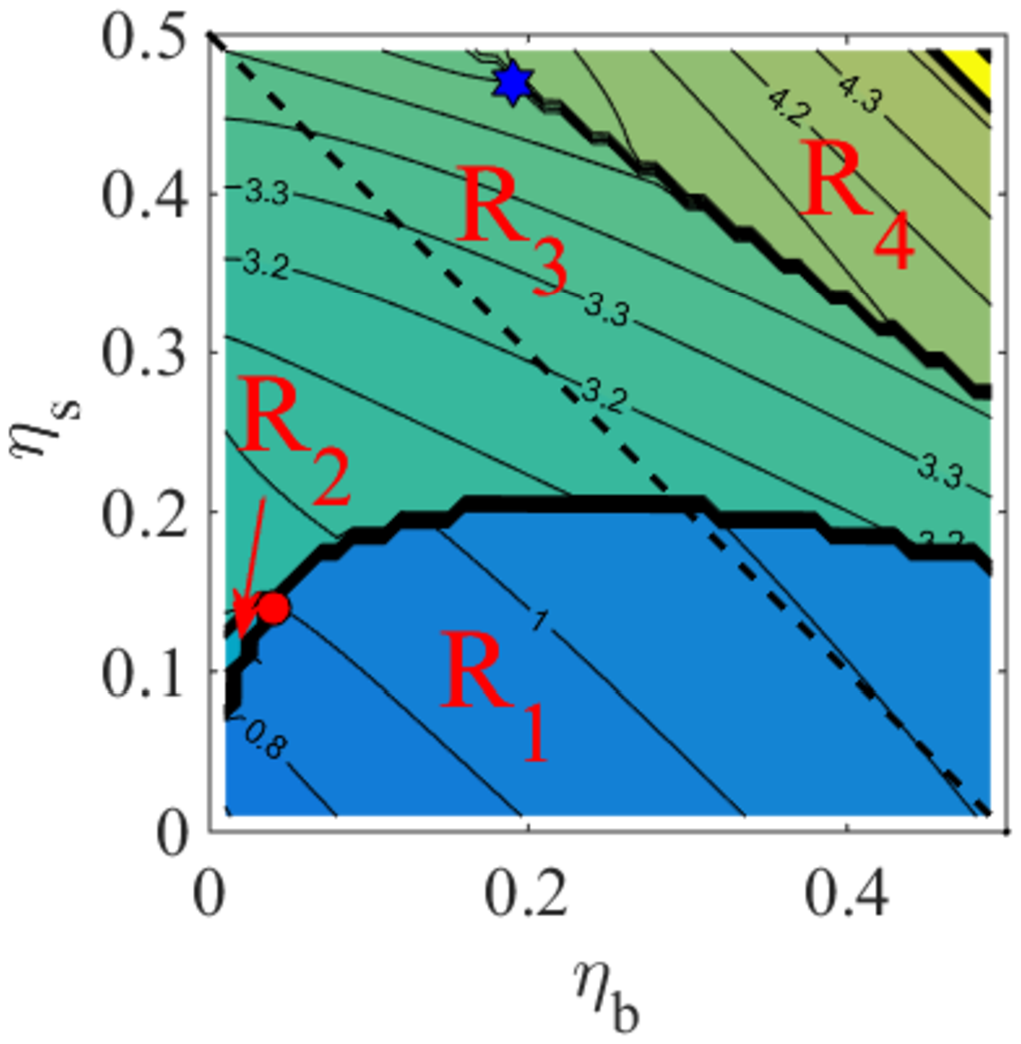}\hspace{0.5cm}
\includegraphics[width=0.22\textwidth]{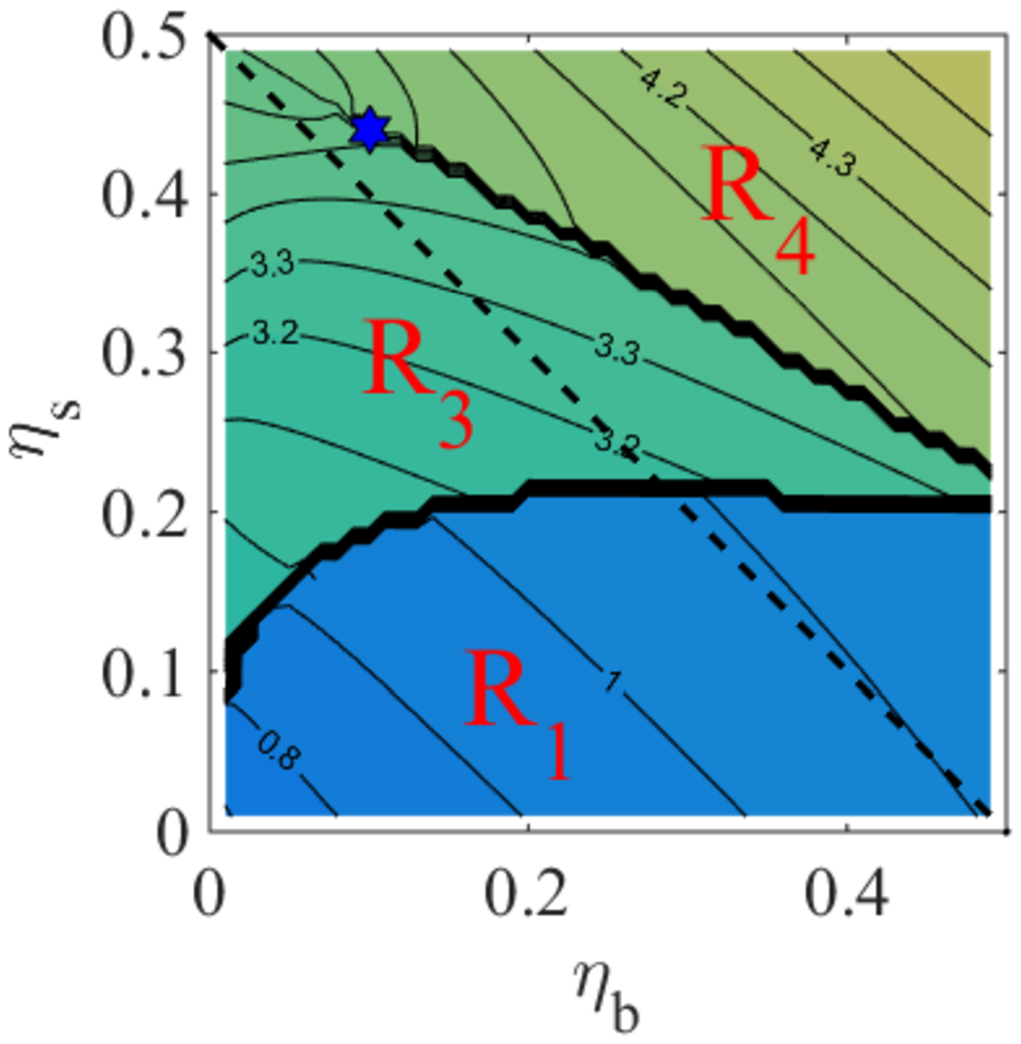}\hspace{0.5cm}
\includegraphics[width=0.22\textwidth]{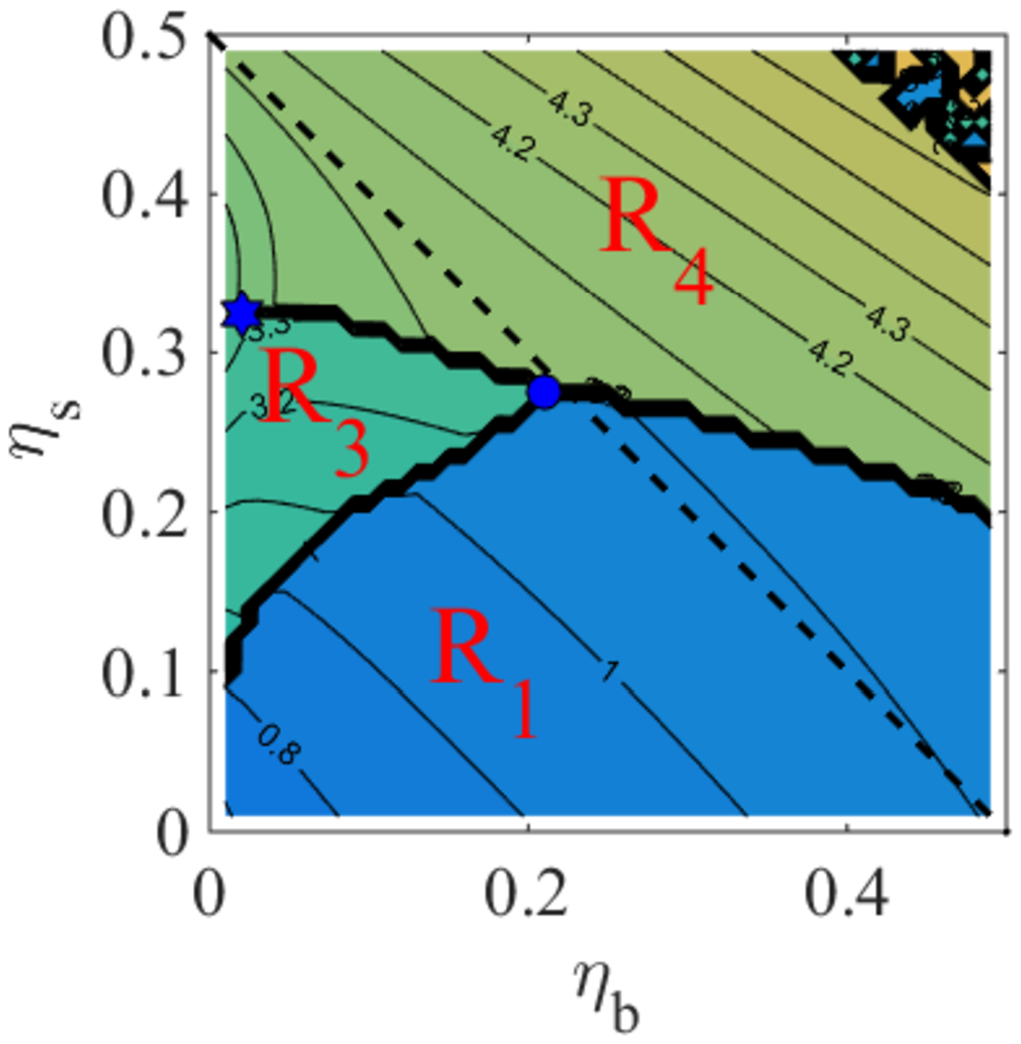}\hspace{0.5cm}
\includegraphics[width=0.22\textwidth]{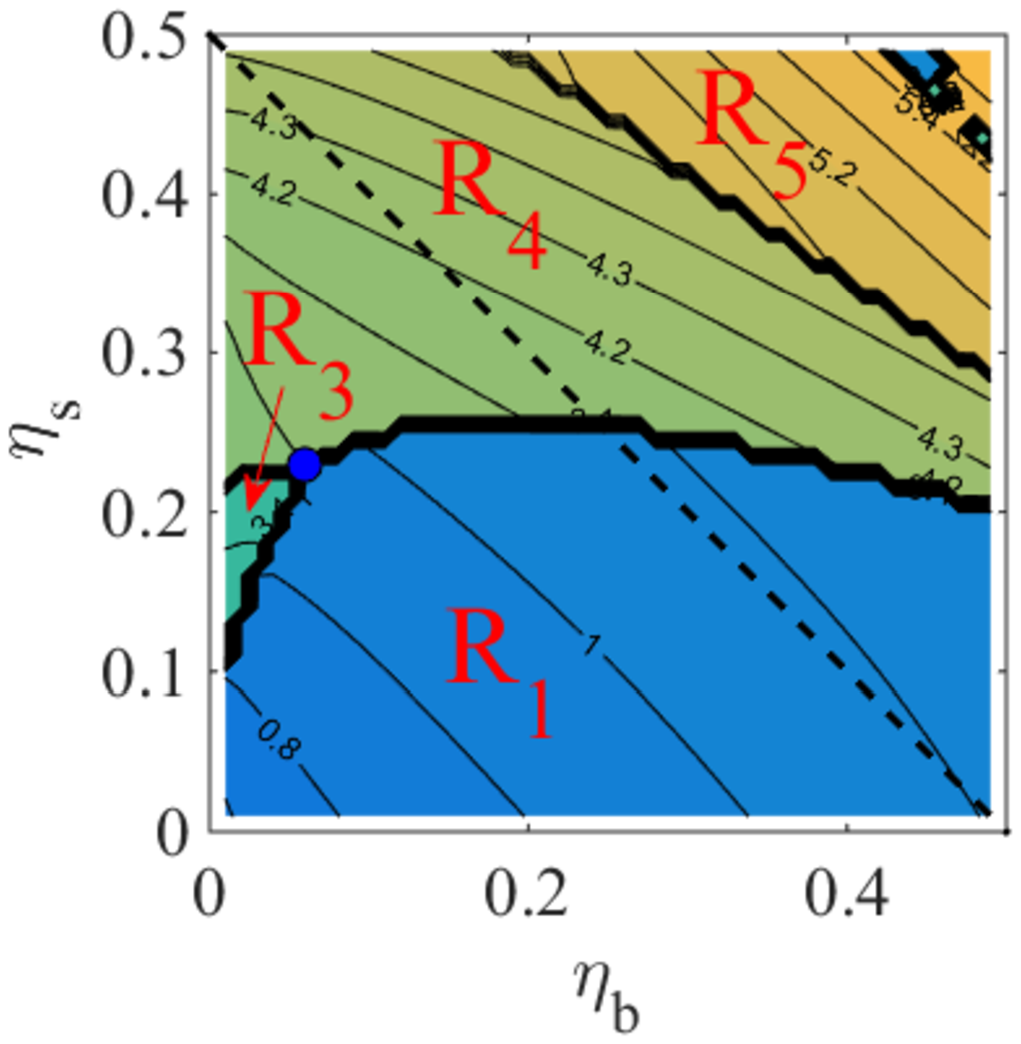}
\vspace{0.2cm}\\
\includegraphics[width=0.3\textwidth]{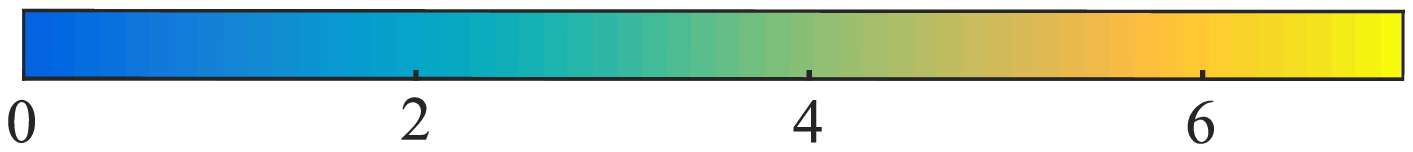}
\caption{
RFA predictions for the contour plots of the  (reduced) oscillation frequency $\omega\sigma_b/2\pi\equiv\sigma_b/\lambda$  corresponding to the leading pole for decreasing representative values of the size ratio (from $q=0.648$ to $q=0.250$).
The colormap in the bottom bar indicates the color code for the values of  $\omega\sigma_b/2\pi$. In each panel, the (black) dashed diagonal line represents the locus $\eta_s+\eta_b=0.5$.	The (colored) stars and circles indicate the end points and the  splitting points, respectively.
Note that $\nq=2$ for $q=0.648$, $0.625$, $0.600$, $0.500$ and $0.425$, $\nq=3$ for $q=0.375$, $0.350$, $0.315$, and $0.300$, and $\nq=4$ for $q=0.275$ and $0.250$.}
\label{fig:evolution}
\end{figure*}

\begin{figure}
\includegraphics[width=0.45\textwidth]{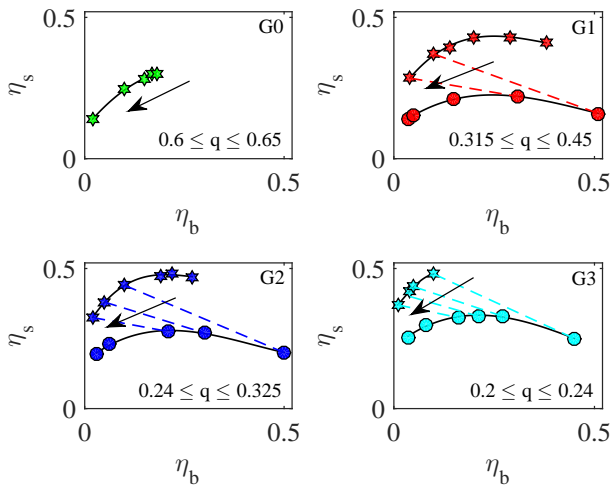}
\caption{
The $q$-evolution of the end point (colored stars) and the splitting point (colored circles) of the generations G0--G3. The arrows signal the evolution as $q$ decreases. The dashed tie lines connect end and splitting points at the same $q$.}
\label{fig:R3}
\end{figure}

\begin{figure}
\includegraphics[width=0.45\textwidth]{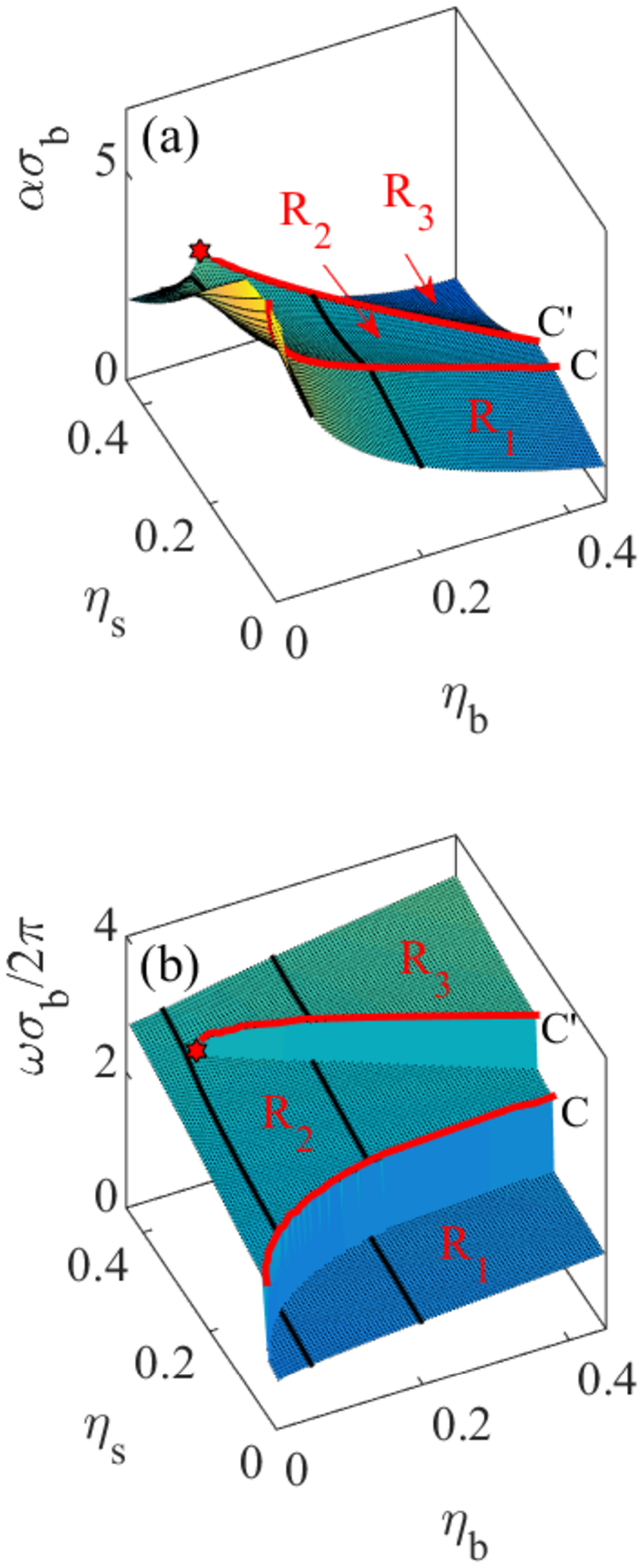}
\caption{3D plots, as predicted by the RFA, for (a) the (reduced) damping coefficient $\alpha \sigma_b$  and (b) the (reduced) oscillation frequency $\omega\sigma_b/2\pi\equiv\sigma_b/\lambda$, corresponding to the leading pole for a size ratio $q=0.4$.
The used colormap is the same as the one in Fig.\ \ref{fig:plane}; it indicates the change in $\alpha \sigma_b$ in the case of panel (a) and  of $\omega\sigma_b/2\pi$ in the case of panel (b). In each panel, the (black) solid lines represent the cuts $\eta_b=0.05$ and $0.20$ considered in Fig.\ \ref{fig:q04}. The (red) thick lines represent the crossover lines and the solid (red) star indicates the end point.}
\label{fig:plane3D}
\end{figure}

\begin{figure}
\includegraphics[width=0.45\textwidth]{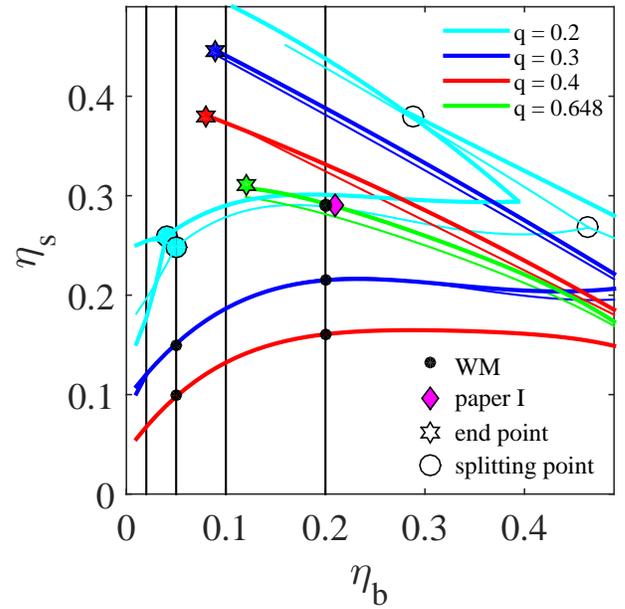}
\caption{RFA (thick lines) and PY (thin lines) predictions for the crossover lines $C$ and $C'$ in the plane $\eta_s$ vs $\eta_b$ for $q=0.2$, $0.3$, $0.4$, and $0.648$.  The solid circles represent results from MD computer simulations (via the WM scheme), which confirm the theoretical RFA
predictions.
The WM points for $\eta_b=0.2$ correspond, from top to bottom, to $q=0.648$, $0.3$, and $0.4$, respectively, whereas for $\eta_b=0.05$ the points correspond to $q=0.3$ and $0.4$. The (magenta) diamond represents a result from paper I \cite{PBYSH20}. The solid (colored) stars indicate the end points (see Figs.\ \ref{fig:evolution}--\ref{fig:plane3D}), while the solid (cyan) circles represent the splitting points of G3 generation (see Figs.\ \ref{fig:evolution} and \ref{fig:R3}). The (black) vertical lines represent the values  $\eta_b=0.02$, $0.05$, $0.10$, and $0.20$ considered in Figs.\ \ref{fig:q02}--\ref{fig:q0648}.}
\label{fig:plane_all}
\end{figure}

\begin{figure*}
\includegraphics[width=0.26\textwidth]{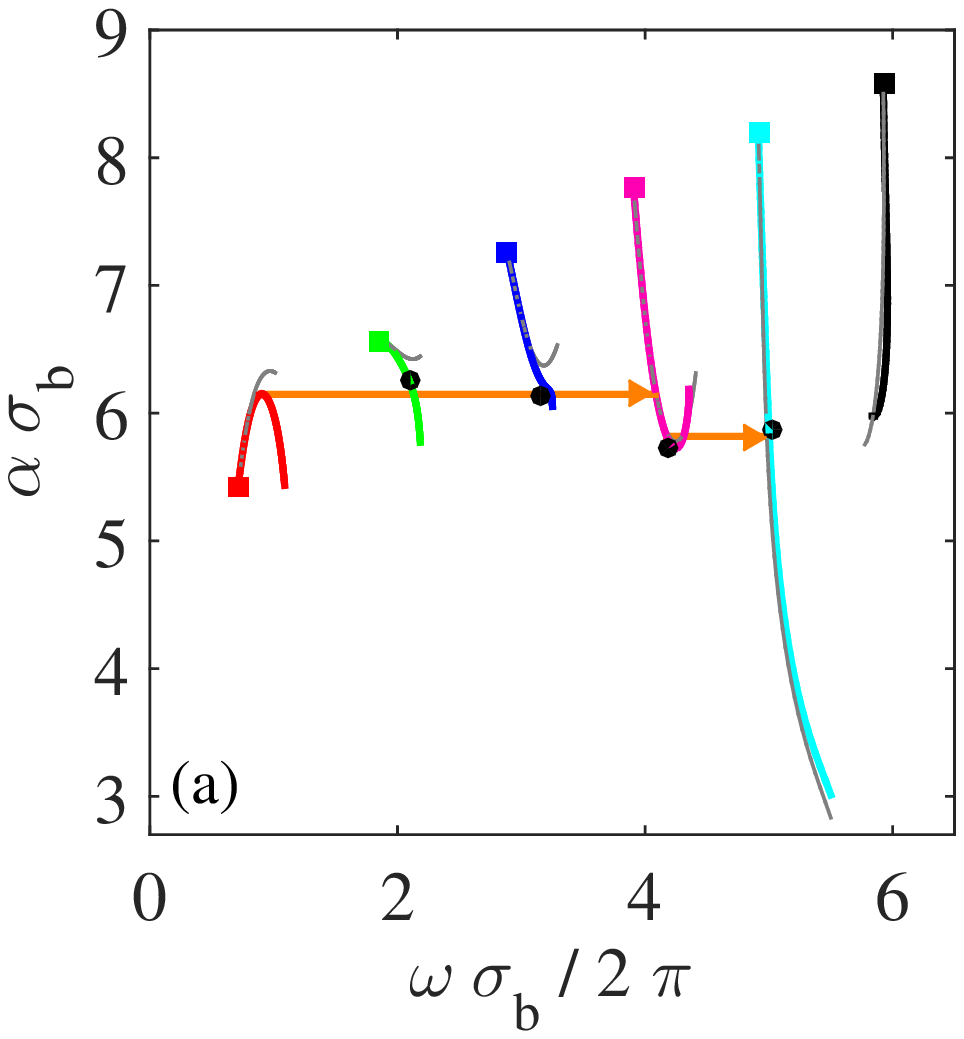}\hspace{0.8cm}
\includegraphics[width=0.26\textwidth]{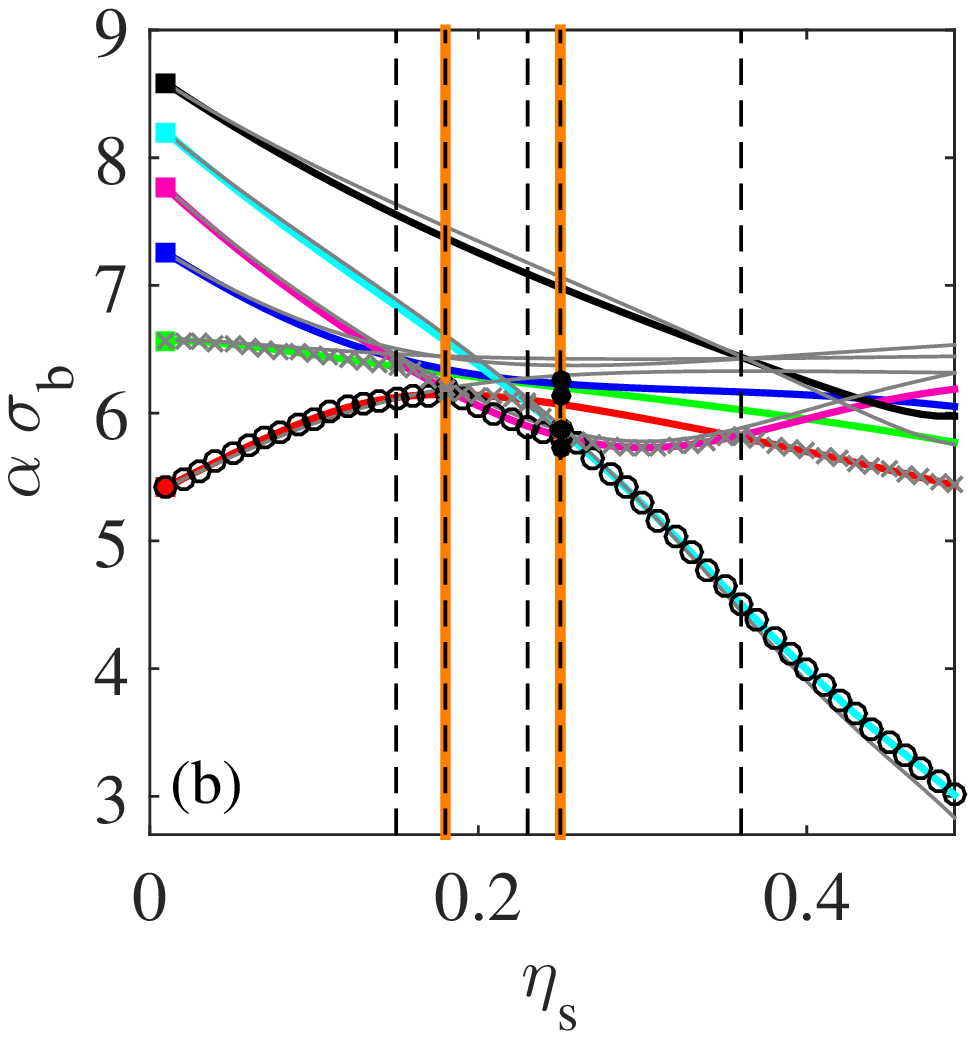}\hspace{0.8cm}
\includegraphics[width=0.26\textwidth]{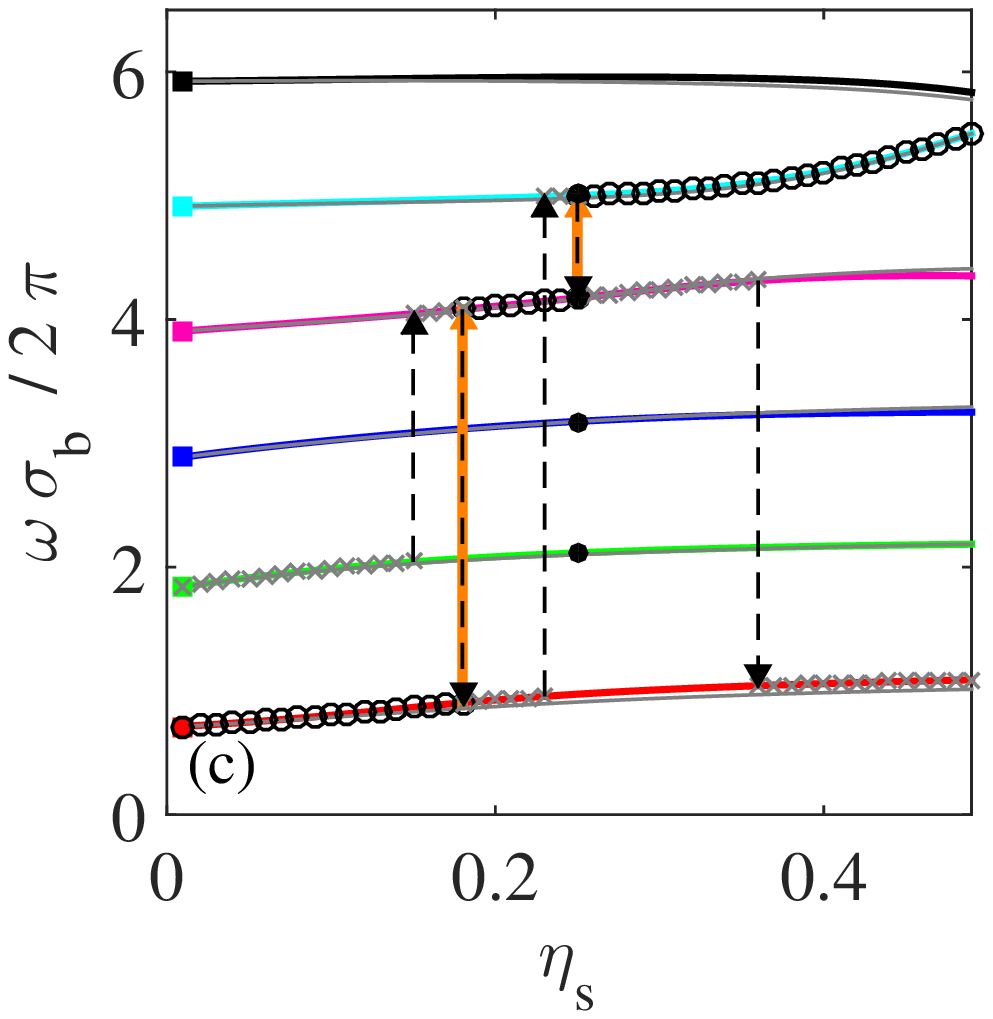}\\
\vspace{0.5cm}
\includegraphics[width=0.26\textwidth]{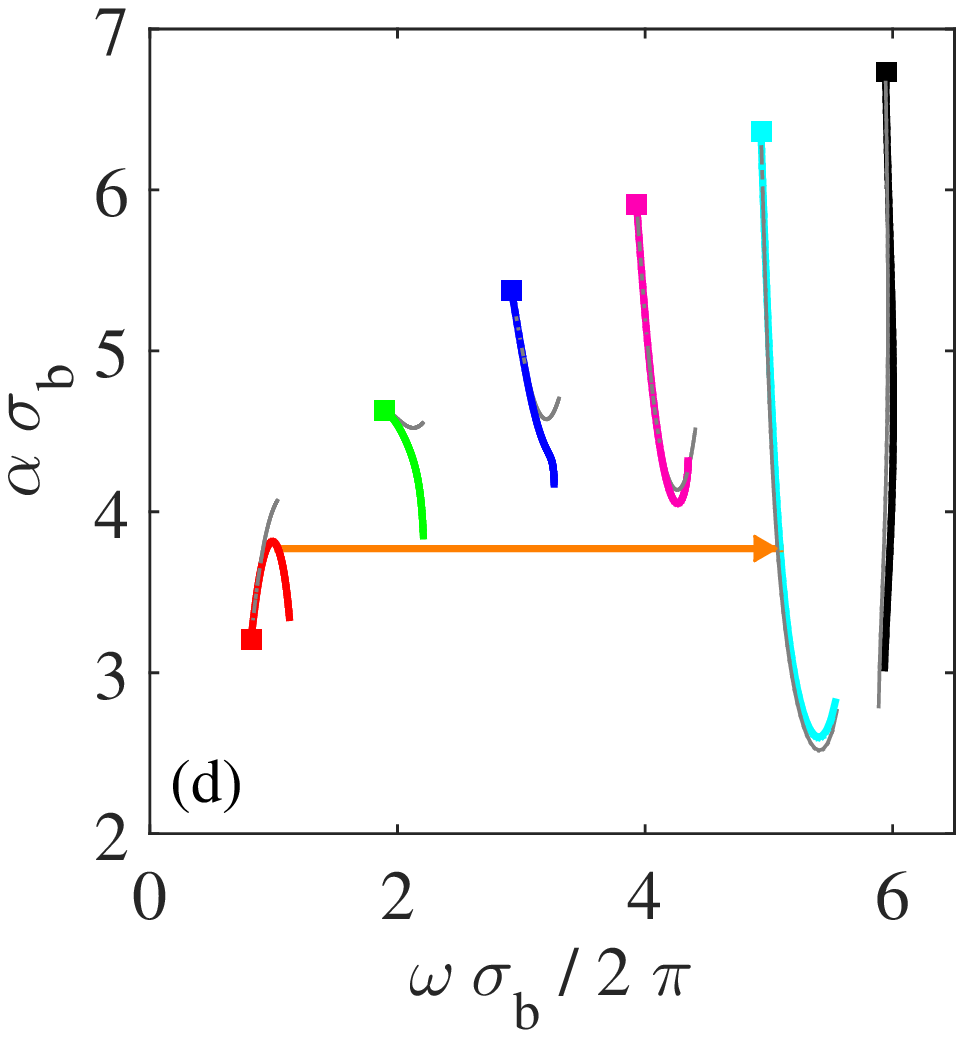}\hspace{0.8cm}
\includegraphics[width=0.26\textwidth]{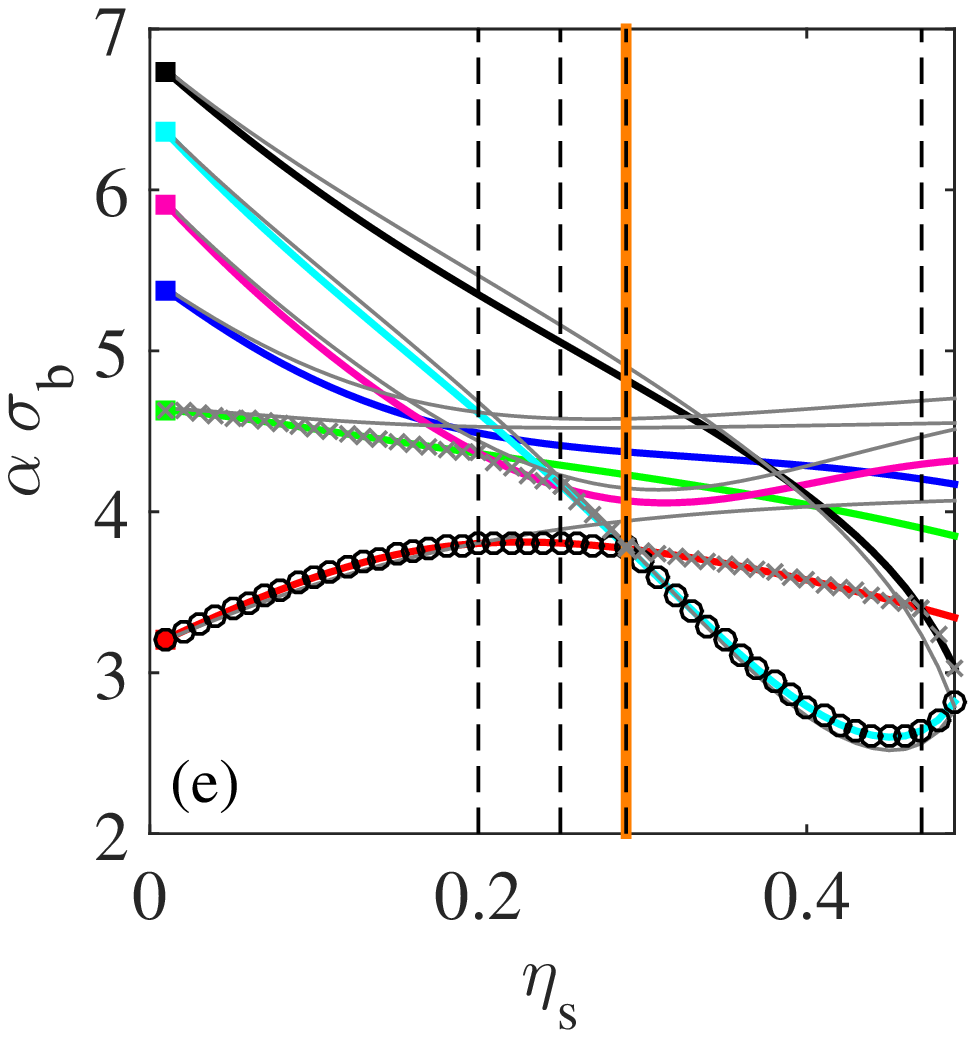}\hspace{0.8cm}
\includegraphics[width=0.26\textwidth]{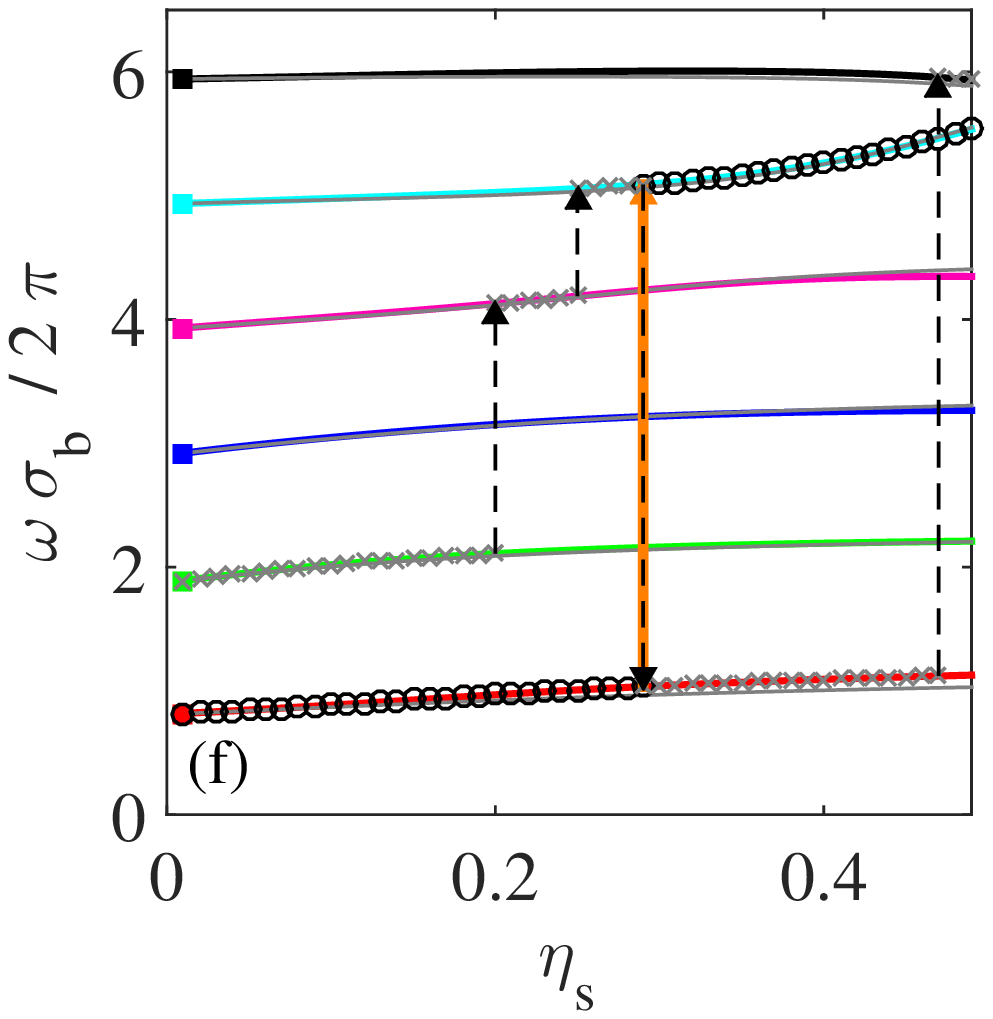}\\
\vspace{0.5cm}
\includegraphics[width=0.26\textwidth]{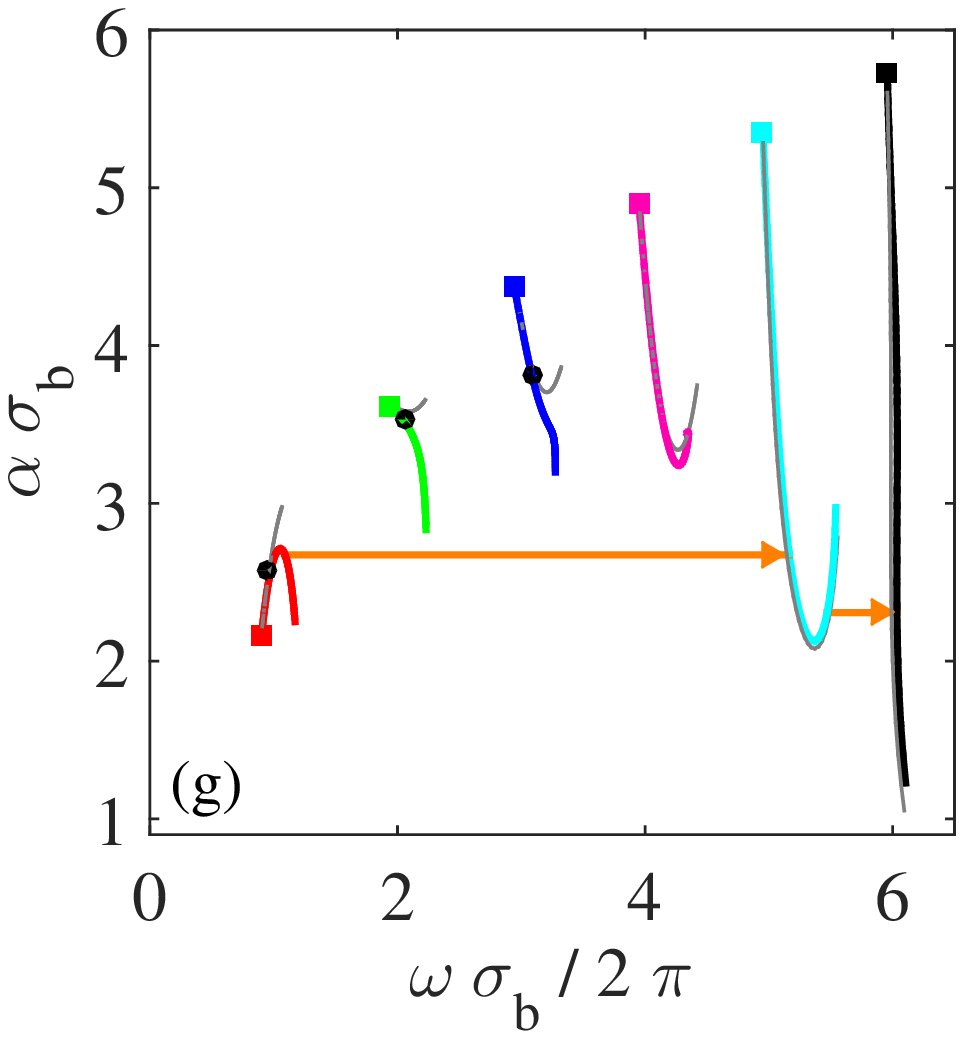}\hspace{0.8cm}
\includegraphics[width=0.26\textwidth]{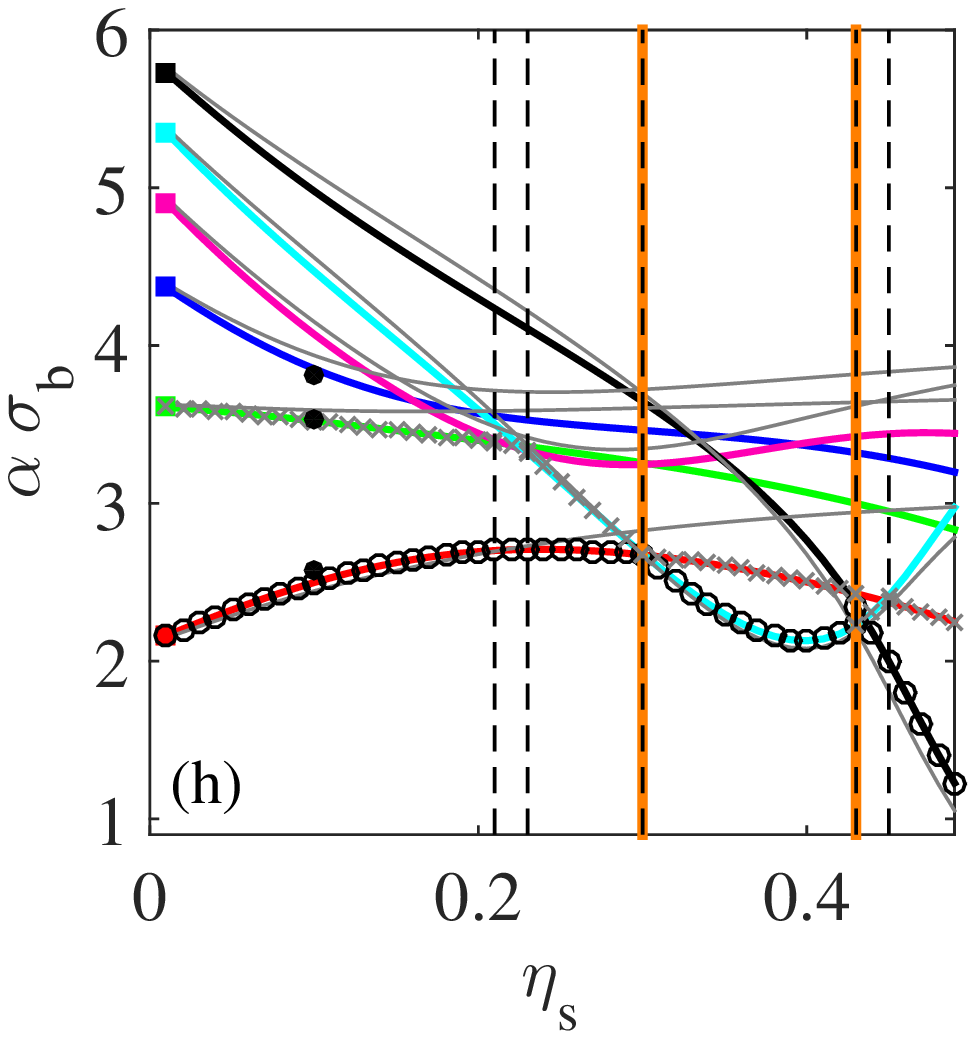}\hspace{0.8cm}
\includegraphics[width=0.26\textwidth]{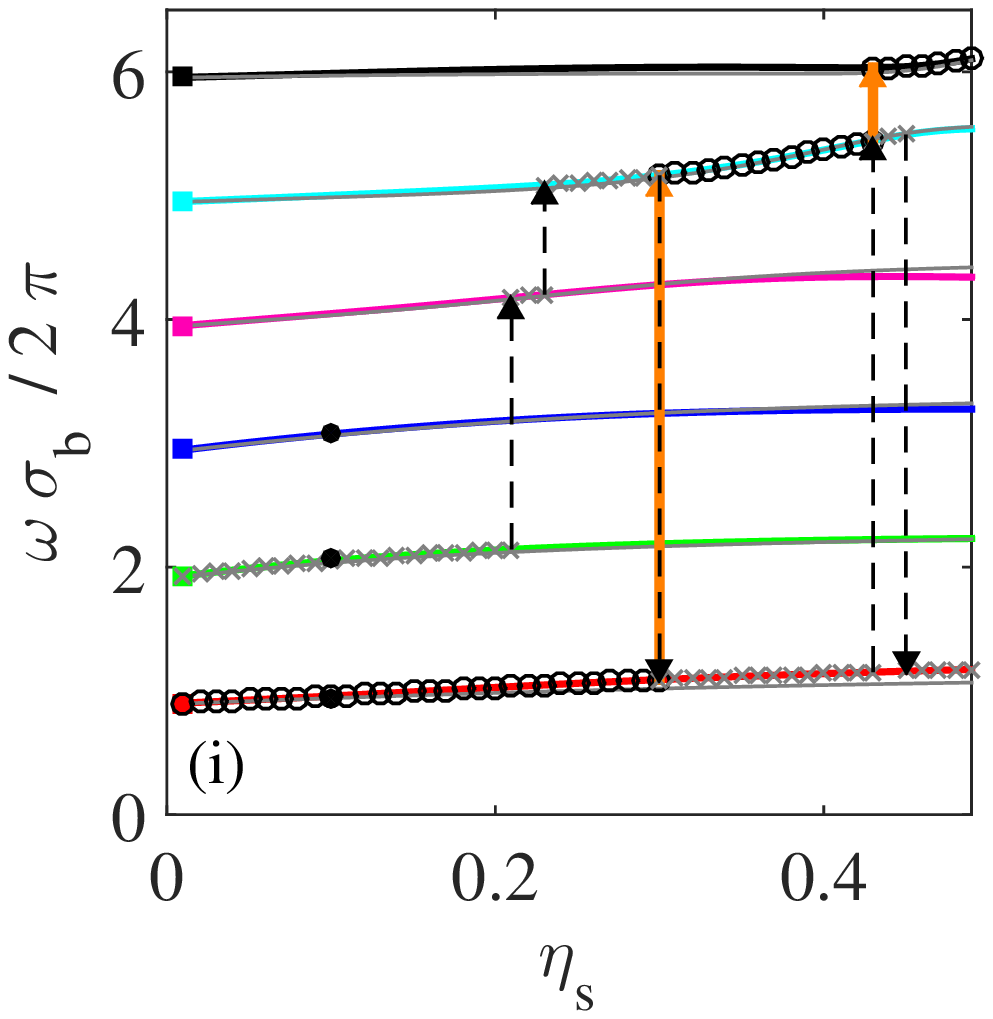}\\
\vspace{0.25cm}
\hspace{0.8cm}\includegraphics[width=1.8\columnwidth]{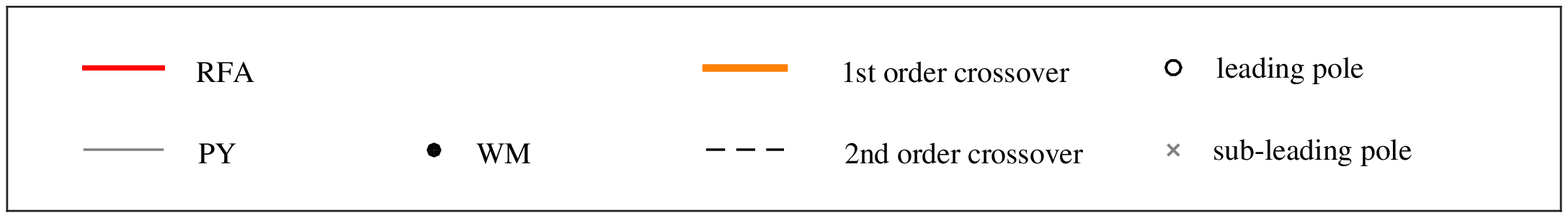}\\
\caption{Dependence of the first six poles on $\eta_s$ for $q=0.2$ and
$\eta_b=0.02$ (top panels),  $\eta_b=0.10$ (middle panels), and $\eta_b=0.20$ (bottom panels).
The thick
{(colored)} and thin
{(gray)} lines correspond to the RFA and PY predictions, respectively, while the solid circles represent the WM values for the cases where MD simulations were performed.
The colored squares denote poles for a small sphere packing fraction $\eta_s=0.01$, and the
lines indicate trajectories for increasing values of $\eta_s$.
The horizontal lines denote the crossovers as $\eta_s$ increases. In the central and right panels, the open circles and crosses represent the leading and subleading poles, respectively.}
  \label{fig:q02}
\end{figure*}

\begin{figure*}
\includegraphics[width=0.26\textwidth]{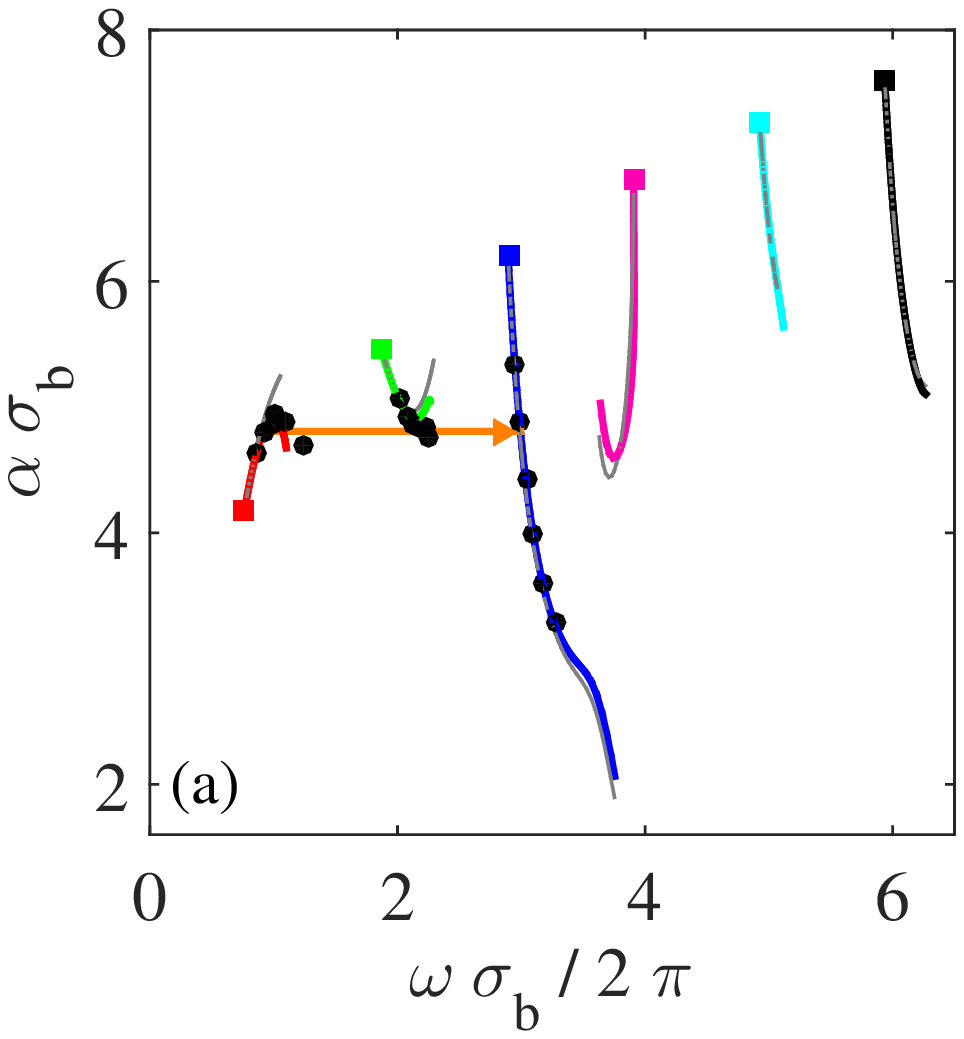}\hspace{0.8cm}
\includegraphics[width=0.26\textwidth]{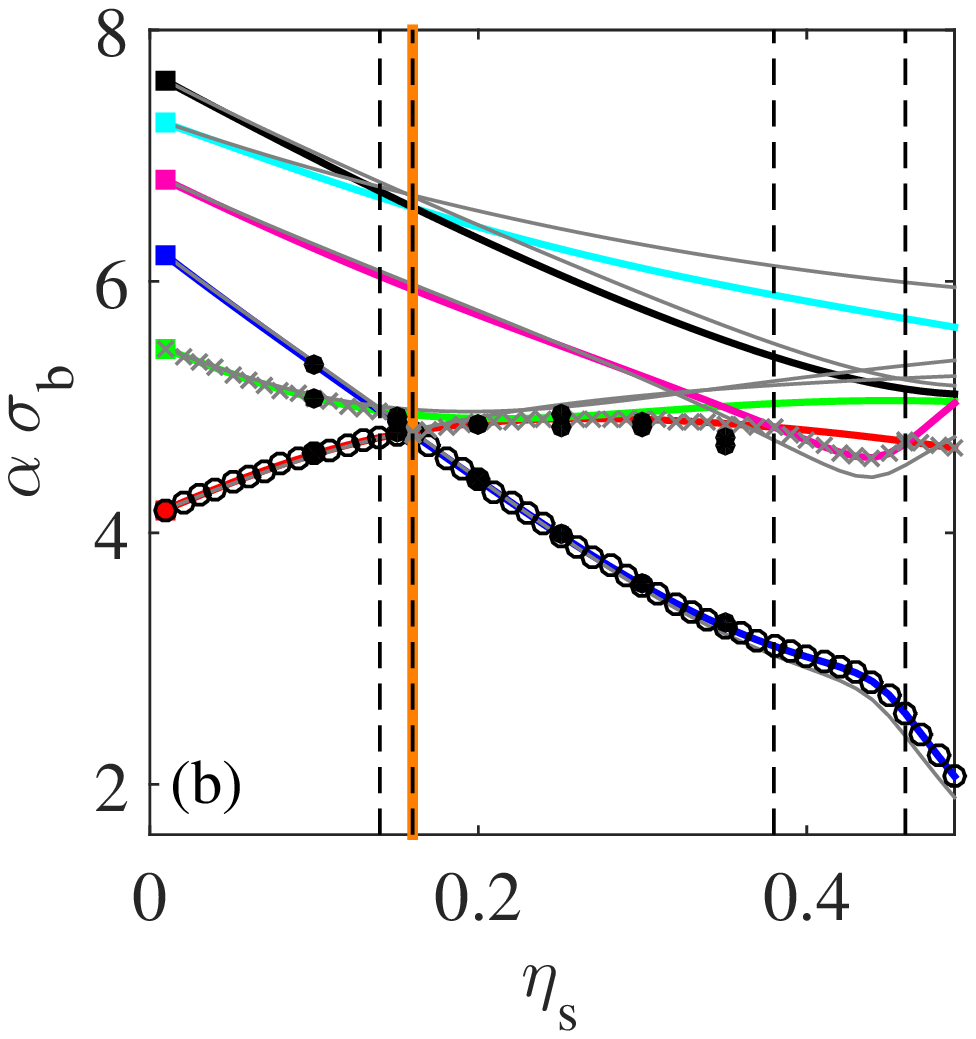}\hspace{0.8cm}
\includegraphics[width=0.26\textwidth]{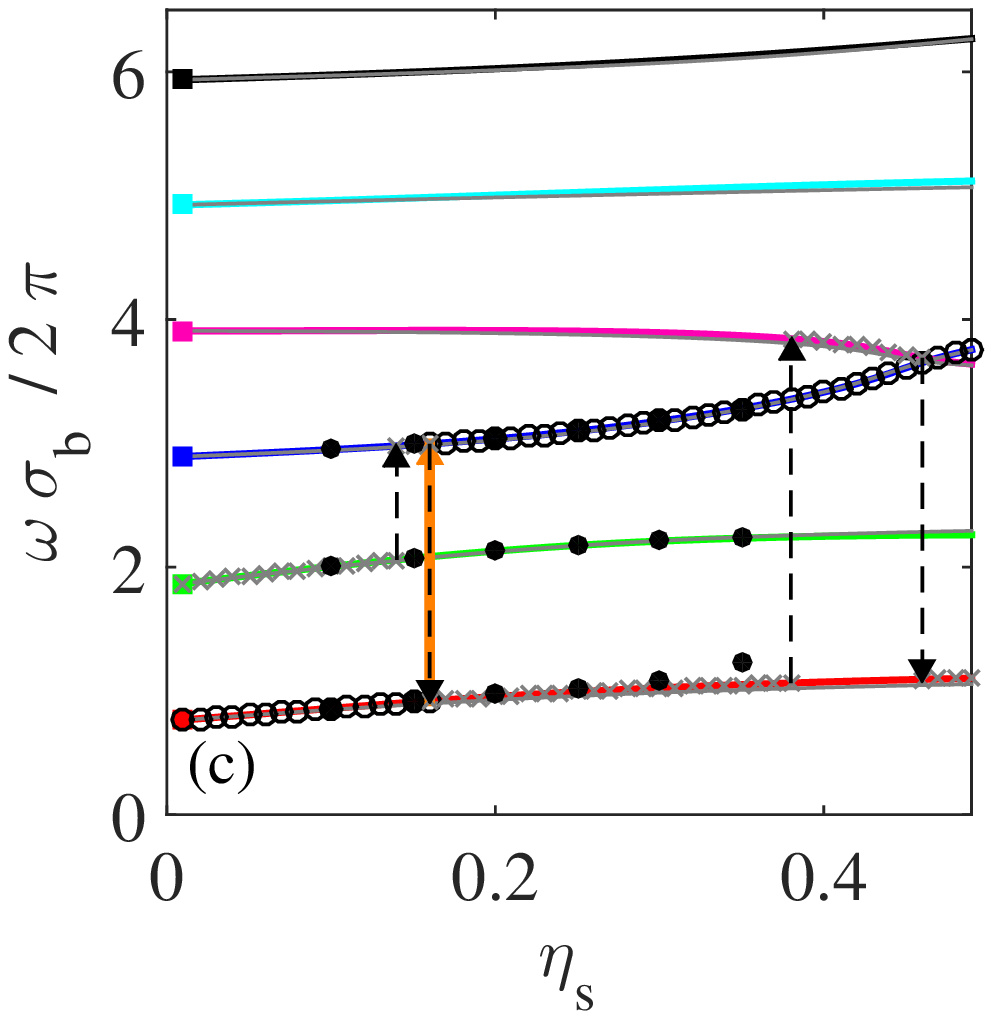}\\
\vspace{0.5cm}
\includegraphics[width=0.26\textwidth]{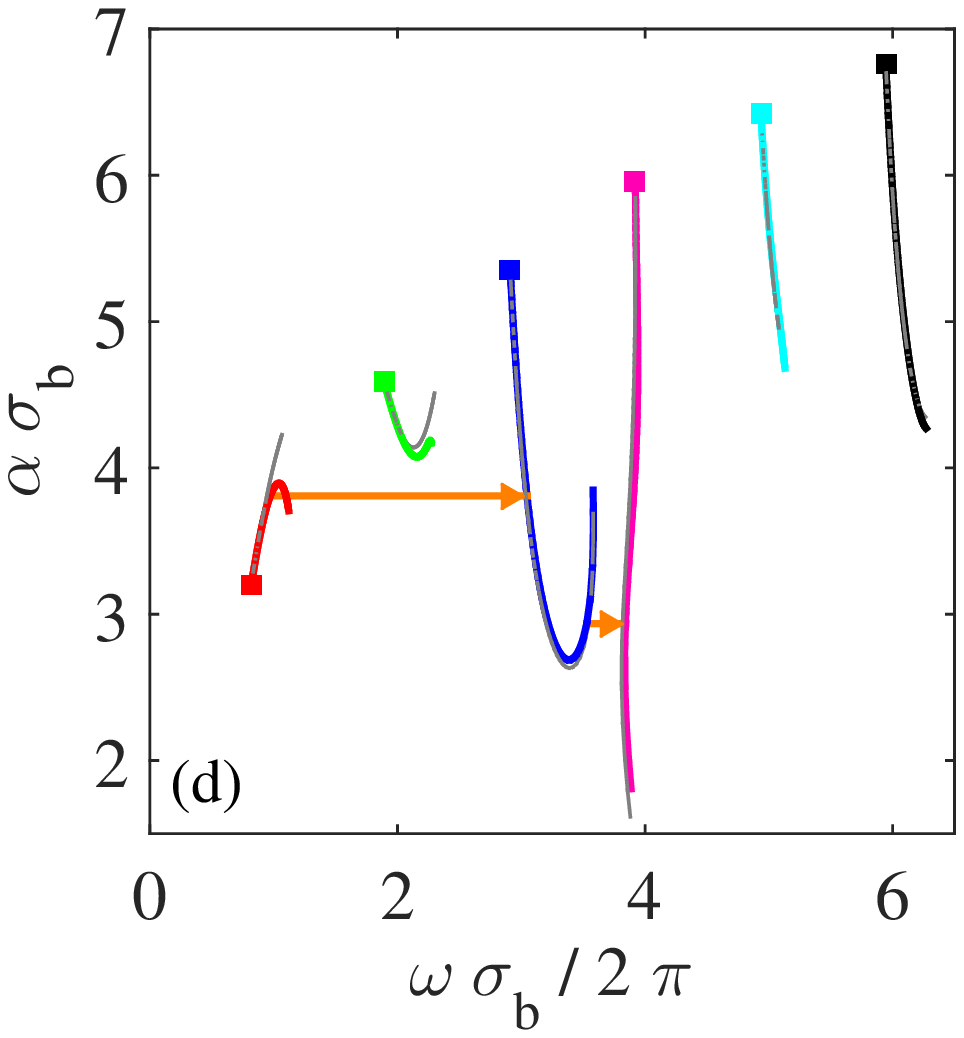}\hspace{0.8cm}
\includegraphics[width=0.26\textwidth]{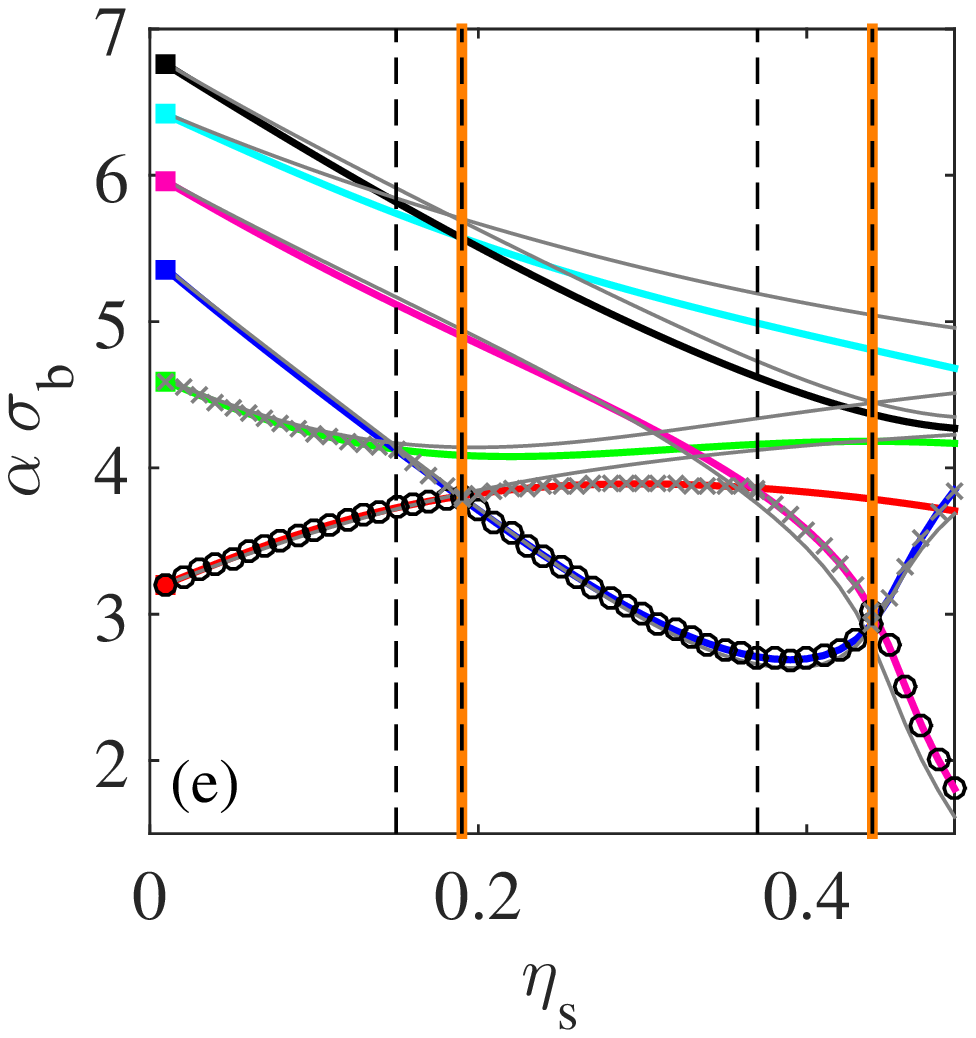}\hspace{0.8cm}
\includegraphics[width=0.26\textwidth]{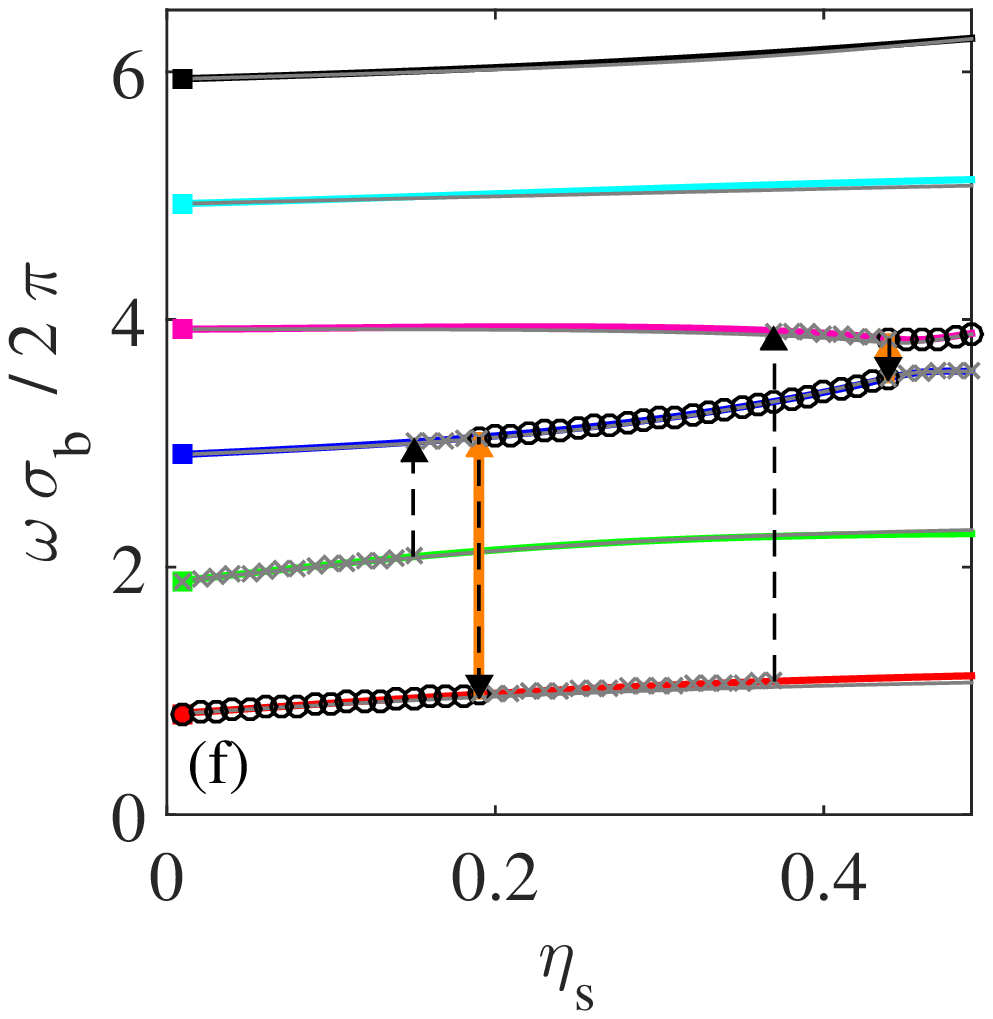}\\
\vspace{0.5cm}
\includegraphics[width=0.26\textwidth]{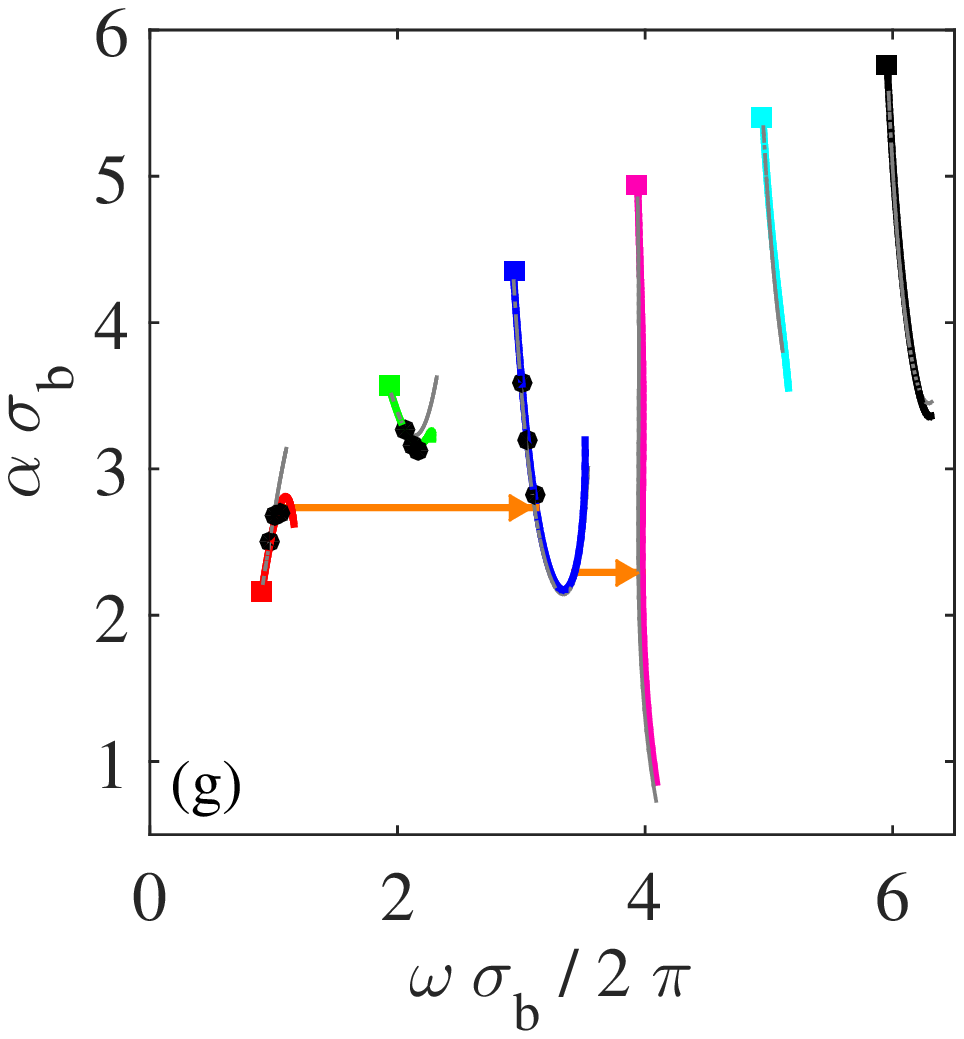}\hspace{0.8cm}
\includegraphics[width=0.26\textwidth]{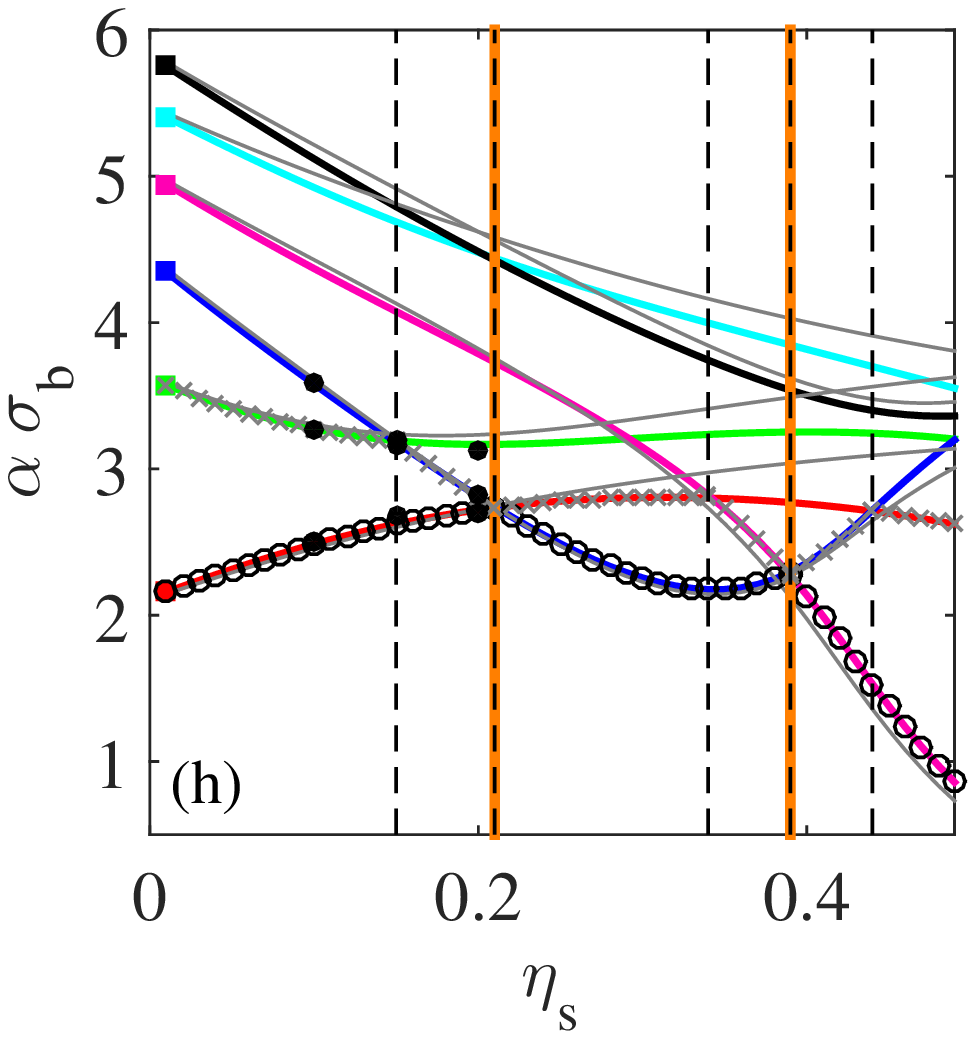}\hspace{0.8cm}
\includegraphics[width=0.26\textwidth]{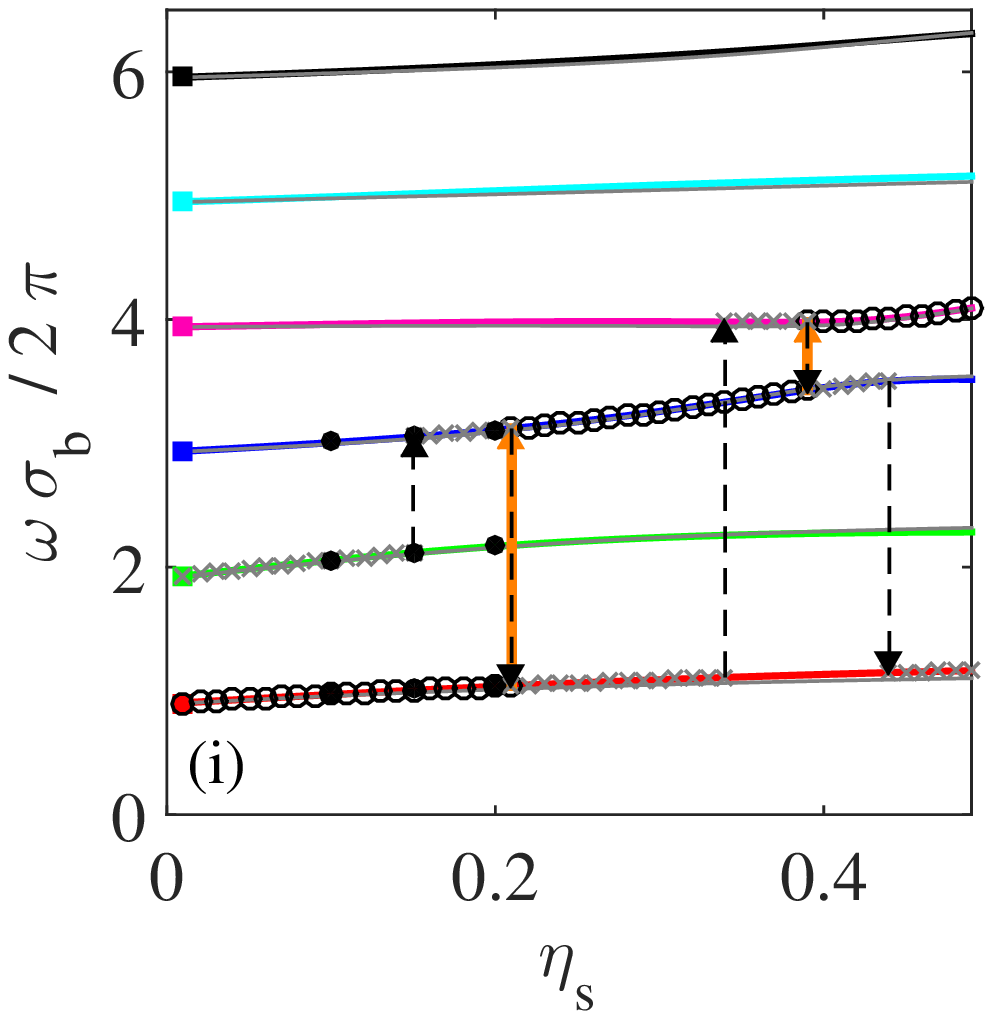}\\
\vspace{0.25cm}
\hspace{0.8cm}\includegraphics[width=1.8\columnwidth]{fig_legend.eps}\\
\caption{Dependence of the first six poles on $\eta_s$ for $q=0.3$ and
$\eta_b=0.05$ (top panels),  $\eta_b=0.10$ (middle panels), and $\eta_b=0.20$ (bottom panels). The thick
{(colored)} and thin
{(gray)} lines correspond to the RFA and PY predictions, respectively, while the solid circles represent the
WM values for the cases where MD simulations were performed.
{The colored squares denote poles for a small sphere packing fraction $\eta_s=0.01$, and the	lines indicate trajectories for increasing values of $\eta_s$.}
The horizontal lines denote the crossovers as $\eta_s$ increases. In the central and right panels, the open circles and crosses represents the leading and subleading poles, respectively.}
  \label{fig:q03}
\end{figure*}

\begin{figure*}
\includegraphics[width=0.26\textwidth]{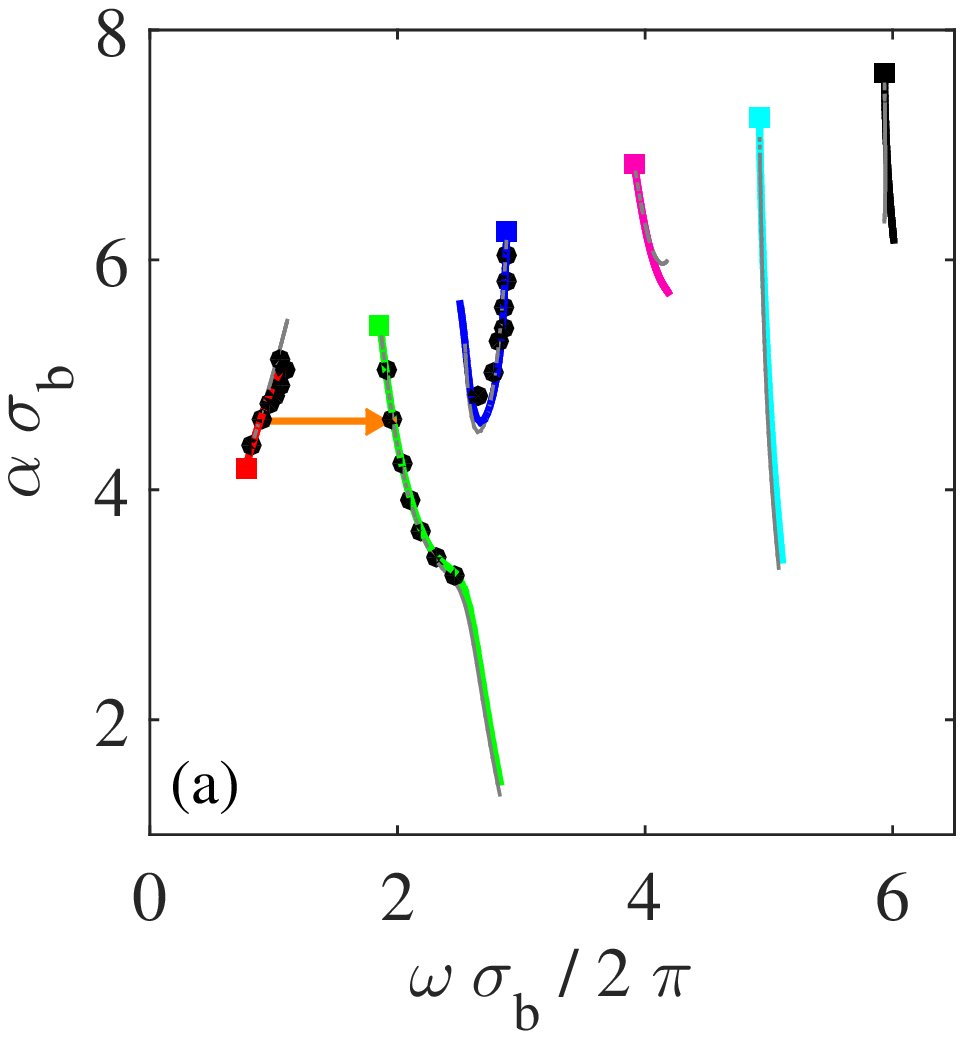}\hspace{0.8cm}
\includegraphics[width=0.26\textwidth]{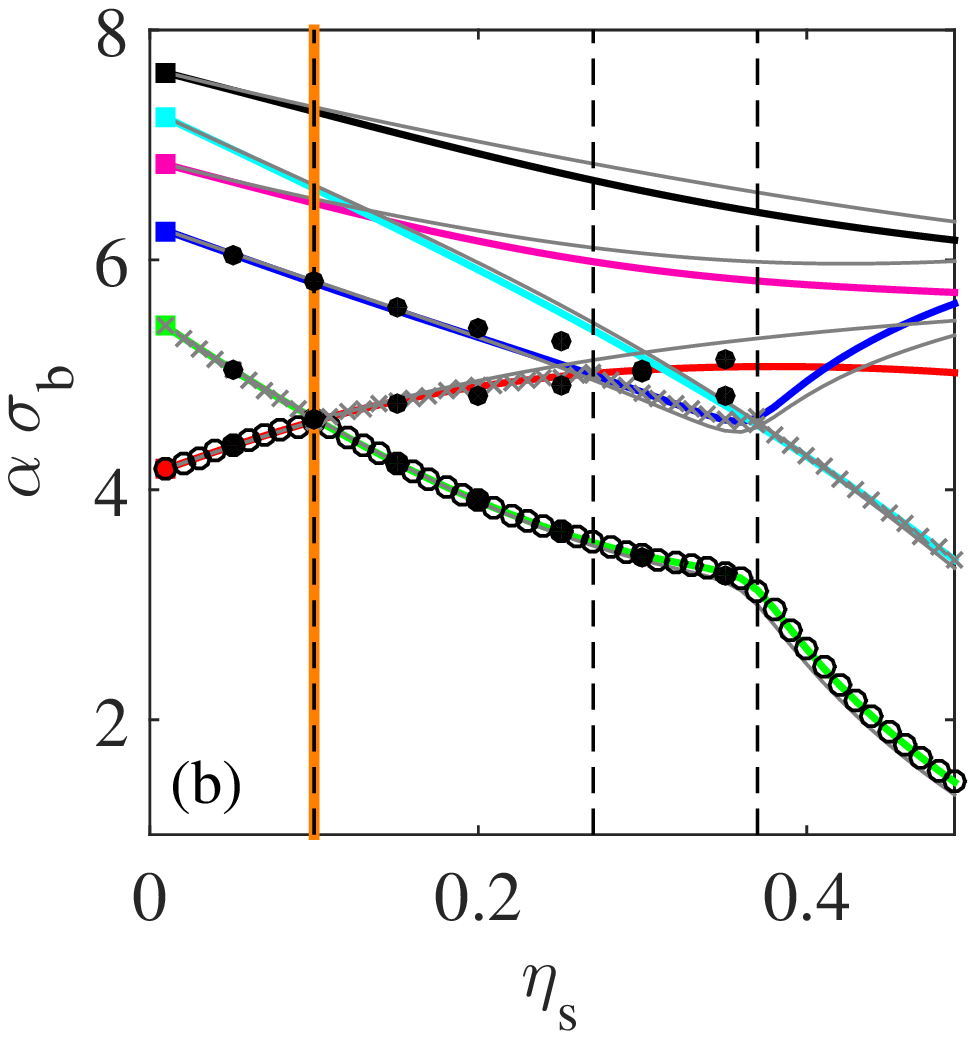}\hspace{0.8cm}
\includegraphics[width=0.26\textwidth]{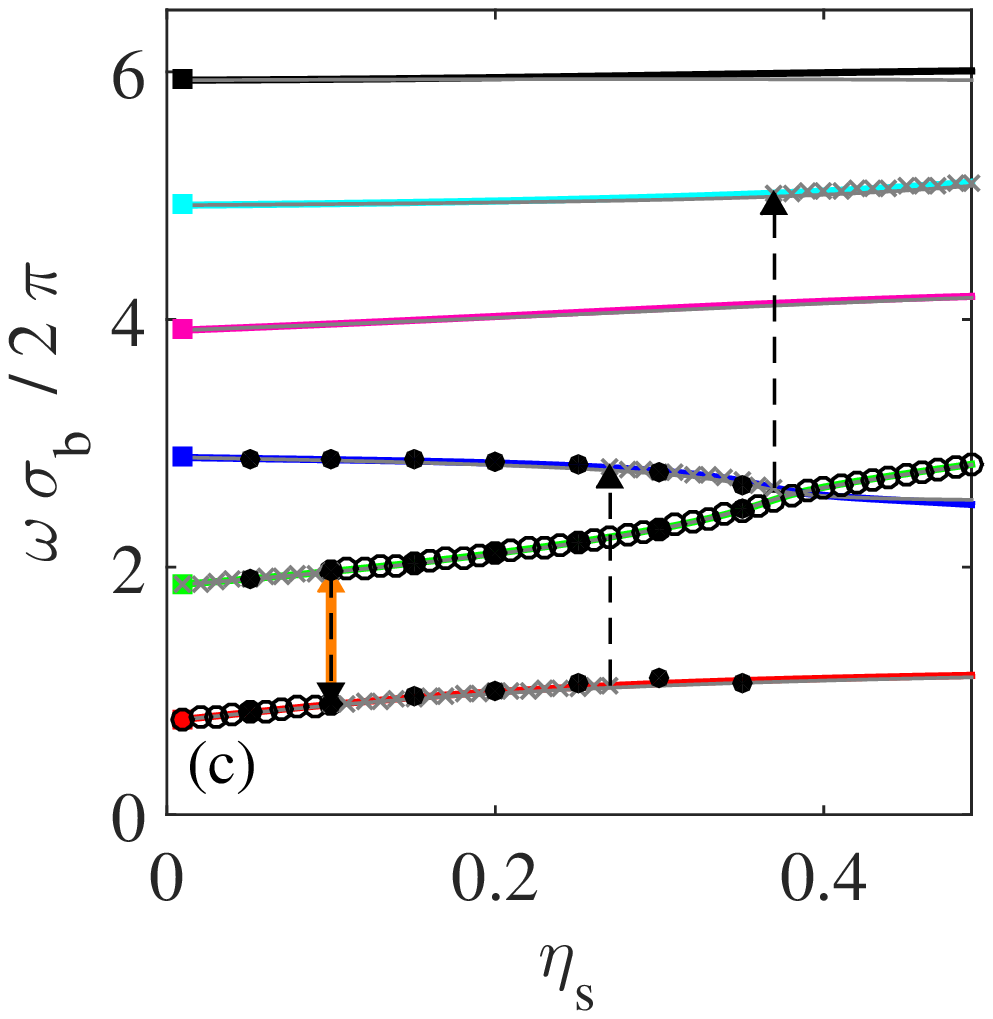}\\
\vspace{0.5cm}
\includegraphics[width=0.26\textwidth]{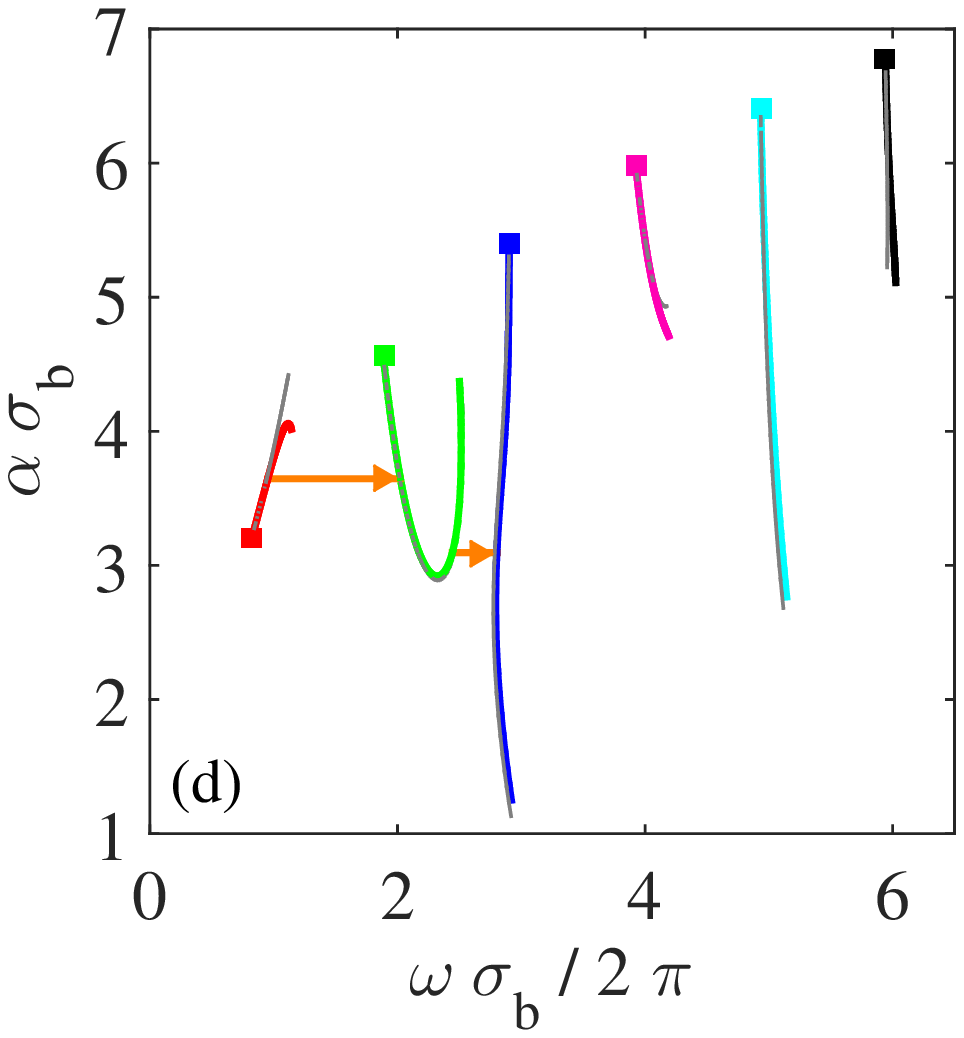}\hspace{0.8cm}
\includegraphics[width=0.26\textwidth]{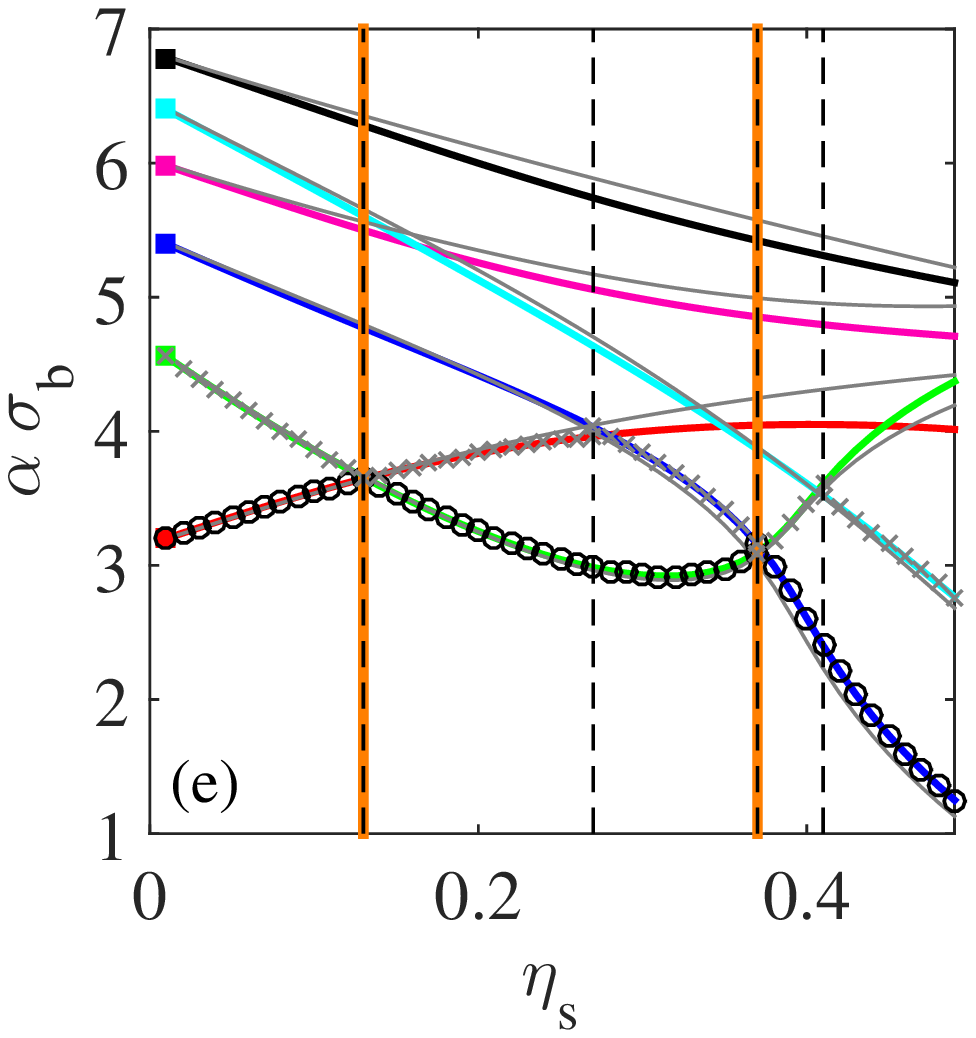}\hspace{0.8cm}
\includegraphics[width=0.26\textwidth]{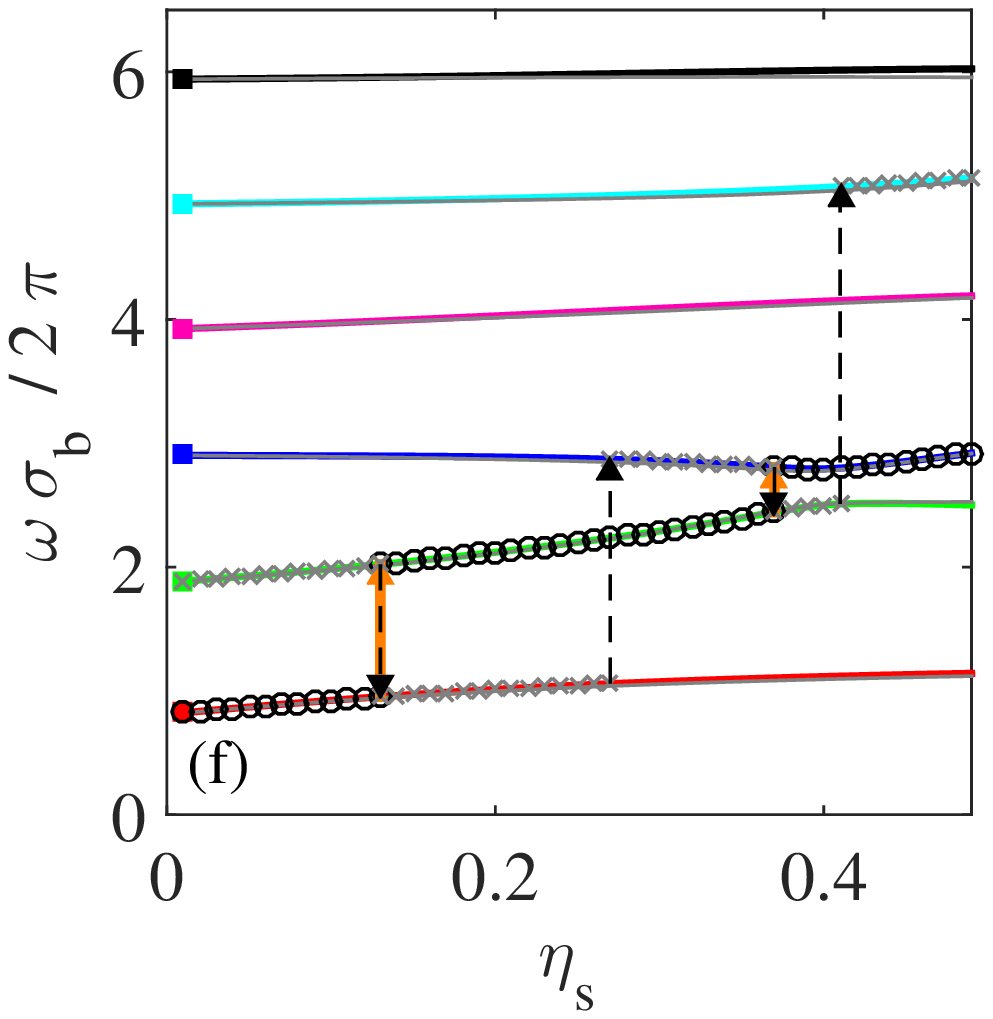}\\
\vspace{0.5cm}
\includegraphics[width=0.26\textwidth]{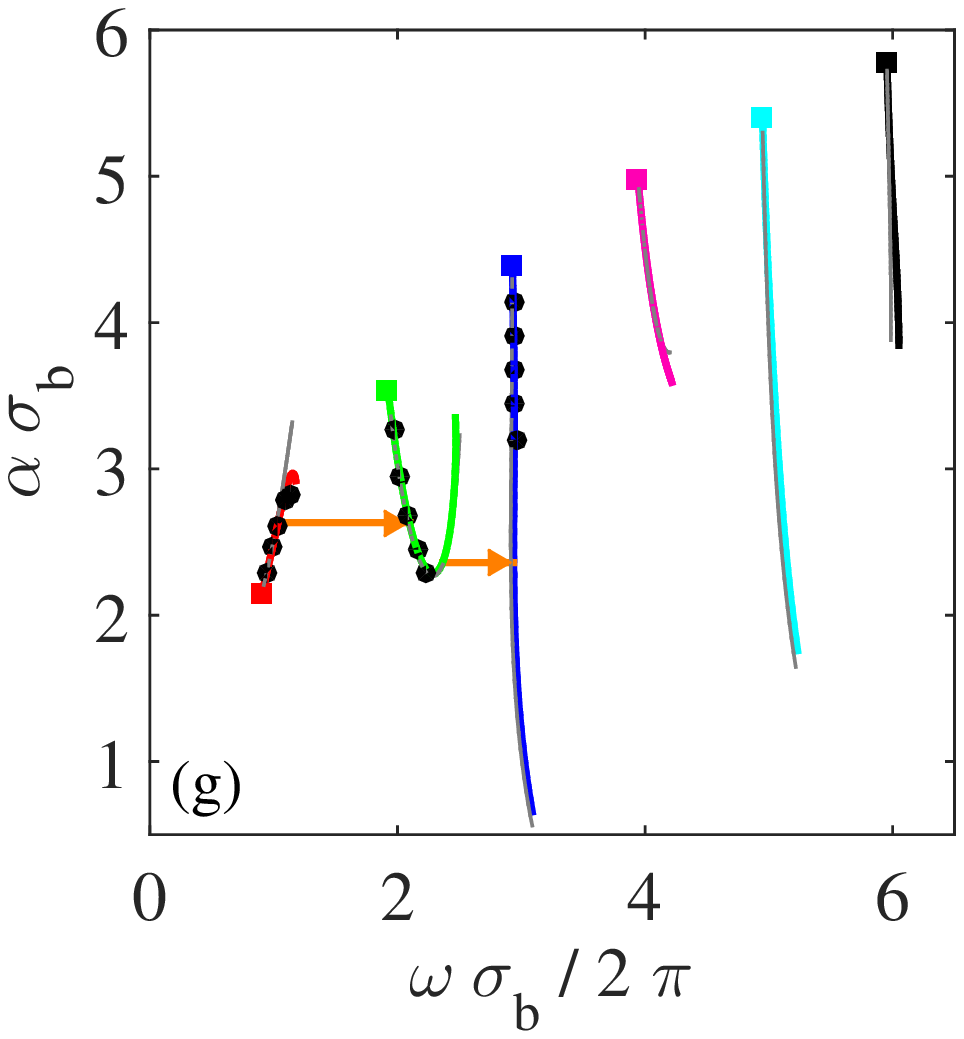}\hspace{0.8cm}
\includegraphics[width=0.26\textwidth]{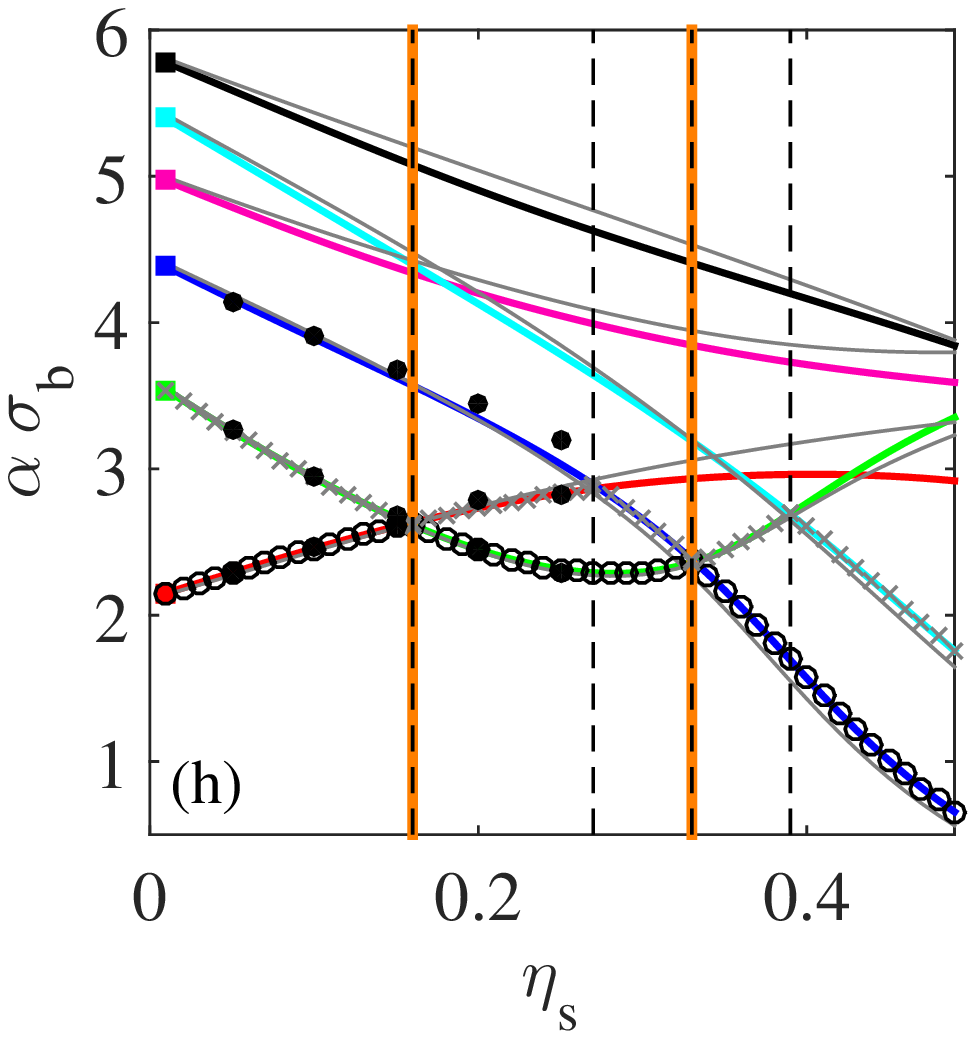}\hspace{0.8cm}
\includegraphics[width=0.26\textwidth]{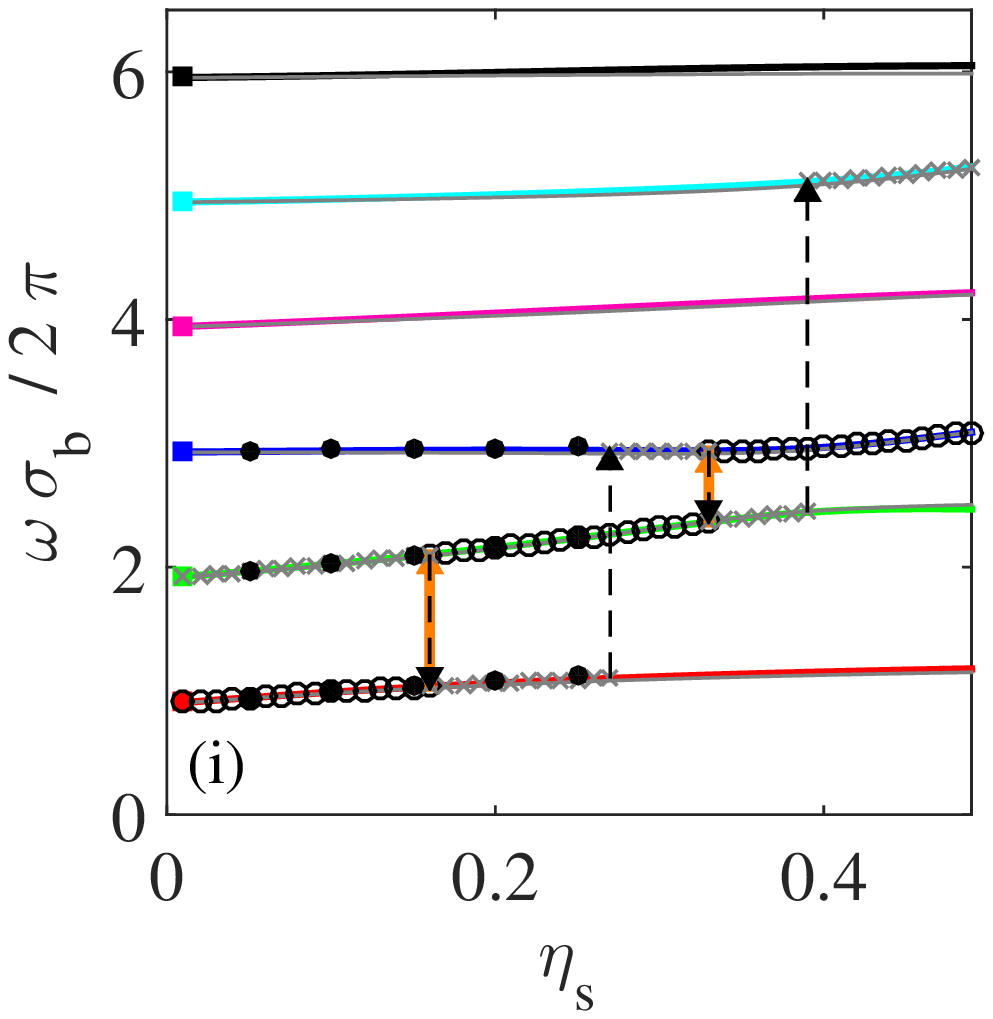}\\
\vspace{0.25cm}
\hspace{0.8cm}\includegraphics[width=1.8\columnwidth]{fig_legend.eps}\\
\caption{Same as in Fig.\ \ref{fig:q03}, except that in this instance $q=0.4$.}
\label{fig:q04}
\end{figure*}

\begin{figure*}
\includegraphics[width=0.26\textwidth]{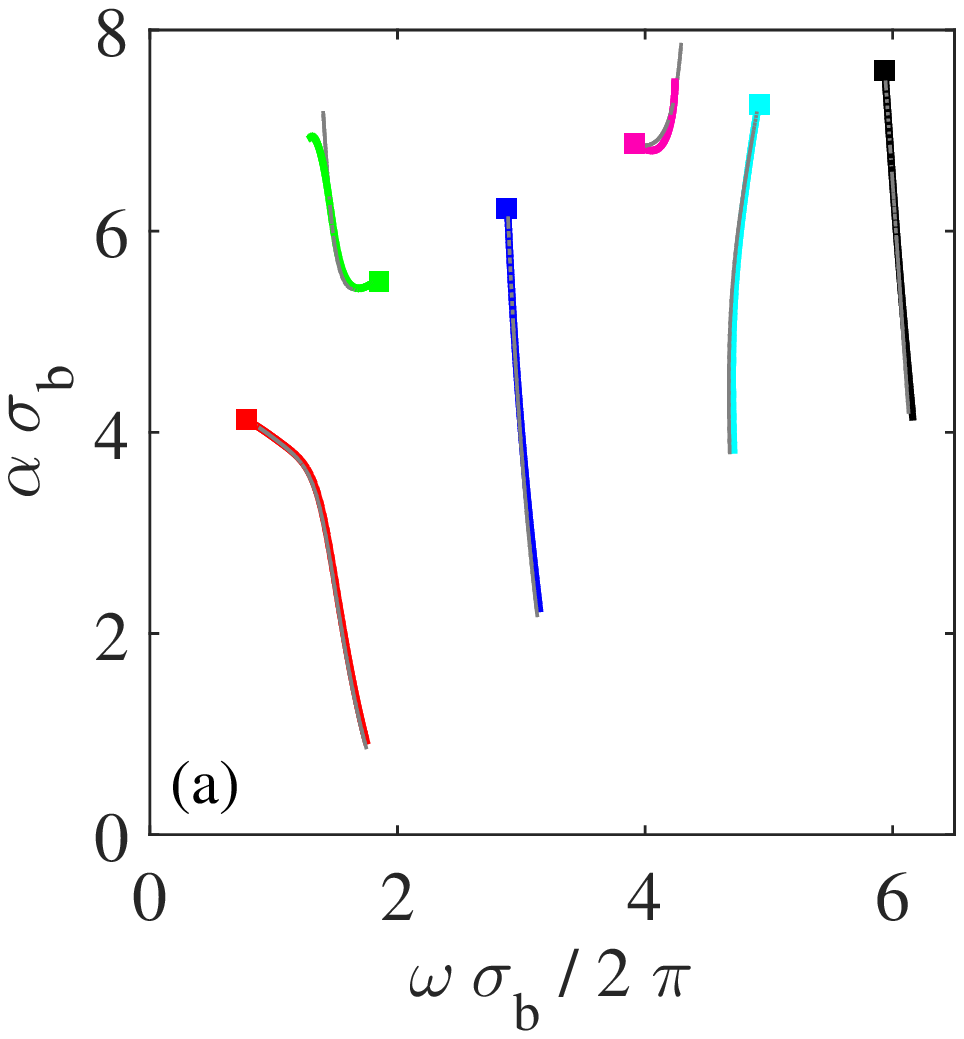}\hspace{0.8cm}
\includegraphics[width=0.26\textwidth]{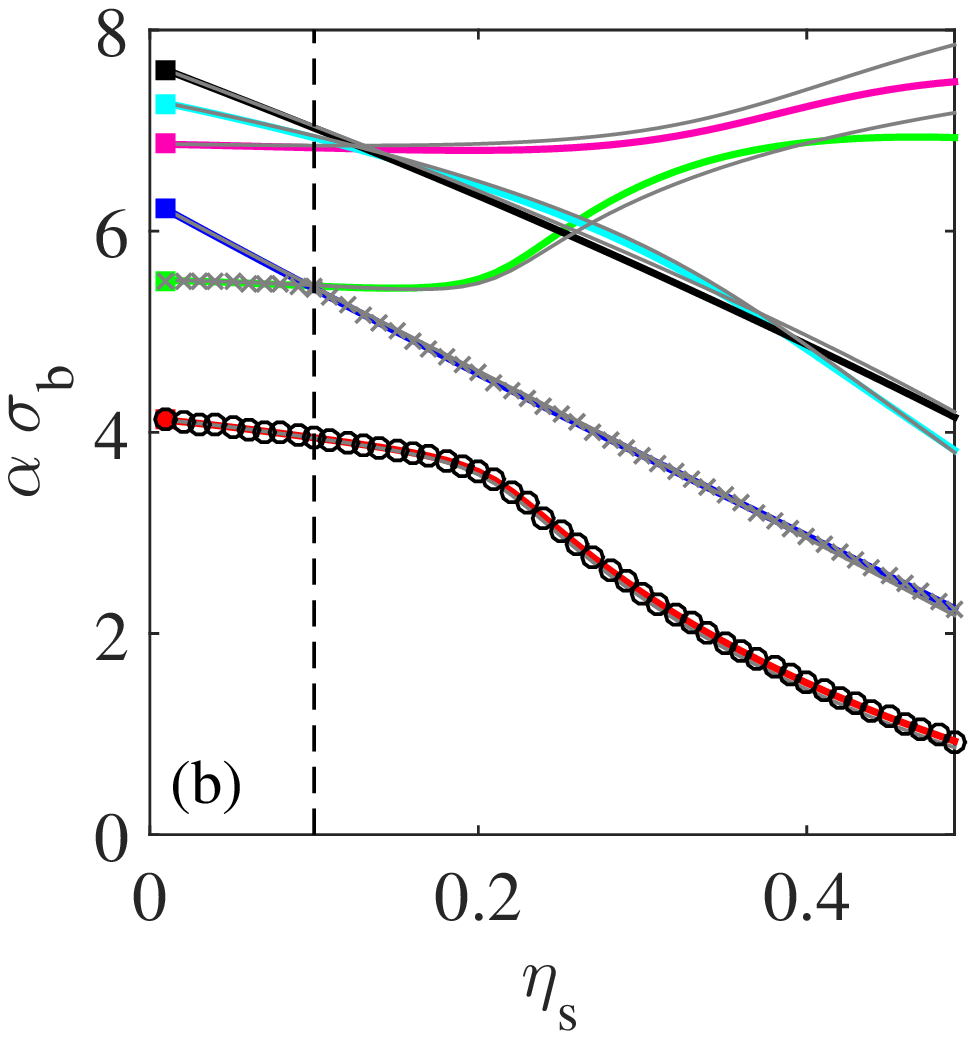}\hspace{0.8cm}
\includegraphics[width=0.26\textwidth]{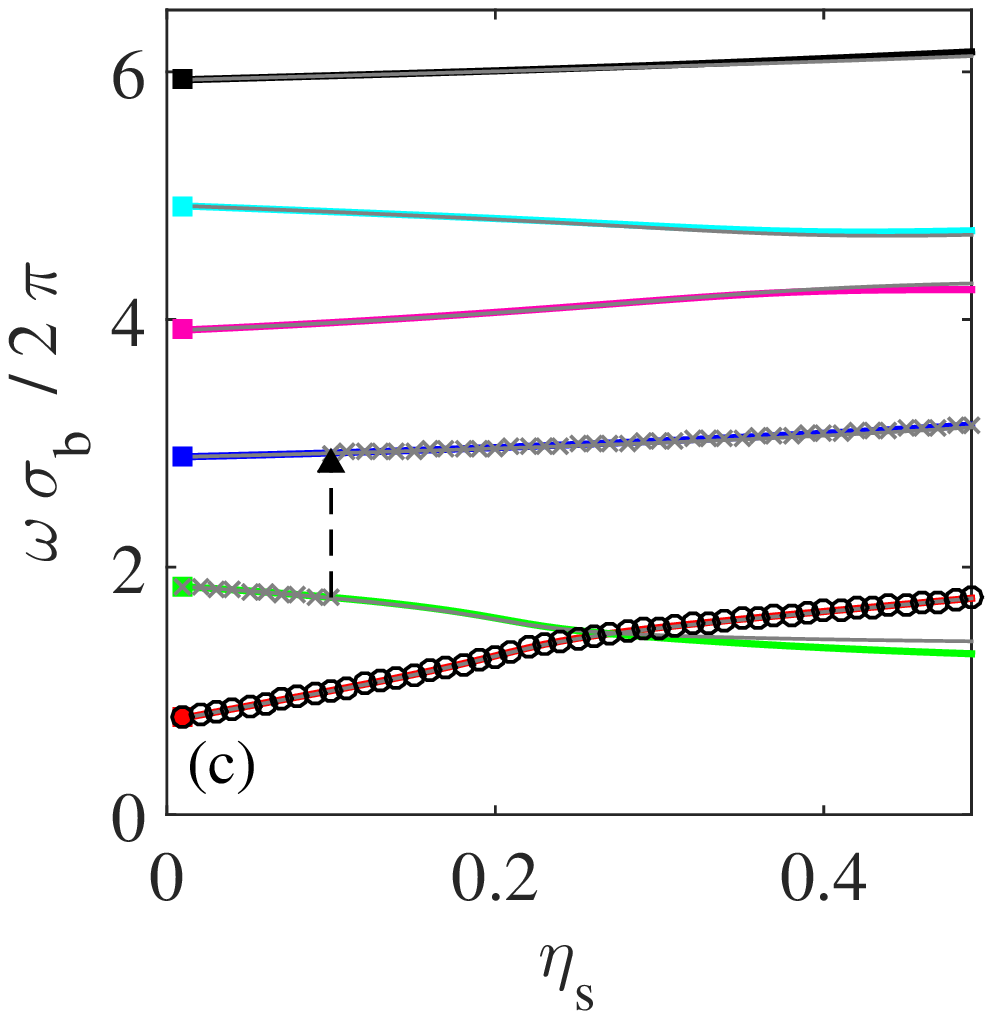}\\
\vspace{0.5cm}
\includegraphics[width=0.26\textwidth]{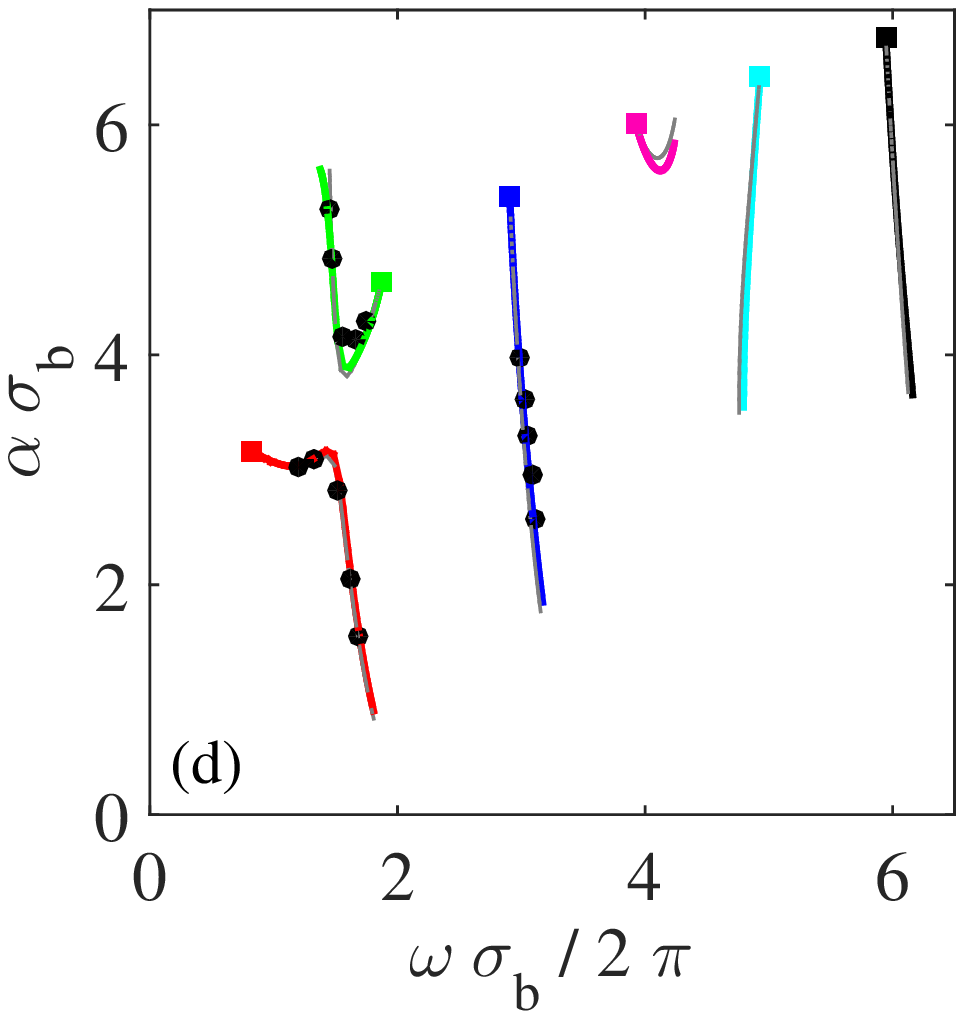}\hspace{0.8cm}
\includegraphics[width=0.26\textwidth]{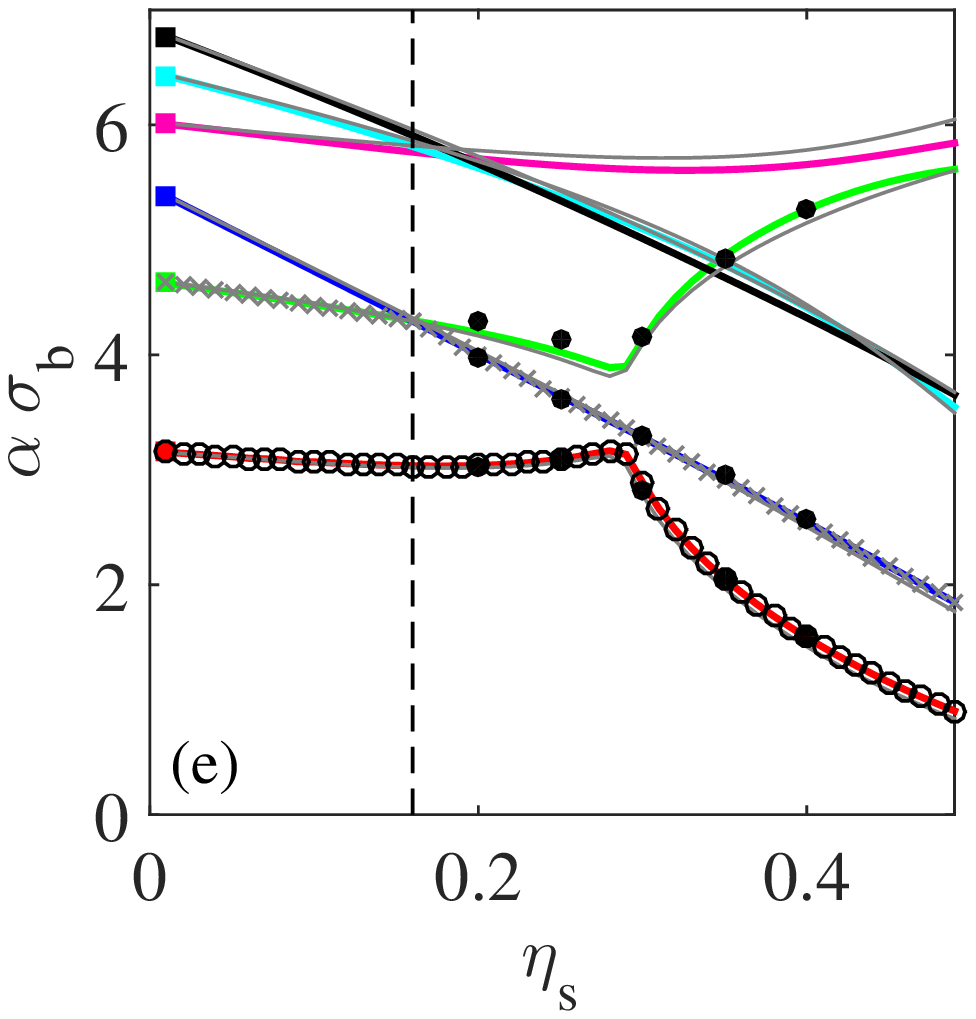}\hspace{0.8cm}
\includegraphics[width=0.26\textwidth]{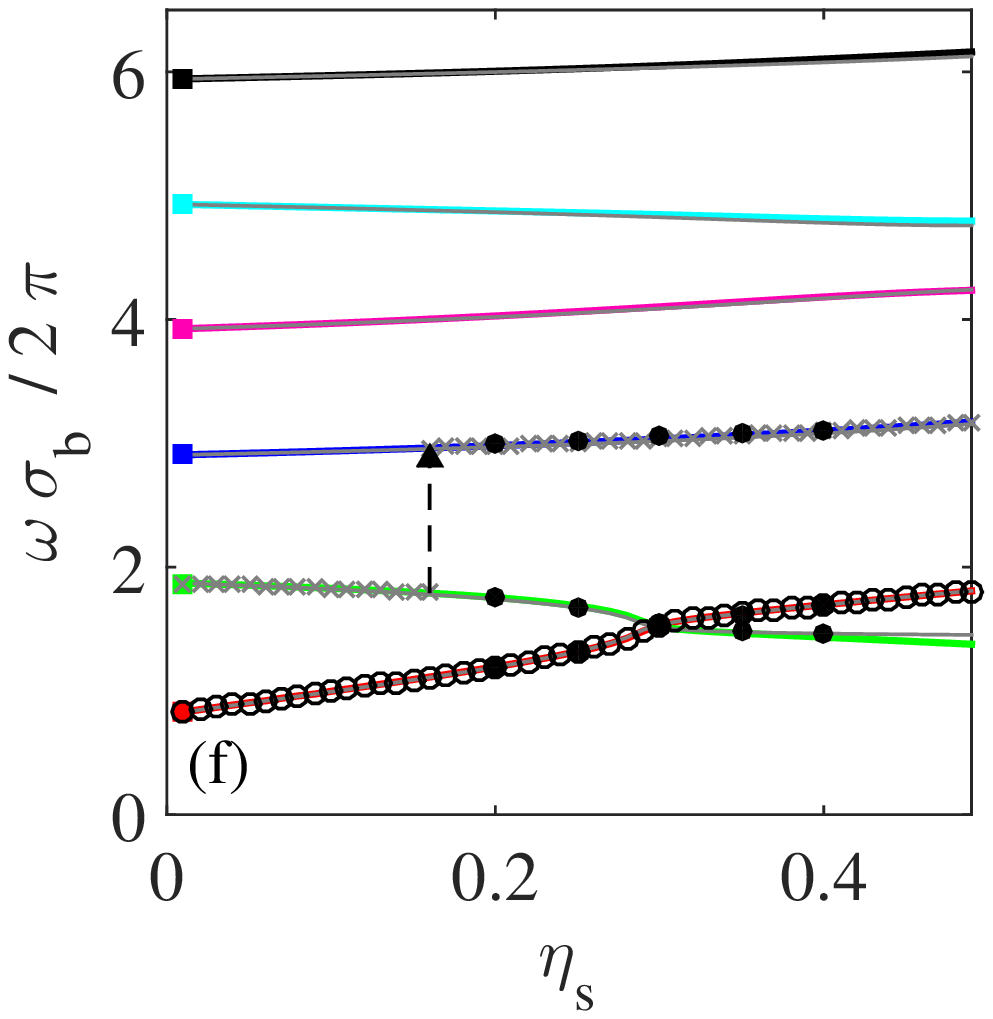}\\
\vspace{0.5cm}
\includegraphics[width=0.26\textwidth]{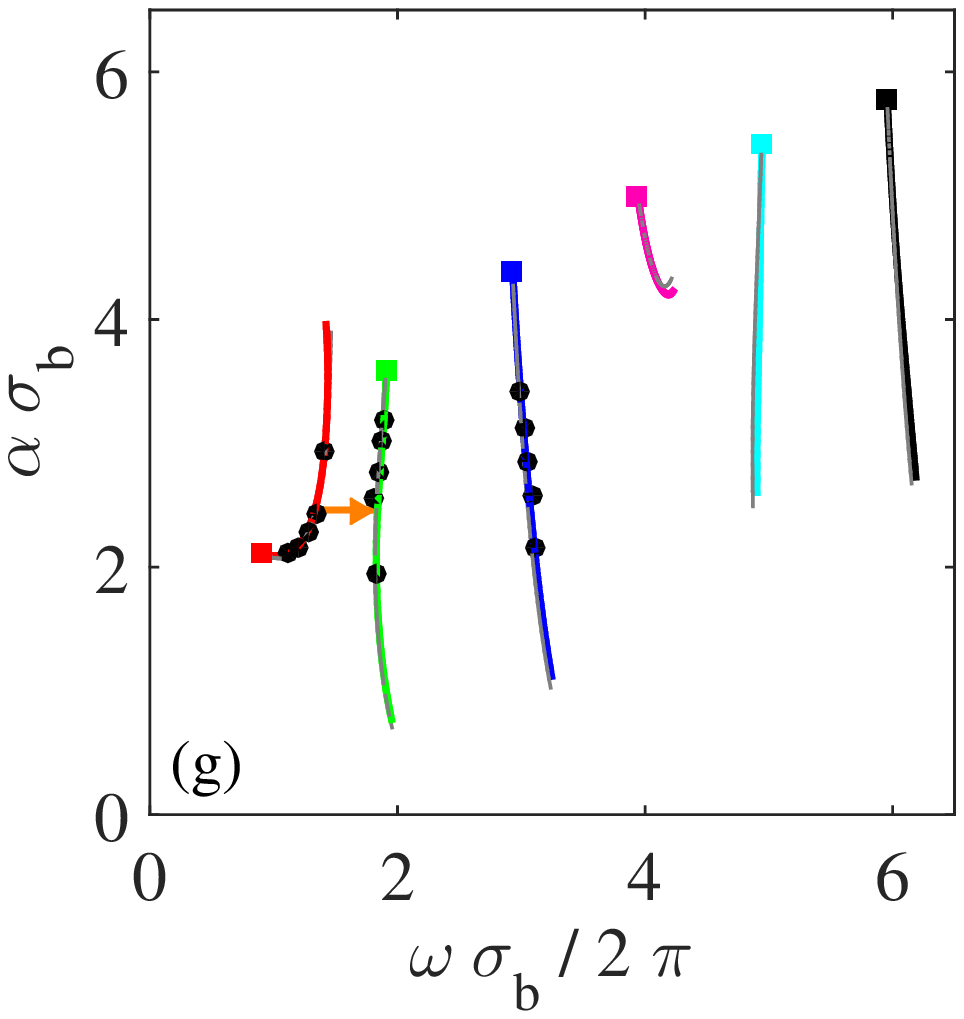}\hspace{0.8cm}
\includegraphics[width=0.26\textwidth]{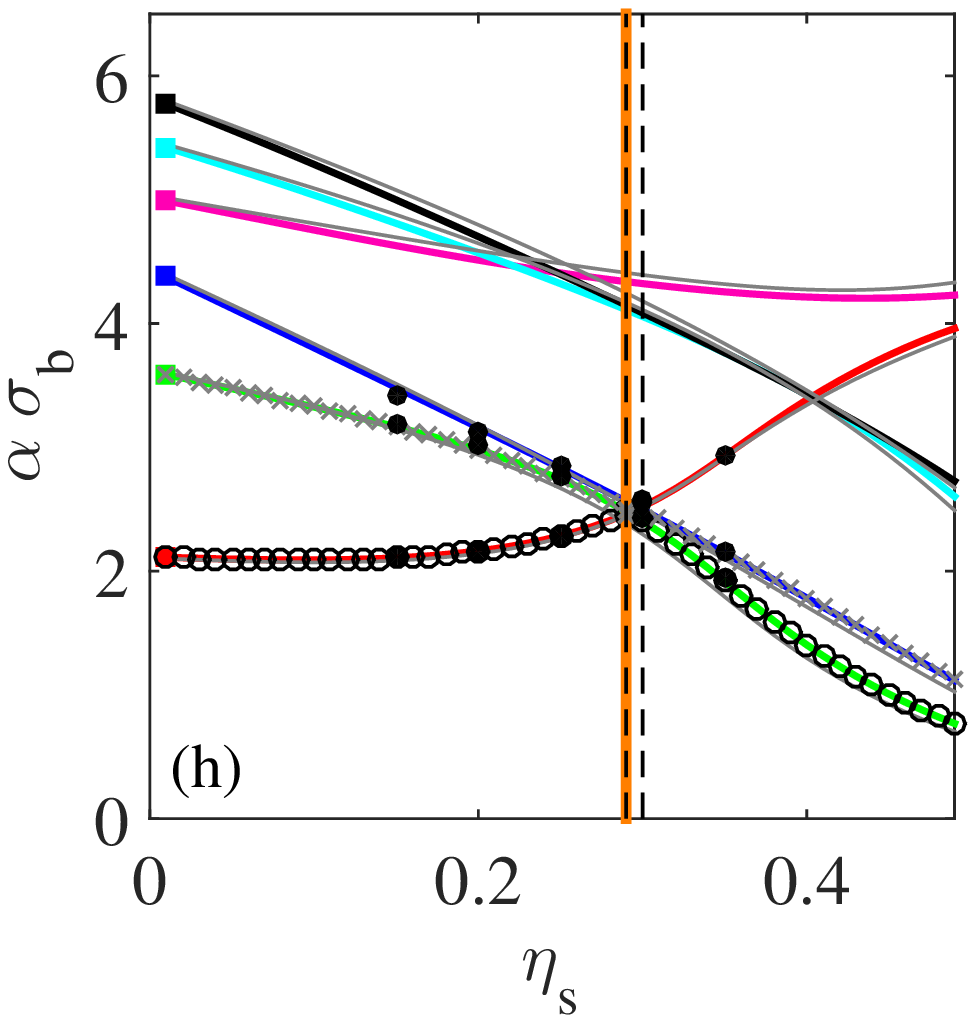}\hspace{0.8cm}
\includegraphics[width=0.26\textwidth]{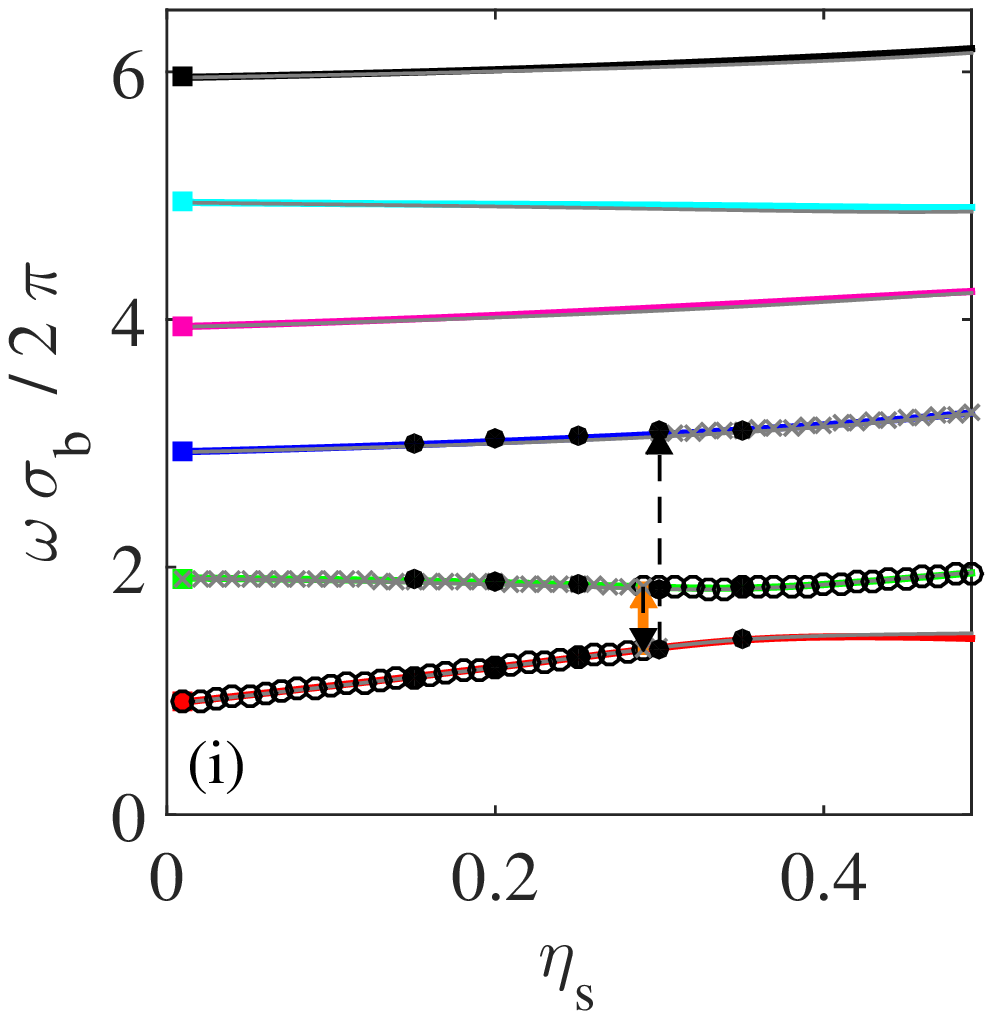}\\
\vspace{0.25cm}
\hspace{0.8cm}\includegraphics[width=1.8\columnwidth]{fig_legend.eps}\\
\caption{Same as in Fig.\ \ref{fig:q03}, except that $q=0.648$.}
  \label{fig:q0648}
\end{figure*}

\begin{figure}
\includegraphics[width=0.45\textwidth]{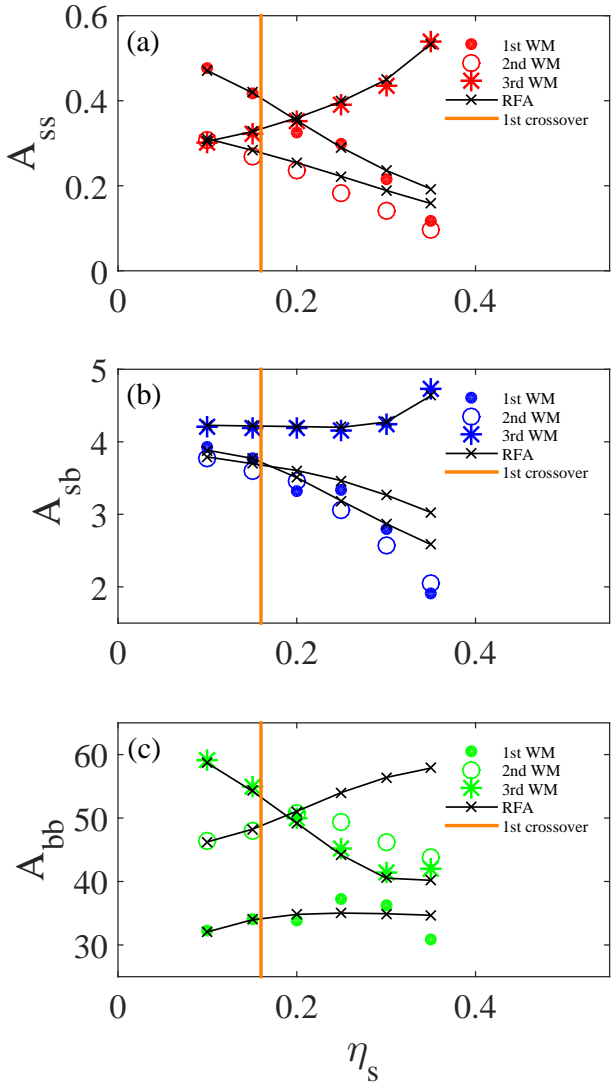}
\caption{Plot of the amplitudes (a) $A_{ss}$, (b) $A_{sb}$, and (c) $A_{bb}$  as functions of $\eta_s$ at $\eta_b=0.05$ for $q=0.3$. The first three poles (in order of increasing $\omega$) are considered. The vertical thick (orange) line at $\eta_s\simeq 0.16$ represents a structural crossover.}
  \label{fig:Aij}
\end{figure}

\begin{figure}
\includegraphics[width=0.4\textwidth]{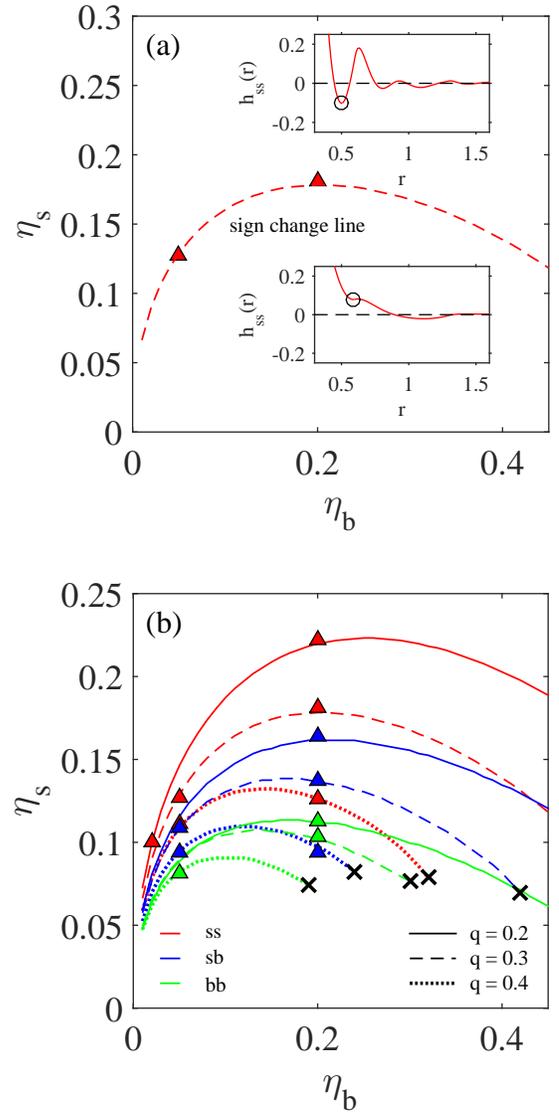}
\caption{(a) Locus $h_{ss}^{\min}=0$ [i.e., line of changing sign in the value of the first minimum of $h_{ss}(r)$], as predicted by the RFA,  for a size ratio $q=0.3$. The insets show representative behaviors of $h_{ss}(r)$ above and below the locus, and the triangles are MD results.
(b) RFA predictions for the loci (from top to bottom for each $q$) $h_{ss}^{\min}=0$ (red), $h_{sb}^{\min}=0$ (blue), and $h_{bb}^{\min}=0$ (green). The size ratios are  $q=0.2$ (solid lines), $q=0.3$ (dashed lines), and $q=0.4$ (dotted lines). In each case, $h_{ij}^{\min}>0$ in the region below the corresponding curve. Some of the curves end at the points marked with crosses. The triangles represent MD results.}
  \label{fig:locus}
\end{figure}

\begin{figure}
\includegraphics[width=0.45\textwidth]{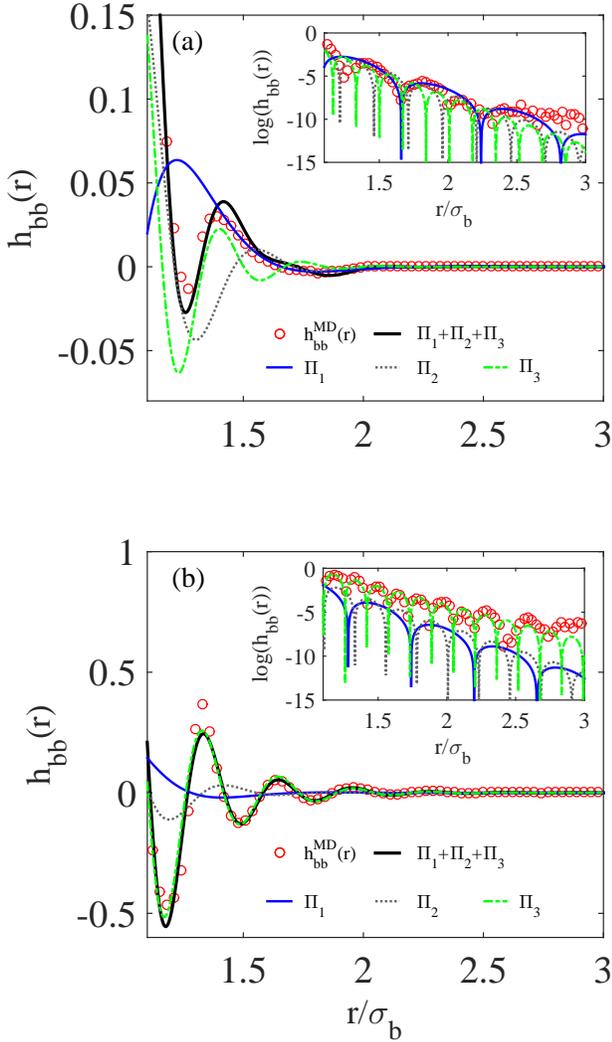}
\caption{Big-big correlation function $h_{bb}(r)$ for $q = 0.3$, $\eta_b=0.05$, and (a) $\eta_s = 0.1$, (b)	$\eta_s = 0.3$. The (red) open  circles are  MD data, while the (blue) solid, (gray) dotted, and (green) dashed lines represent the contribution of the first ($\Pi_1$), second ($\Pi_2$), and third ($\Pi_3$) pole, respectively. Note that at $\eta_s = 0.1$ the leading pole describing the asymptotic decay of $h_{bb}(r)$ is $\Pi_1$ and at $\eta_s = 0.3$ it is $\Pi_3$.}
\label{fig:crossover}
\end{figure}

\section{Pole analysis and structural crossover}
\label{sec3}
In general, the representation of the total correlation functions of additive BHS  mixtures may be expressed as  \cite{HM13,ELHH94,GDER04,PBH17}
\begin{equation}
\label{Eq:hr_reduced}
h_{ij}({r}) =
\begin{cases}
-1, \qquad 0 < \tr < \sigma_{ij}, \\
\sum\limits_{n=1}^{\infty} \frac {\tA_{ij}^{(n)}}{\tr}e^{-\talpha_{n} \tr}\sin\left(\tomega_{n} \tr + \tdelta_{ij}^{(n)}\right) , \qquad
r>\sigma_{ij}.
\end{cases}
\end{equation}

In fact, the functional form of $h_{ij}^{WM}(r)$ for $r>r_{ij}^{\text{min}}$ in Eq.\ \eqref{Eq:hrWM} is based on Eq.\ \eqref{Eq:hr_reduced}. While
the amplitudes, $\tA_{ij}^{(n)}$, and the phase shifts, $\tdelta_{ij}^{(n)}$, are specific for each $h_{ij}(r)$, the damping coefficients, $\talpha_n$, and the  oscillation (angular) frequencies, $\tomega_n\equiv 2\pi/\lambda_n$ ($\lambda_n$ being the associated wavelengths),  are common to all the pairs  \cite{GDER04}.
Although an infinite number of terms is formally considered in Eq.\ \eqref{Eq:hr_reduced}, only a few leading terms (those with the smallest damping coefficients) are needed to characterize the asymptotic behavior of $h_{ij}(r)$.
Note that, while each $\omega_n$ is actually  a wavenumber, we will use throughout this paper the nomenclature ``frequency'' with the proviso that
it does not refer here to time but to space.

A successful way to study  the asymptotic decay behavior of the total correlation functions is based on the pole analysis of their Laplace or Fourier transforms. In Laplace space, the real and imaginary parts of the complex poles $z_n=-\alpha_n\pm \imath \omega_n$ provide the damping coefficient and the oscillation frequencies, respectively. Similarly, in Fourier space the poles are $k_n=-\imath z_n=\pm \omega_n+\imath\alpha_n$.
In order to avoid later confusion, it is convenient at this stage to clarify the nomenclature we adopt in this work. We will refer to leading, subleading, sub-subleading, etc.\ poles to an order in increasing $\alpha_n$, while the nomenclature first, second, third, etc.\ poles will refer to an increasing order in $\omega_n$.
Apart from that, it must be remarked that the infinite set of poles includes roots with $\omega_n=0$ (thus representing contributions decaying monotonically), although they are generally far from the most dominant ones.

Depending on the values of the parameters of the BHS mixture, the position of the poles in the complex plane varies.
In the simplest scenario, given a size ratio $q$, the plane $\eta_s$ vs $\eta_b$ can be split into two main regions
such that the leading pole in one of the regions  has an angular frequency $\omega\approx 2\pi/\sigma_b$, which corresponds to a wavelength in the
oscillatory decay of the total correlation function comparable to the diameter of the big spheres (i.e., $\lambda\approx \sigma_b$), while in the other region the leading pole has $\omega\approx 2\pi /\sigma_s$ (i.e., $\lambda\approx \sigma_s$) \cite{ELHH94,GDER04,PBYSH20}.
The line separating both regions signals a \emph{structural crossover} behavior in which   the  wavelength $\lambda\equiv2\pi/\omega$ of the
oscillations in the large-$r$ asymptotic regime  changes discontinuously from approximately $\sigma_b$ to approximately $\sigma_s$ as the relative
amount of the small spheres is increased. This crossover line in the $\eta_s$ vs $\eta_b$ phase diagram occurs when the corresponding two pairs of poles have the same $\alpha$.  As will be seen in Sec.\ \ref{sec4}, this basic scenario for the structural crossover can become much more complicated as the total packing fraction increases, giving rise to the presence of ``harmonics'' of the  ``fundamental'' oscillation frequency $1/\sigma_b$.

Further, we will talk about a first-order, a second-order, a third-order,
etc.\ crossover to the one involving a change in the leading, subleading, sub-subleading, etc.\ pole, respectively. In particular,  for different values of $q$, $\eta_s$, and $\eta_b$, the decay of $h_{ij}(r)$ is determined by different combinations of poles (i.e., first and second, first and third, first and fifth, and so on).

As said before, in the RFA the poles are obtained from the zeros of $D(z)$ in Eq.\ \eqref{D(s)} and as follows in the case of the WM scheme. Once the total correlation functions $h_{ij}^{WM}(r)$ [see Eq.\ \eqref{Eq:hrWM}] are known after fitting the parameters to the MD data, the direct correlation functions $c_{ij}^{WM}(r)$ are determined via Fourier transforms and the OZ relation in Eq.\ \eqref{Eq:OZ1}, as described in paper I \cite{PBYSH20}. This in turn allows one to find the poles by the method of Evans \emph{et al}.\ \cite{ELHH94} [see Eqs.\ (21) of paper I].

\section{Results}
\label{sec4}
Evidence of the crossover behavior in BHS mixtures was first pointed out  by Grodon \emph{et al.} \cite{GDER04,GDER05}, who used two different formulations of Rosenfeld's fundamental measure theory: the original Rosenfeld functional (which is equivalent to the PY approximation) and the White Bear version. Such crossover behavior was later also mentioned in connection with experiments with colloidal suspensions \cite{BDDRB07,SPTER16}.

Let us now turn to the RFA predictions. We begin with the analysis of the leading pole in the $\eta_s$ vs $\eta_b$ plane.
The detailed landscape as one changes the size ratio is rather complex, so here we provide the most general features \cite{note_21_07}.
By focusing on the behavior of the oscillation frequency $\omega$ associated with the leading pole, one can observe that, given a value of $q$, the $\eta_s$ vs $\eta_b$ plane splits into different regions, in each one of which the (reduced) natural frequency $\omega\sigma_b/2\pi\equiv\sigma_b/\lambda$ is of order of $1$ (region $R_1$) or is of order of $n=2,3,\ldots$ (region $R_n$).  The most relevant regions are $R_1$ (where $\lambda\approx\sigma_b$) and $R_\nq$  (where $\nq$ is the integer closest to $1/q$, so that $\lambda\approx \sigma_s$). Both regions are separated by a crossover line (hereafter labeled as $C$), which is present for any $q$. Interestingly, as $q$ decreases (and thus $\nq$ increases), one can observe a second crossover line ($C'$) separating region $R_\nq$ from either region $R_{\nq+1}$ or region  $R_{\nq-1}$, and even a third line ($C''$) separating region $R_1$ from region $R_{\nq-1}$.

The previous scenario can be observed in Fig.\ \ref{fig:evolution},  which shows the evolution of the different regions and crossover lines as one decreases the size ratio from $q=0.648$ to $q=0.250$. At $q=0.648$ it is quite apparent the existence of the conventional crossover line $C$ separating the regions $R_1$ (below the line) and $R_\nq=R_2$ (above the line). Interestingly, the line $C$ terminates at an ``end point,''  so that one can move continuously between regions $R_1$ and $R_\nq=R_2$ by circumventing the end point from the left. Decreasing the size ratio from $q=0.648$ to $q=0.500$ (top row of Fig.\ \ref{fig:evolution}) produces a downward bending of line $C$ and a left shift of the end point until it eventually disappears at $\eta_b=0$.
Let us now analyze the middle row of Fig.\ \ref{fig:evolution}. At $q=0.425$, region $R_{\nq+1}=R_3$ starts to compete with region $R_\nq=R_2$, thus giving rise to the second crossover line $C'$, which terminates at  a new end point. At the transition value $q=0.400$ (where $\nq$ could  equally be taken as $2$ or $3$) the preceding line $C'$ has started a tendency to bend down and the end point  has moved to the left. Next, at $q=0.375$, $R_3$ has changed from being the old region $R_{\nq+1}$ to being the new region $R_\nq$, while $R_2$ has changed from being the old region $R_{\nq}$ to being the new region $R_{\nq-1}$. Moreover, the old line $C'$ has merged with line $C$ producing a ``splitting point'' (where three different pairs of poles share the same damping coefficient $\alpha$); now, the crossover line $C$ extends to the right of the splitting point, while to the left it experiences a pitchfork bifurcation  into a line $C'$ (separating region $R_\nq=R_3$ from region $R_{\nq-1}=R_2$ and still having an end point) and a line $C''$ (separating regions $R_1$ and $R_{\nq-1}=R_2$). At $q=0.350$, the splitting point has moved to the left, the end point of line $C'$ has disappeared, and the residual
region $R_{\nq-1}$ has significantly shrunk. In the bottom row of Fig.\ \ref{fig:evolution}, we see that, at $q=0.315$, region $R_2$ has almost disappeared, region $R_{\nq+1}=R_4$ starts to compete with region $R_\nq=R_3$, and a new line $C'$ with an end point appears, in analogy with what happened at $q=0.425$. The rest of the evolution is analogous to what has just been described in relation with the middle row: line $C'$ moves down, it eventually merges with line $C$ creating a splitting point, a relay from old region $R_{\nq+1}$ to new region $R_\nq$ and from old region $R_{\nq}$ to new region $R_{\nq-1}$ takes place, and region $R_{\nq-1}$ shrinks until eventually disappearing.

A splitting point is absent in the sequence represented by the top row of Fig.\ \ref{fig:evolution}. In the second row, however, a splitting point (joining regions $R_1$, $R_2$, and $R_3$) is generated and then it disappears, giving rise to the generation of a new splitting point (joining regions $R_1$, $R_3$, and $R_4$) and its later disappearance along the bottom row. Thus, we will refer to the behavior represented by the top, middle, and bottom rows of Fig.\ \ref{fig:evolution} as generations G0, G1, and G2, respectively. As $q$ keeps decreasing beyond $q=0.250$, a new generation G3 appears. The evolution with $q$ of the end point (generations G0--G3) and splitting points (generations G1--G3) are shown in Fig.\ \ref{fig:R3}.

At this stage we should point out two things. On the one hand,  it is worth mentioning that the presence of a second crossover line ({i.e.}, the $C'$ line) was already also reported in Ref.\ \cite{GDER04}, at least for $q = 0.4$. Nevertheless, the distinction between the $C$, $C'$, and $C''$ lines that we have made here is important for understanding and/or predicting the structural crossover behavior, including the appearance of different branches for different $q$ systems. For instance, without such assets, neither the reason for the sequence of the crossover lines in Fig.\ 5 of Ref.\ \cite{GDER04} nor the reasons for the appearance of the second crossover in the case $q=0.4$ or of the  behavior observed in the case $q=0.65$ can be explained. In any case, the overall view of the structural crossovers becomes clearer once one realizes that  the scenario depicted by Figs.\ \ref{fig:evolution} and \ref{fig:R3} takes place. Note in particular that, notwithstanding the theoretical interest of the crossover line $C'$, it must be remarked that, as seen from Fig.\ \ref{fig:evolution}, it generally lies (except for sufficiently asymmetric mixtures) above the region $\eta_s+\eta_b=0.5$, where the fluid phase is  expected to coexist with a solid phase \cite{DRE99b}.
On the other hand, it should also be clear that the present scenario is only a coarse-grained description and, while providing a fair picture of what goes on, is not geared towards addressing all the details pertaining to shrinking regions, the merging of crossover lines, the disappearance and appearance of end points, and the formation of splitting points.

To complete the picture advocated in the present paper, let us now present the results of the RFA and the WM scheme for both the damping coefficient and the oscillation frequency associated with the leading pole for $q=0.2$, $0.3$, $0.4$, and $0.648$. These  results are displayed in Fig.\ \ref{fig:plane}.
We observe that the leading damping coefficient smoothly changes with $\eta_s$, $\eta_b$, and $q$, except for the presence of kinks signaling pole crossings. In general, given a value of $\eta_s$, the reduced damping coefficient $\alpha\sigma_b$ decreases with increasing $\eta_b$. On the other hand,  $\alpha\sigma_b$ exhibits a nonmonotonic dependence on $\eta_s$ at fixed $\eta_b$. In what respects $q$, its influence on $\alpha\sigma_b$ is rather weak, although there is a general tendency for a slight decrease of $\alpha\sigma_b$ with increasing $q$ at fixed $(\eta_b,\eta_s)$.
In contrast to the behavior of the damping coefficient, the oscillation frequency can experience discontinuous changes in the $\eta_s$ vs $\eta_b$ diagram at a given $q$,  as discussed above in connection with Fig.\ \ref{fig:evolution}.
The complexity of the landscape beyond the description of Fig.\ \ref{fig:evolution} is exemplified by Fig.\ \ref{fig:plane}(e) for $q=0.2$, where, apart from the  pitchfork bifurcation  at $(\eta_b,\eta_s)\simeq (0.04,0.26)$ (giving rise to an encapsulated  region $R_4$), a second splitting point is born at $(\eta_b,\eta_s)\simeq (0.29,0.38)$.

As a representative example, Fig.\ \ref{fig:plane3D} shows a 3D view visualizing the overall structure behavior, in particular  the crossover lines $C$ and $C'$ and the specific regions $R_1$, $R_2$, and $R_3$ for the size ratio $q=0.4$.
This helps the understanding of the different features observed  in Figs.\ \ref{fig:evolution} and
\ref{fig:plane}, as well as in Figs.\ \ref{fig:q02}--\ref{fig:q0648} below.
In particular, Fig.\ \ref{fig:plane3D}(b) shows that, as said before, one
can move continuously between regions $R_2$ and
$R_3$ by circumventing  the end point.

For the investigated cases $q=0.2$, $0.3$, $0.4$, and $0.648$, the crossover lines $C$ and $C'$ are plotted in Fig.\ \ref{fig:plane_all}, where RFA and PY predictions, as well as a few points obtained via the WM scheme, are represented. As can be observed, the shape of lines $C$ and $C'$ is qualitatively similar for $q=0.3$ and $q=0.4$, while the cases $q=0.2$ and $q=0.648$ present distinctive features.
We observe that the black solid circle  representing the result from the WM method for $q=0.648$ and $\eta_b=0.20$ lies on the line obtained from the RFA prediction better than on the PY line.
It is worth noting that the differences between RFA and PY grow with increasing density and decreasing size ratio, being especially apparent for $q=0.2$. In that case, for instance, the second splitting point changes from $(\eta_b,\eta_s)\simeq (0.29,0.38)$ in the RFA to $(\eta_b,\eta_s)\simeq (0.46, 0.27)$ in the PY approximation.
In any case, it must be remarked that the main separation between the RFA
and PY lines for $q=0.2$ takes place in regions of the plane $(\eta_b,\eta_s)$ where the total packing fraction is rather large ($\eta>0.6$) and
hence the stable system is expected to consist of coexisting fluid and solid phases \cite{DRE99b}.

The information presented in Figs.\ \ref{fig:plane}  and \ref{fig:plane_all} is complemented by Figs.\ \ref{fig:q02}--\ref{fig:q0648} for $q=0.2$, $0.3$, $0.4$, and $0.648$, respectively, where the dependence on $\eta_s$ of the first six poles is shown for $\eta_b=0.02$ [for $q=0.2$] or $0.05$ [for $q=0.3$, $0.4$, and $0.648$] (top panels), $0.10$ (middle
panels), and $0.20$ (bottom panels). It can be observed that typically the first poles correspond to (reduced) frequencies $\sigma_b/\lambda\approx 1$, $\sigma_b/\lambda\approx 1/q$, and the first few harmonics $\sigma_b/\lambda=2,3,4,\ldots$. Note that  a certain overlap between the pole with $\sigma_b/\lambda\approx 1/q$ and that with the nearest harmonic might exist, in what could be viewed as sort of ``resonance''. This happens for $\sigma_b/\lambda\approx 5$, $\sigma_b/\lambda\approx 3$, $\sigma_b/\lambda \approx 2$, and $\sigma_b/\lambda\approx 2$ in the cases $q=0.2$
(see right panels of Fig.\ \ref{fig:q02}), $q=0.3$ (see right panels of
Fig.\ \ref{fig:q03}), $q=0.4$ (see right panels of Fig.\ \ref{fig:q04}), and $q=0.648$ (see right panels of Fig.\ \ref{fig:q0648}), respectively.

As seen from Figs.\ \ref{fig:q03}(c) and \ref{fig:q04}(c), only the conventional crossover $C$ is present at $\eta_b=0.05$ for  $q=0.3$ and $0.4$, in agreement with what is observed in Figs.\  \ref{fig:plane}(f) and \ref{fig:plane}(g). On the other hand, Figs.\  \ref{fig:q03}(f), \ref{fig:q03}(i), \ref{fig:q04}(f), and \ref{fig:q04}(i) show that the crossover $C$ is followed by the crossover $C'$ as $\eta_s$ increases at $\eta_b=0.10$ and $0.20$. In the case $q=0.2$, Fig.\ \ref{fig:q02}(c) shows two successive crossovers  with $\eta_b=0.02$ when traversing the two branches stemming from the pitchfork bifurcation at the splitting point $(\eta_b,\eta_s)\simeq (0.04,0.26)$. Also for $q=0.2$, the crossovers $C$ and $C'$ are observed in Fig.\ \ref{fig:q02}(i) at $\eta_b=0.20$, while at $\eta=0.10$ the second crossover $C'$ is beyond the range of $\eta_s$
shown [see Figs.\ \ref{fig:plane_all} and \ref{fig:q02}(f)].
On the other hand, according to Figs.\ \ref{fig:q0648}(c), \ref{fig:q0648}(f), and \ref{fig:q0648}(i), a single crossover  exists at $\eta_b=0.20$ and $q=0.648$ since $\eta_b=0.05$ and $\eta_b=0.10$ are located to the right of the end point [see Fig.\ \ref{fig:plane}(h)].

Figures \ref{fig:q02}--\ref{fig:q0648} also show that some quantitative differences between the RFA predictions and those of the PY approximation are present, as already mentioned in connection with Fig.\ \ref{fig:plane_all}. We can observe as well a good agreement of the results of the RFA method with the poles obtained from the WM scheme for those cases where MD simulations were carried out. This shows (via WM) that the RFA can effectively be used to predict the structural properties of BHS mixtures over
a wide range of the phase diagram.

It should be noted that, apart from the damping coefficients and the oscillation frequencies, the amplitudes $A_{ij}^{(n)}$ [see Eq.\ \eqref{Eq:hr_reduced}] can be obtained from both the RFA and the WM scheme. As an illustration, Fig.\ \ref{fig:Aij} shows the $\eta_s$-dependence of the amplitudes corresponding to the first three poles in the case $\eta_b=0.05$ and $q=0.3$. For this rather disparate mixture, it can be  observed that $A_{bb}\sim 10 A_{sb}\sim 100 A_{ss}$. The general agreement between the RFA and WM results is fair, except for the amplitude $A_{bb}$ associated with the second pole ($\omega\sigma_b/2\pi\equiv\sigma_b/\lambda\approx 2$) when $\eta_s$ increases. Note, however, that this second pole is never the leading one. If one focuses on the leading and subleading  poles ($\sigma_b/\lambda\approx 1$ or $\sigma_b/\lambda\approx 3$), the agreement is very good.

Obviously, this complex behavior concerning the asymptotic decay of $h_{ij}(r)$ and the associated structural crossovers emerge as a consequence of the competition between the three basic length scales of the problem, namely $\sigma_s$, $\sigma_b$, and $\sigma_{sb}$. Another manifestation of
this competition appears when dealing with the sign of the first (local) minimum of $h_{ij}(r)$, here denoted as $h_{ij}^{\min}$, typically located at $r_{ij}^{\min}\approx2\sigma_{ij}$. The conventional expectation is $h_{ij}^{\min}<0$, thus signaling the beginning of the oscillations around $h_{ij}(r)=0$. In fact, this is what happens in monocomponent fluids.
However, given $\eta_b$ and $q$, it turns out that  $h_{ij}^{\min}>0$ if $\eta_s$ is smaller than a certain transition value. The loci $h_{ij}^{\min}=0$ separating the conventional behavior $h_{ij}^{\min}<0$ (above the locus) from the peculiar property $h_{ij}^{\min}>0$ (below the locus) are shown in Fig.\ \ref{fig:locus} for $q=0.2$, $0.3$, and $0.4$. Given a value of $q$, the locus $h_{ss}^{\min}=0$ envelops the locus $h_{sb}^{\min}=0$, and the latter envelops the locus  $h_{bb}^{\min}=0$. Additionally, the region with $h_{ij}^{\min}>0$ shrinks as $q$ increases. Moreover,  the curves $h_{sb}^{\min}=0$ and $h_{bb}^{\min}=0$ for the cases $q=0.3$ and $0.4$, as well as the curve $h_{ss}^{\min}=0$ for $q=0.4$, lose their meaning to the right of the crosses. This is because at a larger value of $\eta_b$, $h_{ij}^{\min}=h_{ij}(r\approx2\sigma_{ij})$ changes from being a negative local minimum to being negative, but not an extremum, as $\eta_s$ decreases.

The structural crossover phenomenon is illustrated in Fig.\ \ref{fig:crossover}, where the decay of $h_{bb}(r)$ at $\eta_b=0.05$ and (a) $\eta_s
= 0.1$ and (b) $\eta_s = 0.3$ for $q=0.3$ is shown. In agreement with the top rightmost panel of Fig.\ \ref{fig:q03}, the leading pole changes from being the first one (wavelength   $\lambda\approx\sigma_b$) at $\eta_s = 0.1$ to being the third one (wavelength   $\lambda\approx\sigma_s$) at $\eta_s = 0.3$, the transition taking place at $\eta_s\simeq 0.18$.
Additionally, Fig.\ \ref{fig:crossover}(a) shows that, at least for intermediate distances (say $1<r/\sigma_b<2$), the three leading poles are needed to capture the (large-wavelength) oscillations of $h_{bb}(r)$ at $\eta_s=0.1$. This situation becomes even more relevant as one approaches the transition value $\eta_s\simeq 0.18$ since the damping coefficients associated with the three first poles almost coincide near $\eta_s\simeq 0.18$ [see Fig.\ \ref{fig:q03}(b)]. However, at $\eta_s=0.3$, Fig.\ \ref{fig:crossover}(b) shows that the leading pole is enough to account for the (small-wavelength) oscillations, even for intermediate distances.

\section{Concluding remarks}
\label{concl}
The results we have presented in this paper deserve more consideration. First of all, it must be emphasized that the good agreement found between the results of the RFA method and those of the WM scheme in paper I \cite{PBYSH20} for a single value of the total packing fraction ($\eta=0.5$)
and size ratio ($q=0.648$) has been hereby confirmed. Therefore, we now have a powerful theoretical (almost completely analytic) tool to examine the complex behavior of the structural properties of BHS mixtures, including their asymptotic decay. In particular, we have found that, in the case
of the leading pole of the total correlation functions, given a value of $\eta_s$, the (reduced) damping coefficient $\alpha\sigma_b$ generally decreases with increasing $\eta_b$, while it exhibits a nonmonotonic dependence on $\eta_s$ at fixed $\eta_b$. Also, the influence of $q$ on $\alpha\sigma_b$ appears to be rather weak, although there is a general tendency
for a slight decrease  with increasing $q$ at fixed $(\eta_b,\eta_s)$.

On the other hand,
{in agreement with the work of Grodon \emph{et al.} \cite{GDER04,GDER05} (which considered a different approximation and a particular value of $q$), we have confirmed that there exists a  crossover line ($C$) in the plane $\eta_s$ vs $\eta_b$ separating a
region ($R_1$)  where the (reduced) natural oscillation frequency $\omega\sigma_b/2\pi\equiv\sigma_b/\lambda$ is of the order of $1$ from another region ($R_\nq$, with $\nq\approx 1/q$) where $\sigma_b/\lambda$ is of the order of $1/q$. The former extends to smaller values of $\eta_s$, while  the latter extends to larger values of $\eta_s$. Further, we have also found that, if $q$ is small enough, there is a second crossover line ($C'$) separating $R_\nq$ from a third region ($R_{\nq+1}$) where $\sigma_b/\lambda$ is of the order of the first harmonic of $\sigma_b/\lambda=1$ that turns out to be larger than $\nq$. This line $C'$ may terminate at an end point located at a large value of $\eta_s$ and a small value of $\eta_b$, which implies that one can move continuously between regions $R_\nq$ and $R_{\nq+1}$ by circumventing the end point from the left. However, except for small $q$, this line tends to lie above the region $\eta_s+\eta_b=0.5$, where the fluid phase is expected to coexist with the solid one. Finally, we have also shown that the above scenario can even be more complex, with additional crossover lines ($C'$ separating $R_\nq$ from $R_{\nq-1}$ and $C''$ separating $R_1$ from $R_{\nq-1}$) and splitting points,  as the total packing fraction $\eta=\eta_s+\eta_b$ increases or the size ratio $q$ decreases.}
One important issue, which remains to be assessed, is to understand the behavior of the crossover lines in the limiting region $\eta_b \to 0$.  It should be remarked that this region is hardly accessible by simulations but certainly needs to be studied in more depth.

To close this paper, two other outcomes of our work are worth pointing out. The first one concerns the value of the first local minimum in the total correlation  function $h_{ij}(r)$, which changes sign as one crosses a
certain transition line which depends on $q$ and the pair under consideration. This point is dealt with by Fig.\ \ref{fig:locus}. The second is that the direct correlation function $c_{sb}(r)$ is not monotonic in the region $0<r<\sigma_{sb}$ and presents a well-defined minimum. To our knowledge, this feature has not been pointed out before. Due to the fact that we are persuaded that this finding is relevant, we will address this point
and analyze it in detail in the following (third) paper of this series.\\

\begin{acknowledgments}
S.B.Y. and A.S. acknowledge financial support from  Grant No.\ PID2020-112936GB-I00/AEI/10.13039/501100011033 and from the Junta de Extremadura (Spain) through Grants No.\ IB20079 and No.\ GR18079, all of them partially financed by Fondo Europeo de Desarrollo Regional funds. S.P. is grateful to the  Universidad de Extremadura, where most of this work was carried out during his scientific internship, which was supported by Grant No. DEC 2018/02/X/ST3/03122 financed by the National Science Center, Poland. Some of the calculations were performed at the Pozna\'{n} Supercomputing and Networking
Center (PCSS).
\end{acknowledgments}



%

\end{document}